\documentclass[a4paper,11pt]{article}

\usepackage[left=2.5cm,right=2.5cm,bottom=2.5cm,top=2.5cm]{geometry}
\usepackage[english]{babel}
\usepackage[utf8]{inputenc}
\usepackage{microtype}
\usepackage{graphicx}
\usepackage{fancyhdr}
\usepackage{amsmath}
\usepackage[T1]{fontenc}
\usepackage[justification=RaggedRight, singlelinecheck=false]{caption}
\usepackage{float}
\usepackage[rightcaption]{sidecap}
\usepackage{siunitx}
\usepackage[citestyle=numeric-comp,sorting=none,backend=bibtex]{biblatex}
\usepackage{csquotes}
\usepackage{tabularx}
\usepackage{multirow}
\usepackage{subcaption}
\usepackage{wrapfig}
\usepackage{sidecap}
\usepackage{gensymb}
\usepackage{enumitem}
\usepackage{relsize}

\usepackage[symbol]{footmisc}

\makeatletter
\RequireBibliographyStyle{numeric}
\makeatother

\usepackage{pdfpages}

\usepackage{hyperref}
\hypersetup{
	pagebackref=true,
	colorlinks=false, %set true if you want colored links
	linktoc=section,     %set to all if you want both sections and subsections linked 
	citecolor=black, %choose some color if you want links to stand out
	filecolor=black,
	linkcolor=black,
	urlcolor=blue!80!black,
	urlbordercolor=blue,
	citebordercolor=red,
	linkbordercolor=red,
	pdfborder={0 0 0.5},
	pdfmenubar=false,
	pdftoolbar=false
	}

\makeatletter
\def\thickhline{%
	\noalign{\ifnum0=`}\fi\hrule \@height \thickarrayrulewidth \futurelet
	\reserved@a\@xthickhline}
\def\@xthickhline{\ifx\reserved@a\thickhline
	\vskip\doublerulesep
	\vskip-\thickarrayrulewidth
	\fi
	\ifnum0=`{\fi}}
\makeatother

\makeatletter
\renewenvironment{abstract}{%
	\if@twocolumn
	\section*{\abstractname}%
	\else %% <- here I've removed \small
	\begin{center}%
		{\bfseries \Large\abstractname\vspace{\z@}}%  %% <- here I've added \Large
	\end{center}%
	\quotation
	\fi}
{\if@twocolumn\else\endquotation\fi}
\makeatother

\newlength{\thickarrayrulewidth}
\title{Theoretical and Experimental Investigation into the flight of an X-Zylo}
\author{Nils Wagner}
\date{\today}

\begin{document}
	
%------------------------------------------------------------------
	
\begin{titlepage}
	\begin{center}
		\vspace*{1cm}
		
		\huge
		\textbf{Theoretical and Experimental Investigation into the flight of an X-Zylo}
		\vspace{0.8cm}
		%\large 
		%(PrePrint)
		
		\vspace{1.5cm}
		
		\Large
		%\textbf{Nils Wagner}\footnote[1]{Email address for correspondence: nils.wagner@tum.de}
		\textbf{Nils Wagner}\footnote[1]{This work begun as a project for the national science fair in Germany (Jugend forscht) and was later overhauled for publication. Email address for correspondence: nils-wagner-98@t-online.de}
		
		\vspace{0.6cm}
		%\large
		%\emph{Karl-Kneidl-Weg 2a, D–85386 Eching, Germany}
		
		\vfill
	
		\begin{abstract}
			\large
			\noindent 
			Flying Gyroscopes are fascinating flight objects, which, due to gyroscopic stabilization, can achieve surprisingly long flight distances when thrown with rapid spin. The most common example hereby is a traditional Frisbee disc. This paper focuses on a similar object called X-Zylo, that shows a remarkable straight flight despite its simple geometry. 
			
			The main aim of the present study is to investigate the flight behavior of the X-Zylo and to build a reliable groundwork for further quantitative parameter studies on ring wing configurations. To achieve this goal, a six degree of freedom model to predict the flight trajectory was developed. The trajectory computation uses interpolated high-fidelity CFD simulation data to calculate the acting moments and forces on the object during flight. A launch contraption was built to be able to validate the theory systematically and reproducible in experiments without human factors involved in the launch.
			
			Despite the complexity of the flight, the theoretical simulations match the real world data qualitatively, however quantitative differences still prevail. The investigation shows that the deviation between theory and experiment mostly stems from uncertainties in the CFD data as well as the optical recording of the experimental data. Despite the methods outperforming those of prior studies, advancements still have to be made in those areas in order to obtain better quantitative accordance between theory and experiment.
			
			%It is expected that advancements in those areas would entail an exceptionally accordance between theory and experiment.  
		\end{abstract}
		\vspace{1.5cm}
		
		
		%Trajectory simulation, Computational Fluid Dynamics, Frisbee, ring wing, gyroscopic stability
	\end{center}
\end{titlepage}
\thispagestyle{plain}
%\pagenumbering{Roman}

%------------------------------------------------------------------
	
\newpage
	
{\large \tableofcontents}
\thispagestyle{plain}
\pagenumbering{gobble}

%------------------------------------------------------------------

\newpage
\pagenumbering{arabic}

\section{Introduction}
Besides paper airplanes several flying toys have been developed, which, thrown the right way, can travel long distances through the air in a controlled manner. The ``X-zyLo\texttrademark'' (from now on simply called \emph{X-Zylo}) is a lesser-known example of such toys. Thrown like a football spinning along its long axis can result in flight distances up to $\SI{100}{m}$ and above. The distance record given by the manufacturer is $\SI{655}{ft}$ or $\SI{199.6}{m}$ \parencite{WMC600feet}.

\vspace{-0.15cm}
\noindent 
\begin{minipage}[t]{0.36\textwidth}
	\begin{figure}[H]
		\includegraphics[width=0.76\textwidth]{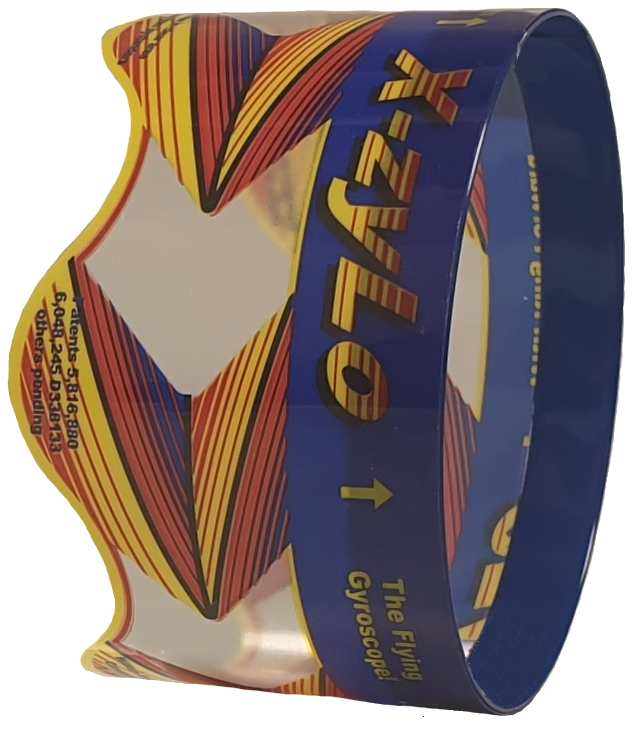}
		\caption{Picture of the X-Zylo}
		\label{fig:X-Zylo}
	\end{figure}
\vspace{0.25cm}
\end{minipage}
\hfill
%\begin{minipage}[t][5cm]{0.6\textwidth}
\begin{minipage}[t]{0.6\textwidth}
    \vspace{0.9cm}
	$\quad$ \textbf{\large Relevant Data:}\\
	
	%\vspace{0.3cm}
	$\quad$ mass: $m=\SI{22.73\pm0.16}{g}$
	
	$\quad$ length: $l=48.0-\SI{61.0}{mm}$ ($l_\text{avg}=\SI{54.5}{mm}$)
	
	$\quad$ diameter: $d_\text{outer}=\SI{97.0}{mm}$
	
	$\quad$ thickness: \SI{1}{mm} (leading edge), \SI{0.25}{mm} (trailing edge)
	
	$\quad$ center of mass: $\SI{1.22\pm0.01}{mm}$ behind leading edge
	
	$\quad$ special feature: weighted front, sinusoidal trailing edge
	        
	$\quad$ Reynolds number\footnotemark[1]: $\text{Re}\approx 7.5 \cdot 10^4$
	\vspace{0.3cm}
\end{minipage}
\footnotetext[1]{reference values: $v=\SI{17.35}{m/s}$, characteristic length is the ring's chord length $l_\text{avg}$}

\begin{figure}[H]
	\includegraphics[width=\textwidth]{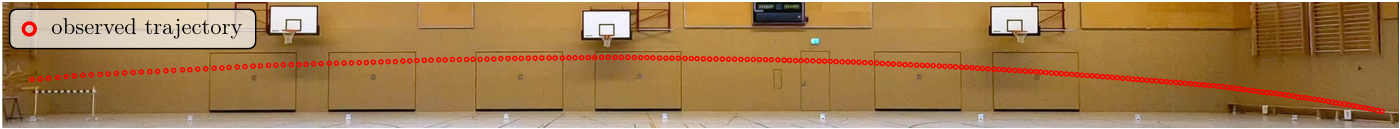}
	\caption{Typical trajectory for an X-Zylo launched with a small launch angle. Even though the ring flies more than \SI{40}{m}, the flight path is extraordinary flat.}
	\label{fig:TrajectoryFlatPic}
\end{figure}

\vspace{-0.15cm}
The object flies in an almost straight line and seems to lose very little height while airborne (see picture \ref{fig:TrajectoryFlatPic}). With the X-Zylo being simply a thin hollow cylinder as seen in figure \ref{fig:X-Zylo}, this flight characteristic is quite impressive. To understand the flight behavior in great detail, a six degree of freedom model was developed to compute the trajectory of an X-Zylo. To be able to accurately calculate the aerodynamic forces and moments acting on the object during flight, CFD simulations were conducted on a computing cluster. In order to test the prediction systematically, a launch device with the ability to launch the X-Zylo in a controlled manner was developed and build. The flight of the object is tracked using several cameras to obtain detailed flight information; this data is then corrected for camera induced errors. In the end the theoretical predictions are extensively compared to the observed data.  

Many papers already simulated the flight of a spin-stabilized disc, widely known as Frisbee\texttrademark, which shows a quite similar flight behavior \parencite{FrisbeeCrowtherPotts,FrisbeeHubbardHummel}. However, a less rigorous approach on the experimental part mostly defies a good comparison between theoretical and experimental trajectory. This work aims to resolve this issue by using a dedicated launch mechanism for better control on the initial flight parameters. The X-Zylo itself was also subject of former investigations \parencite{HirataXZylo,TarrXZylo}, but in less depth than in the present study. Future applications could involve the optimization of such toys as well as potential insights in annular airfoil technology experimentally used for coleopters in the past. Furthermore the understanding of the aerodynamics of such elementary objects could yield insight into the flow past more complex structures. 

In the following work those thin hollow cylinders investigated are referred to as \emph{throwing rings} or simply \emph{rings}. The ``X-Zylo'' is hereby only a particular, commercially available ring with the special features stated above, which was used for all experiments conducted.

%------------------------------------------------------------------

\section{Theory of Flight}
\label{sec:TheoreticalModel}

Two technical terms have to be differentiated, as they are of great importance for the initial flight and can be misunderstood easily. At first there is a specific \emph{launch angle}, which is the angle between the velocity vector at launch and its projection onto the $xy$-plane (note the used coordinate system in figure \ref{fig:coordinate-system}). It captures how steep the X-Zylo is thrown in respect to the ground. The second important angle is the \emph{Angle of Attack} (AoA). The AoA is the angle between the ring's symmetry axis and the velocity vector. It has to be emphasized that the initial AoA at launch and the launch angle are independent of each other; one can throw the ring very flat, however with its axis inclined to give it a high initial AoA.

As stated before, the characteristic of the X-Zylo is the stable and straight flight. This holds for every initial launch angle. The ring typically is thrown without an initial AoA since the axis of the ring matches the direction, in which the ring is launched at start. This is especially true for the launch system built, see section \ref{sec:HumanInducedLaunch}. This begs the question why the ring does not fall to the ground as any other object would, considering that the ring is virtually rotationally symmetrical, therefore generating no lifting force. The sinusoidal tracing edge of the X-Zylo breaks this symmetry, but even without this wavy edge the ring will fly nonetheless. For simplicity only idealized hollow cylinders with a straight trailing edge will be discussed in the following. Also the air is seen as stationary, therefore the ideal scenario is windless. %For the explanation on how lift emerges, the body-fixed reference frame proves useful, where the ring can be seen as stationary while the air moves past it.

	\subsection{First Drop and Equilibrium Phase}
	\label{sec:EarlyFlight}
	
\begin{figure}[h]
		\begin{subfigure}{0.5266\textwidth}
			\includegraphics[width=\textwidth]{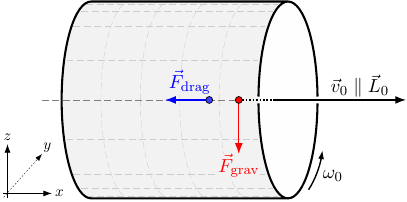}
			\caption{Initial launch condition (idealized, $\alpha_\text{launch}=\ang{0}$).}
			\label{fig:flight-start}
		\end{subfigure}
		%\hspace*{\fill} % separation between the subfigures
		\begin{subfigure}{0.4734\textwidth}
			\includegraphics[width=\textwidth]{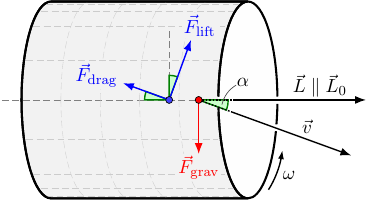}
			\caption{Ring mid-flight (equilibrium).}
			\label{fig:flight-later}
		\end{subfigure}
		\captionsetup{justification=justified}
		\caption{Simplistic model of the flight of a throwing ring without consideration of torques acting. While the idealized initial conditions show no angle between velocity vector and ring axis, during flight an AoA $\alpha$ is formed due to the gravitational acceleration.} 
		\label{fig:flight}
	\end{figure}

	At launch it holds true that the ring has an AoA of \ang{0}, meaning the symmetry axis as well as the flight direction are parallel (see figure \ref{fig:flight-start}). As the ring generates no lift force, it gets accelerated towards the ground due to gravity. This results in a direction change of the velocity vector, while the gyroscopic stabilization---due to the rapid spin imparted at launch---keeps the axis direction of the ring (nearly) constant. Therefore, even after a short amount of time an increasing AoA between the ring's flight direction as well as ring's axis will form shown in figure see figure \ref{fig:flight-later}. This imparts a linearly increasing lift force until the AoA is great enough to support the weight of the ring. This initial flight phase is further denoted as the \emph{first drop}. At a specific angle this lift force compensates the gravitational pull and the ring encounters an \emph{equilibrium phase}, where the AoA is stable. If the lift force exceeds the gravitational force, the ring gets accelerated upwards and therefore the AoA decreases, decreasing lift; and vice versa. Hence the ring flies straight for an elongated time since this equilibrium state is maintained. Drag decreases the velocity of the ring, and will slowly increase the \emph{equilibrium AoA} since a higher AoA is needed to generate the lift force equivalent to gravity. Ever-increasing AoA yields flow separation later during flight so that the ring plummets quickly.
	\newpage
	It has to be noted that when thrown by a human, the initial condition with \ang{0} AoA is not perfectly met, additionally the ring wobbles a lot and stabilizes only after a short amount of time \parencite{HirataXZylo}. This makes the flight hard to predict since the launch conditions are hard to measure accurately. The influence of an initial AoA is further investigated in section \ref{sec:HumanInducedLaunch}.
	
	Even though the flight can be explained well using those descriptions, which were formerly known (see \parencite{KaempfXZylo}), it can be observed that the ring drifts sideways later during flight. This effect will be explained further in the following section since it was mentioned in former publications (e.g. Tarr \parencite{TarrXZylo}) but not analyzed in detail.
	
	\subsection{Late Flight and Second Drop}
	\label{sec:lateFlight}
	
	Later during flight torques on the flying ring, that impact the direction in which the ring flies, become more important. There are several key components that decide how much the direction of the ring changes in the sideways $y$-direction:
	
	\begin{itemize}
		\item[i)] \textbf{Position of the Center of Pressure:} As \emph{Center of Mass} (COM) and \emph{Center of Pressure} (COP) do not fall together in a single point, a torque is imparted which lets the ring precess. For traditional airfoils it is known that the COP moves upstream for higher AoA \parencite[pp. 385+386]{schlichting2000aerodynamik}, which complicates the torque calculation. The CFD results in section \ref{sec:CFD_results} also confirm that the COP is not fixed for the X-Zylo during flight (see figure \ref{fig:results}). As the COP is usually found to be at approximately a quarter of the chord length for a flat plate (quarter chord point), the X-Zylo is designed to counter this by having a weighted front using a thin metal band. This shifts the COM towards the quarter chord point so that the acting torques are of small magnitude.
		\item[ii)] \textbf{Angular Velocity:} During launch the ring spins rapidly, but friction decreases this spin midst flight, so that the translatoric as well as the rotational speed decreases while airborne. This results in torques becoming more prominent later on since the angular momentum of the ring decreases over time. 
		\item[iii)] \textbf{Aerodynamic Forces:} Since the only forces which produce a net torque on the rotating cylinder are aerodynamic forces, the torque is directly proportional to the magnitude of lift and drag. As those forces are very small at launch due to the small AoA, the change in angular momentum is not visible. However, the AoA increases steadily over time magnifying aerodynamic forces. This is coupled to the decreasing translatoric velocity of the ring which in contrast decreases lift and drag. 
	\end{itemize}

	\begin{figure}[h]
		%\hspace*{\fill} % separation between the subfigures
		\begin{subfigure}{0.55\textwidth}
			\includegraphics[width=\textwidth]{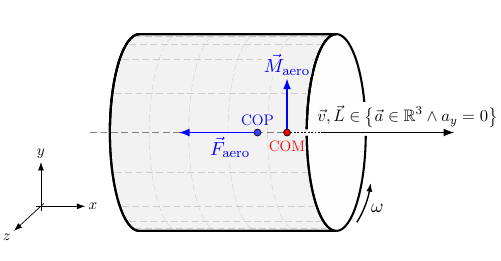}
			\caption{Initial cruising condition with torqe acting on the COM due to aerodynamic forces.}
			\label{fig:flight-start-drift}
		\end{subfigure}
		\hspace*{\fill} % separation between the subfigures
		\begin{subfigure}{0.39\textwidth}
			\includegraphics[width=\textwidth]{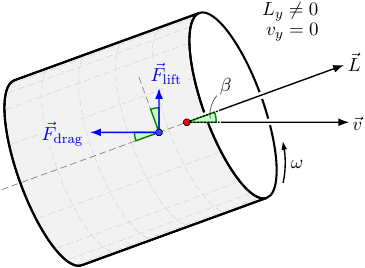}
			\caption{Ring condition after torque changes the direction of angular momentum.}
			\label{fig:flight-later-drift}
		\end{subfigure}
		%\hspace*{\fill} % separation between the subfigures
		\captionsetup{justification=justified}
		\caption{Simplistic model of how torques act on the ring during flight shown in birds-eye perspective. The sideways drift angle $\beta$ is exaggerated for better visibility. All vectors depicted in the picture are projections on the $xy$-plane, the $z$-component is not shown.} 
		\label{fig:flight-drift}
	\end{figure}
	
	As the aerodynamic forces act in the $xz$-plane at launch, the torque will purely act in the $y$-direction at first (see figure \ref{fig:flight-start-drift}). This torque slowly lets the ring precess, turning the ring sideways. However the velocity vector remains unchanged at first, only the ring precesses (see figure \ref{fig:flight-later-drift}). The tilt of the symmetry axis towards the flight direction and therefore towards the oncoming air generates a sideways lift force. Only then does the velocity vector follow the ring axis vector, letting the ring drift sideways, which can be observed. The direction in which the ring drifts is therefore dependent on the spin direction imparted at launch along with the location of the COP (behind or in front of the COM). It will become visible in section \ref{sec:ComparisonResults} that as the location of the COP changes, also the drift direction changes midst flight.
	
	A force not accounted for in this approach is the Magnus force acting on the spinning cylinder when swerving sideways, effectively creating a sideways incident flow component. This component however is negligibly small, only when dealing with stronger sideways winds, those forces have to be considered. In the ideal windless case, this factor can therefore be neglected.\\

%------------------------------------------------------------------

\section{Trajectory Calculation}

To predict the flight behavior a program was developed which approximates the trajectory numerically. A forward Euler method is chosen in which the solution is propagated using small discrete time steps; the implementation was done in MATLAB \parencite{MatLab}. The Euler method showed to be sufficient for this problem, as a test using Adams-Bashforth methods of the second and third order showed insignificant discrepancy between the results.

%\footnote{The commented source code as well as an executable can be downloaded using the following OneDrive link: \url{https://1drv.ms/u/s!AiJ_kf9sOZpEikttBzS5SXy-j16G?e=oYPDqf}}

	\subsection{Basic Mechanics}
	\label{sec:theoretical-calculation}
	
	Starting the calculation, some initial parameters have to be specified, for example the velocity magnitude at launch $v_\text{launch}$, the launch angle $\alpha_\text{launch}$, the launch height $h(t_0)$, and the angular velocity $\omega(t_0)$. Furthermore, the initial time is set to $t_0=0$ and a discrete time step $\Delta t$ chosen for the iterative forward Euler method. In addition, some ring parameters have to be specified, e.g. mass $m$, inner radius of the hollow cylinder $r_i$, outer radius $r_a$, and length $l$. Assuming a uniform mass distribution, the inertia of the hollow cylinder is then calculated to be $I=\frac{1}{2}m\left(r_i^2+r_a^2\right)$.
	\noindent
	\begin{minipage}[c]{0.48\textwidth}
		%\raggedright
		\hspace*{1em} The position of the tip of the ring (point on its axis in the plane of the leading edge) is named $P_\text{tip}$. All other used ring locations are named using $P$ with the point specified as subscript. The normalized ring axis direction vector is called $\vec{R}_\text{axis}$. For the launch conditions one gets
		\begin{align}
			\vec{v}(t_0)&=\begin{pmatrix} v_{\text{launch}}\cdot\cos(\alpha_{\text{launch}})\\0\\ v_{\text{launch}}\cdot\sin(\alpha_{\text{launch}}) \end{pmatrix}\, ,\\[8pt]
			\vec{P}_{\text{tip}}(t_0)&=\begin{pmatrix} 0\\0\\ h(t_0) \end{pmatrix}\, ,\\[8pt]
			\vec{R}_{\text{axis}}(t_0)&=\frac{\vec{v}(t_0)}{\left\lvert \vec{v}(t_0) \right\rvert} \, .\label{eq:RingAxisInitial}
		\end{align}
		
	\end{minipage}
 	\hfill
	\begin{minipage}[c]{0.43\textwidth}
		\begin{figure}[H]
			\includegraphics[width=\textwidth]{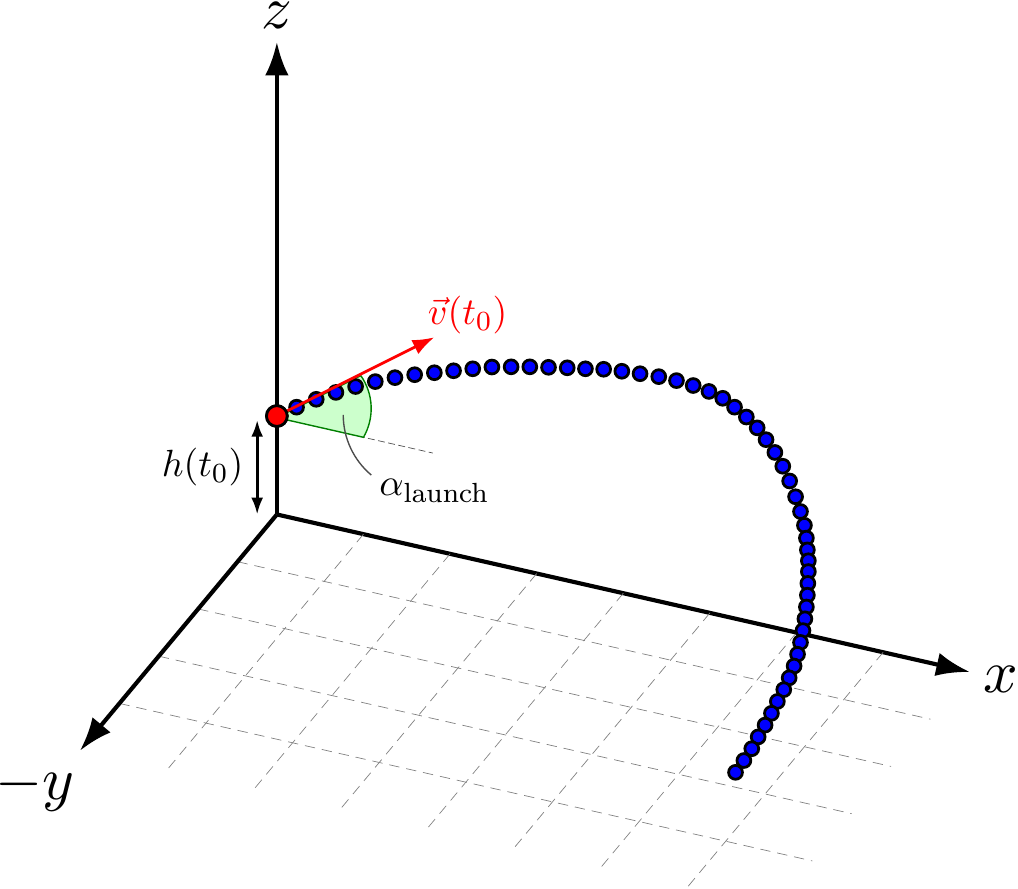}
			\captionsetup{justification=justified}
			\caption{Fixed coordinate system used throughout the work with a schematic trajectory.}
			\label{fig:coordinate-system}
		\end{figure}
	\end{minipage}
	\vspace{0.6cm}
	
	In equation \eqref{eq:RingAxisInitial} the simplification that at start $\vec{R}_\text{axis}$ is parallel to $\vec{v}(t_0)$ is used. Note that this is sufficiently true for the launch construction, in a human induced launch with its perturbations this will not hold true (see section \ref{sec:HumanInducedLaunch}). With the given distance between the COM and the tip, the position of the COM can be calculated by
	\begin{equation}
		\vec{P}_{\text{COM}}(t_0)=\vec{P}_{\text{tip}}(t_0)-\left\lvert \vec{P}_{\text{COM}}(t_0)-\vec{P}_{\text{tip}}(t_0) \right\lvert \cdot\vec{R}_{\text{axis}}(t_0) \, .
	\end{equation} 
	A loop variable $n$ is introduced and set to 0 initially. As long as $P_{\text{tip, \emph{z}}}(t_{n})>0$ holds, the ring is considered airborne and equations \eqref{eq:TotalAcc} to \eqref{eq:RotDecreaseEquation} are repeatedly solved for the next iteration until this condition is not satisfied any more.
	
	All current forces acting on the ring are summed up and used to calculate the velocity and positions for the next iteration. For the acceleration one gets
	\begin{equation}
		\vec{a}_{\text{tot}}(t_n)=\frac{\vec{F}_{\text{tot}}(t_n)}{m}=\frac{\vec{F}_{\text{lift}}(t_n)+\vec{F}_{\text{drag}}(t_n)+\vec{F}_{\text{grav}}(t_n)}{m} \label{eq:TotalAcc}
	\end{equation}
	with $\vec{F}_{\text{grav}}(t_n)=-mg\cdot \vec{e_z}$ using $g=\SI{9.81\pm 0.02}{m/s^2}$. The velocity and position of the COM for the next time step are then given by
	\begin{align}
		\vec{v}(t_{n+1})&=\vec{v}(t_n)+\Delta t\cdot\vec{a}_{\text{tot}}(t_n)\;\;\, ,\\
		\vec{P}_{\text{COM}}(t_{n+1})&=\vec{P}_{\text{COM}}(t_n)+\Delta t\cdot\vec{v}(t_{n}) \, . 
	\end{align}
	Section \ref{sec:aero} will focus on the magnitude of lift and drag as well as the COP; those quantities are interpolated from the CFD results. The direction of the lift and drag vector can be calculated using the Gram-Schmidt process resulting in
	\begin{subequations}
	\begin{align}
		\vec{F}_{\text{lift}}(t_n)&=\frac{\vec{R}_{\text{axis}}(t_n)-\frac{\vec{v}(t_n)}{\lvert \vec{v}(t_n) \rvert ^2}\cdot\left\langle \vec{R}_{\text{axis}}(t_n), \vec{v}(t_n)\right\rangle}{\left\lvert \vec{R}_{\text{axis}}(t_n)-\frac{\vec{v}(t_n)}{\lvert \vec{v}(t_n) \lvert ^2}\cdot\left\langle\vec{R}_{\text{axis}}(t_n), \vec{v}(t_n)\right\rangle \right\rvert} \cdot  \left\lvert \vec{F}_{\text{lift}}\big(\alpha(t_n), \left\lvert \vec{v}(t_n) \right\rvert\big) \right\rvert \, ,  
		\label{eq:liftDirection} \\[5pt]
		\vec{F}_{\text{drag}}(t_n)&=-\frac {\vec{v}(t_n)} {\lvert \vec{v}(t_n) \lvert} \cdot  \left\lvert \vec{F}_{\text{drag}}\big(\alpha(t_n), \left\lvert \vec{v}(t_0) \right\rvert\big) \right\lvert \, ,
		\label{eq:dragDirection}	
	\end{align}  
	\end{subequations}
	\noindent
	where $\langle \vec{a},\vec{b} \rangle$ denotes the standard scalar product in euclidean space of vectors $\vec{a}$ and $\vec{b}$. The lift and drag forces are dependent on the angle $\alpha(t_n)$ between the ring axis direction vector and the velocity vector. One gets
	\begin{equation}
		\alpha(t_n)=\arccos \left(\frac{\left\langle \vec{R}_{\text{axis}}(t_n), \vec{v}(t_n)\right\rangle} {\left\lvert \vec{R}_{\text{axis}}(t_n) \right\rvert \cdot \left\lvert \vec{v}(t_n) \right\rvert} \right) \, .
		\label{eq:angleOfAttack}
	\end{equation}
	Note that as stated in section \ref{sec:lateFlight}, the lift force does not always act in the $xz$-plane. After the ring tilts sideways this will be accounted for in the direction of the aerodynamic forces (\hyperref[ref{eq:liftDirection}]{\ref{eq:liftDirection}+\ref{eq:dragDirection}}) as well as the AoA \eqref{eq:angleOfAttack} used in the calculation of lift and drag.
	
	Due to torques acting on the ring the direction of the ring axis $\vec{R}_\text{axis}$ changes over time. With the angular momentum $\vec{L}(t_n)$ being parallel to the axis, one deduces
	\begin{equation}
		\vec{L}(t_n) = \left\lvert \vec{L}(t_n) \right\rvert \cdot\vec{R}_{\text{axis}}(t_n)= I\cdot \omega(t_n) \cdot \vec{R}_{\text{axis}}(t_n) \, .
	\end{equation}
	Of significant relevance is the position of the COP, which is evaluated using the interpolated CFD results. A stated before, COM and COP do not fall together, therefore creating a torque $\vec{M}(t_n)$ acting on the ring: 
	\begin{equation}
		\vec{M}(t_n)=\left( \vec{P}_{\text{COM}}(t_n)-\vec{P}_{\text{COP}}(t_n)\right) \times \left(\vec{F}_{\text{lift}}(t_n)+\vec{F}_{\text{drag}}(t_n)\right) \, . \label{eq:TorqueChange}
	\end{equation}
	This torque changes the angular momentum so that one gets $\vec{L}(t_{n+1})=\vec{L}(t_n)+\Delta t \cdot \vec{M}(t_n)$. The new direction of the ring axis $\vec{R}_{\text{axis}}(t_{n+1})$ is then the normalized angular momentum vector.
	
	For the sake of simplicity the magnitude of the angular frequency is seen as decoupled from the complicated motion of the X-Zylo. Using Sliding Mesh simulations (see section \ref{sec:CFD_Rotation}), the mean Wall Shear Stress $\tau_\text{w}$ is calculated for different rotational frequencies and interpolated for the trajectory simulation. Using only the angular velocity one can then calculate the additional torque acting on the ring, which in turn reduces its angular momentum using the simplified relation
	\begin{equation}
		\left\lvert \vec{L}(t_\text{n+1})\right\rvert=\left\lvert \vec{L}(t_\text{n})\right\rvert-\tau_\text{w}\big(\omega(t_\text{n})\big)\cdot A_\text{ring}\cdot \frac{\omega(t_\text{n})\cdot r_\text{a}}{\sqrt{\big (\omega(t_\text{n})\cdot r_\text{a}\big)^2+\left\lvert \vec{v}(t_\text{n})\right\rvert^2}}
		\label{eq:RotDecreaseEquation}
	\end{equation}
	\noindent
	with the rings surface area $A_\text{ring}$ and the ratio of the rotational velocity $\omega(t_\text{n})\cdot r_\text{a}$ to the total velocity on the rings surface. Mind that equation \eqref{eq:RotDecreaseEquation} only holds for small AoA as the ratio of rotational velocity and total velocity only holds for zero AoA flight. Therefore it is only a rough approximation for larger AoA scenarios later in flight.
	
	At last the loop variable $n$ gets incremented by one, the time incremented by $\Delta t$, and after that all steps will repeat.

	\subsection{Aerodynamic Forces}
	\label{sec:aero}
	
	As can be seen in the theoretical calculation in section \ref{sec:theoretical-calculation}, the magnitude of the lift and drag force as well as the COP is needed for any arbitrary angle $\alpha(t_n)$ and any flow velocity $\lvert \vec{v}(t_n) \rvert$. As all magnitudes change gradually, the approach will be to calculate lift and drag coefficient as well as the COP for discrete AoA using CFD (\textbf{C}omputational \textbf{F}luid \textbf{M}echanics) and then to interpolate the results. The change in Reynolds number and therefore the change in drag and lift coefficient for different flow velocities is neglected. As the velocity magnitude of the ring lies between \SI{6}{m/s} and \SI{18}{m/s} for the whole flight duration in most standard cases, this assumption should yield fairly accurate results. Nonetheless, this is a potential source of error one has to keep in mind, see section \ref{sec:OpenIssues}.
	
		\subsubsection{Theoretical Approximations}
		
		Several sources have derived analytical approximations for the lift and drag forces of hollow cylinder configurations which will be shortly mentioned and later compared to the CFD results in section \ref{sec:CFD_results}. For clarity, the lift and drag coefficients ($C_\text{D}$ and $C_\text{L}$) are written in square brackets as only the force equations are shown.
		
		One of the first analytical descriptions was done by Ribner \parencite{RibnerRingAirfoil}, who used concepts of Prandtl lifting-line theory in order to derive the lift force for a thin ring airfoil of diameter $d$ and (chord) length $l$. This resulted in the expression
		\begin{equation}
		F_\text{lift}=\underbrace{\left[\frac{\frac{2d}{\pi l}}{1+\frac{2d}{\pi l}}\cdot \pi^2 \alpha\right]}_{C_\text{L } [9]} \cdot \frac{\rho v^2}{2}  \cdot d l = \underbrace{\left[\frac{\pi^2 \alpha}{1+\frac{\pi\lambda}{2}}\right]}_{C_\text{L } [10]} \cdot \frac{\rho v^2}{2}  \cdot d l \,, \label{eq:liftRibnerPIVKO}
		\end{equation} 
		where $\lambda=l/d$ is the \emph{aspect ratio} between chord length and diameter of the ring, $S=dl$ the reference surface area for the coefficient calculation, $v$ the velocity magnitude of the oncoming air, and $\rho$ the air density. Mind differences in the definition of the reference wing surface area $S$ in different publications, sometimes $S'=\pi dl$ as well as $S''=2 dl$ are used. Moreover, in other literature the definition for the aspect ratio $\lambda$ is sometimes defined as the reciprocal fraction. Formula \eqref{eq:liftRibnerPIVKO} was also found by Pivko \parencite{PivkoRingfluegel}, who additionally calculated the induced drag force to be
		\begin{equation}
		F_\text{drag, induced}=\left[\frac{\lambda}{2}\cdot \left(\frac{\pi^2 \alpha}{1+\frac{\pi\lambda}{2}}\right)^2\right]\cdot \frac{\rho v^2}{2}  \cdot d l \label{eq:dragPIVKO}
		\end{equation} 
		\noindent
		using the found correlation $C_\text{D}=\lambda/2\,\cdot \,{C_\text{L}}^2$.
		
		Weissinger \parencite{WeissingerRingWing} developed a refined theory on general ring wing configurations and found the formulae \eqref{eq:liftRibnerPIVKO} and \eqref{eq:dragPIVKO} to be a special case for $\lambda\xrightarrow{} 0$. A more accurate approximation for the lift force for small aspect ratios ($\lambda < 5$) was found to be
		\begin{equation}
		F_\text{lift}= \left[\frac{\pi^2 \alpha}{1+\frac{\pi\lambda}{2}+\lambda\cdot \arctan(1.2\lambda)}\right] \cdot \frac{\rho v^2}{2}  \cdot d l \,. \label{eq:liftWeissinger}
		\end{equation} 
		For rotationally symmetric rings the drag can be calculated analogical as in formula \eqref{eq:dragPIVKO} using $C_\text{D}=\lambda/2\cdot {C_\text{L}}^2$ \parencite{WeissingerRingWing}. 
	
		Tarr derived another analytic expressions for lift and drag forces in his book ``What Makes The Amazing X-Zylo Fly'' \parencite{TarrXZylo}, specifically to examine the X-Zylo, yielding
		\begin{subequations}
			\begin{align}
			F_\text{lift}&=\left[4\pi\alpha\right] \cdot \frac{\rho v^2}{2} \cdot d l \, , \label{eq:liftTarr}\\
			F_\text{drag}&=\underbrace{4\pi \cdot \rho \cdot v^2 \cdot l^2 \cdot \alpha^2}_\text{induced drag}+\underbrace{\frac{\pi\cdot\mu\cdot v^{3/2}\cdot d \cdot l^{1/2}}{\nu^{1/2}}\cdot (S_\text{upper}+S_\text{lower})}_\text{viscous drag} \label{eq:dragTarr}
			\end{align}
		\end{subequations}
		where $\mu$ is the dynamic viscosity and $\nu$ the kinematic viscosity of air, $S_\text{upper}$ and $S_\text{lower}$ are the velocity gradients on the upper and lower surface. Those gradients were numerically computed by Tarr for an AoA of \ang{1.32} and a flow velocity of $\SI{15}{m/s}$, yielding $S_\text{upper}=0.3172$ and $S_\text{lower}=0.3466$. The same values were used for the comparison seen in figure \ref{fig:results} while neglecting changes in the velocity gradients based on the angle of attack $\alpha$ and the flow velocity $v$. 	
		
		For later comparison with the CFD results, the coefficients were calculated for a temperature of \SI{20}{\celsius}, yielding $\rho=\SI{1.204}{\kilogram\metre\tothe{-3}}$, $\mu=\SI{1.825e-5}{\kilogram\per\metre\per\second}$ and $\nu=\SI{1.516e-5}{\kilogram\per\square\metre}$.

		%----------------------------------------------------------

		\subsection{Computational Fluid Dynamics}
		
		To get reliable values for the aerodynamic forces as well as the COP, CFD simulations are used. It is undoubted that todays CFD solvers (when used correctly) can outperform even the best analytical models due to complex turbulence modeling and consideration of viscous effects. Therefore even for higher AoA beyond flow separation, approximate results can be obtained.
		
		\subsubsection{Geometry and Mesh}
\begin{figure}[H]
			\centering
			\includegraphics[width=.92\textwidth]{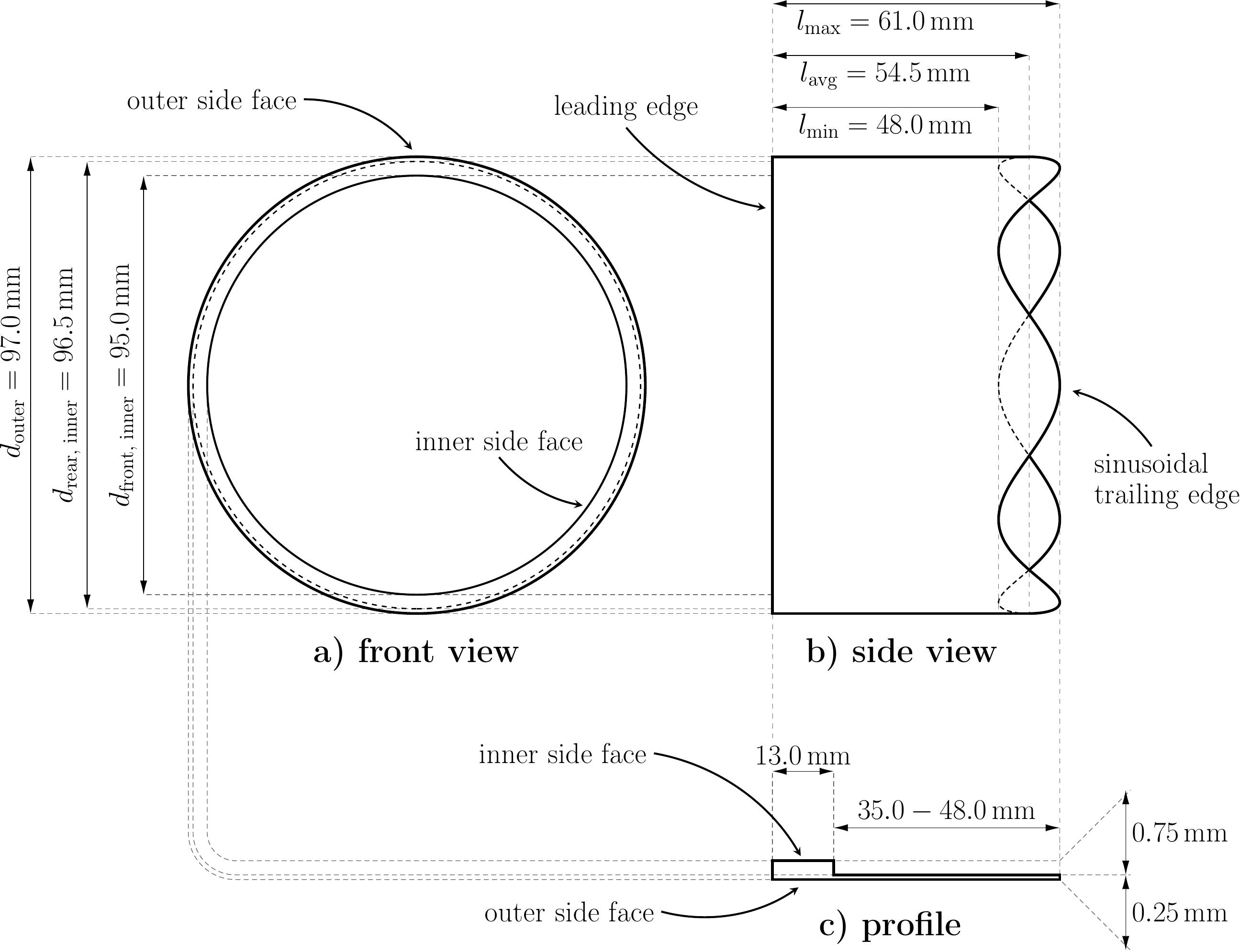}
			\caption{Details of the geometry of an X-Zylo. The thickness is scaled by a factor of 4 in the drawings for better visibility. The sinusoidal trailing edge consists of 5 full sine waves with ampitude \SI{13}{mm}. The profile drawing shows the lower part of the cross-section of the ring.}
			\label{fig:GeometryXZylo}
		\end{figure}

\begin{figure}[h]
			\begin{subfigure}{0.26\textwidth}
				\includegraphics[width=\textwidth]{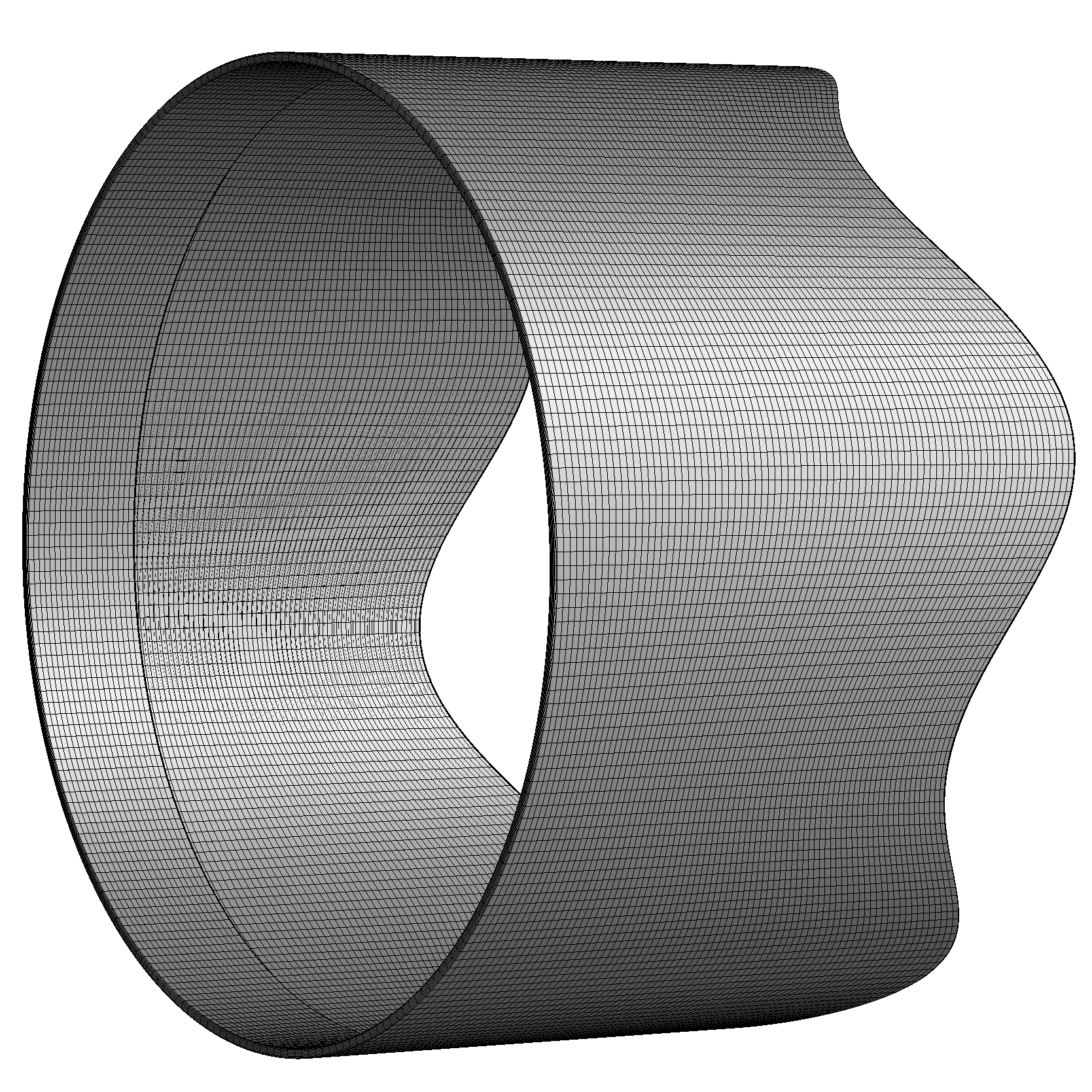}
				\caption{Surface mesh}
				\label{fig:2D-Mesh}
			\end{subfigure}
			%\hspace*{\fill} % separation between the subfigures
			\begin{subfigure}{0.43\textwidth}
				\includegraphics[width=\textwidth]{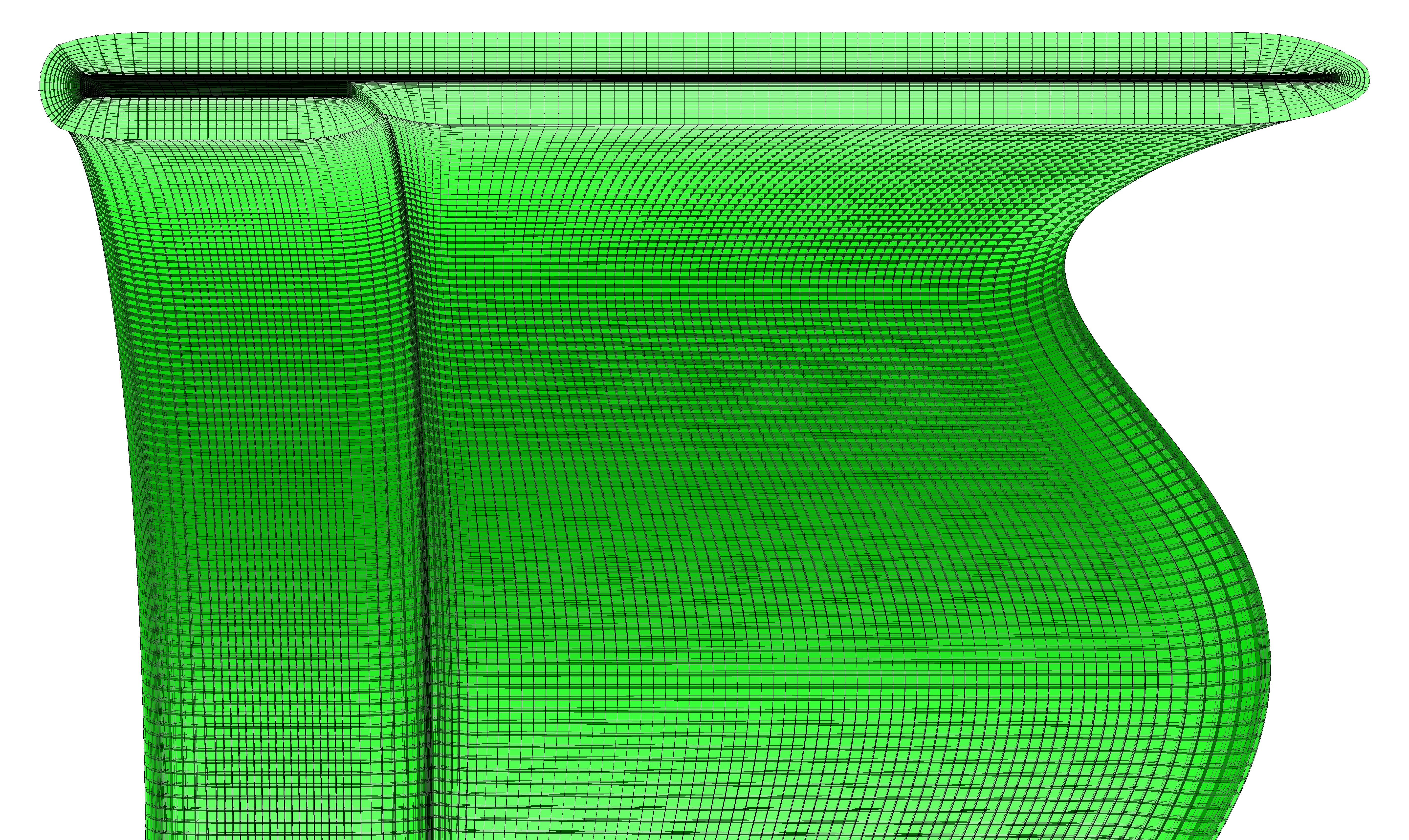}
				\caption{Boundary prism layers}
				\label{fig:Boundary-Mesh}
			\end{subfigure}
			%\hspace*{\fill} % separation between the subfigures
			\begin{subfigure}{0.26\textwidth}
				\includegraphics[width=\textwidth]{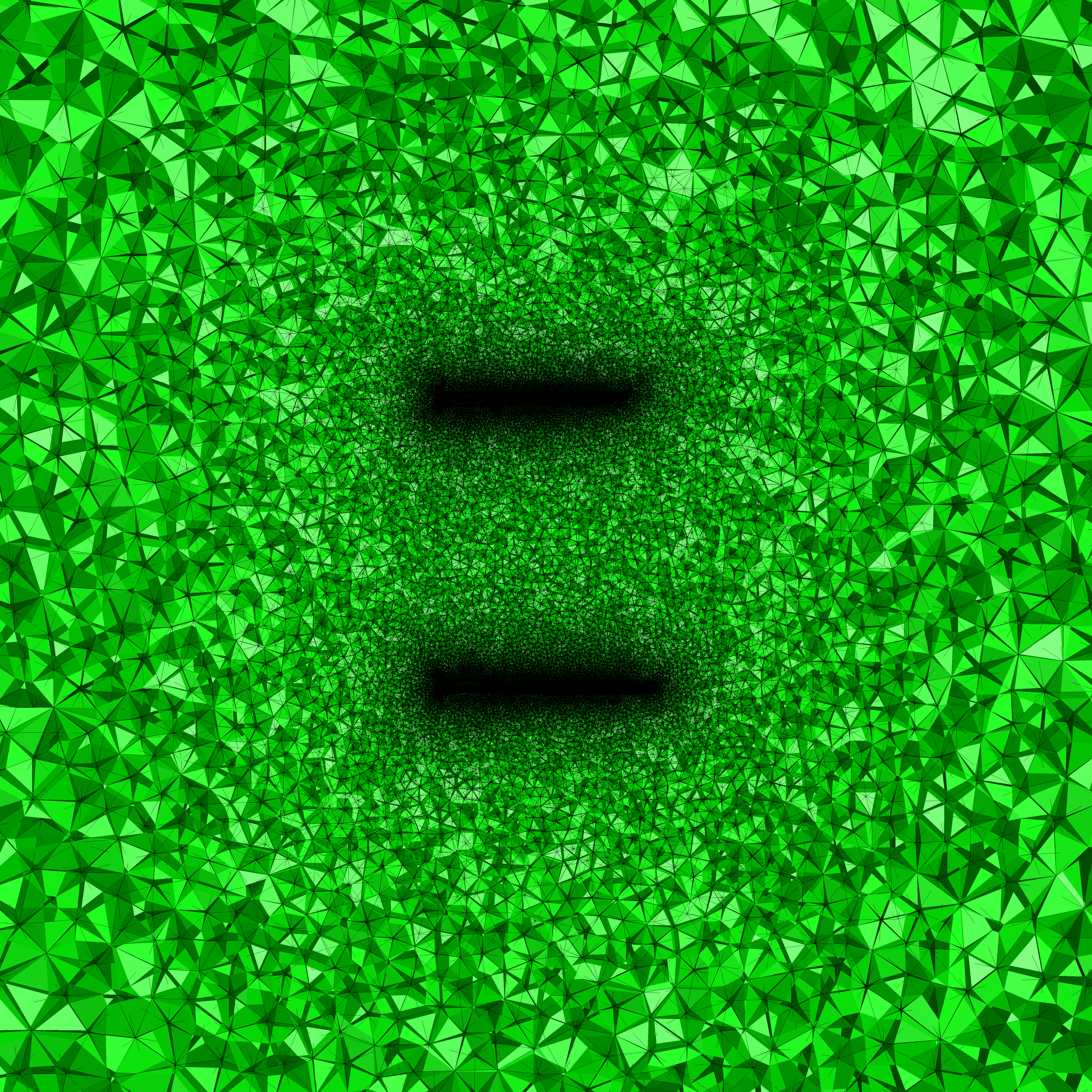}
				\caption{Volume mesh}
				\label{fig:3D-Mesh}
			\end{subfigure}
			\caption{Mesh for the X-Zylo (\textbf{M2}), which showed small error from finer meshes as well as a justifiable computing time.} 
			\label{fig:MESH}
		\end{figure}

		The geometry of an X-Zylo, which is used for all simulations other than the validation cases, is presented in detail in figure \ref{fig:GeometryXZylo}. All meshes are created in the same process. SALOME \parencite{SALOME} is used to create a structured quad surface mesh using quadrangle mapping, which can be seen in figure \ref{fig:2D-Mesh} for the X-Zylo. From there ANSYS Fluent \parencite{ANSYS_Fluent} is used to generate the volume mesh (see figure \ref{fig:3D-Mesh}) as well as the CFD calculations itself. The boundary prism layer consists out of 20 layers using a geometric layer height increase of 1.2. Hereby the initial height is set to satisfy $y+\approx0.5$, which for the X-Zylo results in an initial layer height of $\SI{0.01}{mm}$ resulting in a total boundary layer height of approximately $\SI{1.87}{mm}$ (see figure \ref{fig:Boundary-Mesh}). A sphere of radius $\SI{2.5}{m}$ is used as far-field; the volume mesh is an unstructured tetrahedral mesh. The simulations were conducted on the Linux-Cluster of the LRZ (\textbf{L}eibnitz-\textbf{R}echen\textbf{Z}entrum Garching, DE), utilizing the HPC resources to cut computing times.
		
		\subsubsection{Simulation Settings}
		
		The simulation settings were kept constant throughout the whole work, only specific setting that change from case to case (e.g. the validation cases) are mentioned separately in the respective paragraphs.
		For the simulation itself the transient pressure-based RANS (\textbf{R}eynolds-\textbf{A}veraged \textbf{N}avier \textbf{S}tokes) solver is used. The Pressure-Velocity-Coupling was set to use the SIMPLE (\textbf{S}emi-\textbf{I}mplicit \textbf{M}ethod for \textbf{P}ressure-\textbf{L}inked \textbf{E}quations) solver, the relevant discretisation schemes (pressure, density, momentum and energy) were set to second order. The ring itself was modeled to be a no slip wall while the sphere served as a pressure-far-field as inlet---therefore the air is modeled as an ideal gas. To initialize the flow field, ANSYS Fluent's standard initialization from the pressure-far-field was used. As all simulations were conducted transient, the time-step was set for the maximum CFL (Courant) number to be between 1 and 10, depending on the convergence of the problem. All obtained values are for a temperature of $\SI{20}{\celsius}$. 
		As turbulence modeling is a key component using the RANS solver, and many different models are implemented in ANSYS Fluent, two validation cases were simulated using a variety of turbulence models.
	
		\subsubsection{Validation}
		\label{sec:CFD_validation}
		
		There exist several sources that experimentally evaluate different forms of annular airfoils. Early experiments done by Fletcher \parencite{NASAAnnularWing} were performed using annular airfoils with Clark-Y cross-section and aspect ratios of 1/3, 2/3, 1, 3/2, and 3. Chord length and diameter varied from $\SI{0.235}{m}$  to $\SI{0.704}{m}$, the flow velocity used was $\SI{44.6}{m/s}$. Therefore the dimensions of the used models as well as the flow velocity greatly exceed the operating conditions of an X-Zylo with Reynolds numbers 11 to 33 times greater than in the present case. A more recent paper by Traub \parencite{TraubAnnularAirfoil} examines annular wings with an Eppler-68 section and aspect ratios 1/2 and 1. Those wings were merged into a NACA 0012 cross-section on the vertical sides allowing the Eppler-68 profile to be normal at the bottom and the top section. Also a revolution of a NACA 0012 profile ($\lambda=1/2$) was tested by Traub \parencite{TraubAnnularAirfoil,TraubAnnularAirfoil2}. The chord length used for every model was $\SI{0.1}{m}$, therefore the dimensions resemble the studied case in this work better. The free-stream velocity used in the wind tunnel was $\SI{40}{m/s}$, which is still well above the flow velocities for the X-Zylo, yielding a Reynolds number still 4.2 times greater than needed. Another experimental investigation was done by Latoine \parencite{LatoineRingWing} on small circular hollow cylinders with flat-plate cross-section. The largest cylinder had a diameter of $6\;\text{in}$ $(\approx\SI{0.152}{m})$ and an aspect ratio of $1/8$. Since the used flow velocity was only $\SI{12.2}{m/s}$, the Reynolds number based on the chord length is in this case 4 times smaller than that of an X-Zylo in flight. Additionally, in this work the aspect ratio of $\lambda=1/8$ is far from the ratio of an X-Zylo with $\lambda=0.56\,$. 
		
		Due to the lack of experimental data in the exact Reynolds number regime of the X-Zylo, both data sets by Traub (closed NACA0012 revolution) and Latoine were chosen to test the setup. The Reynolds number of the studied case then lies between the numbers of both validation cases. 
		\begin{figure}[h]
			\begin{subfigure}{0.49\textwidth}
				\includegraphics[width=\textwidth]{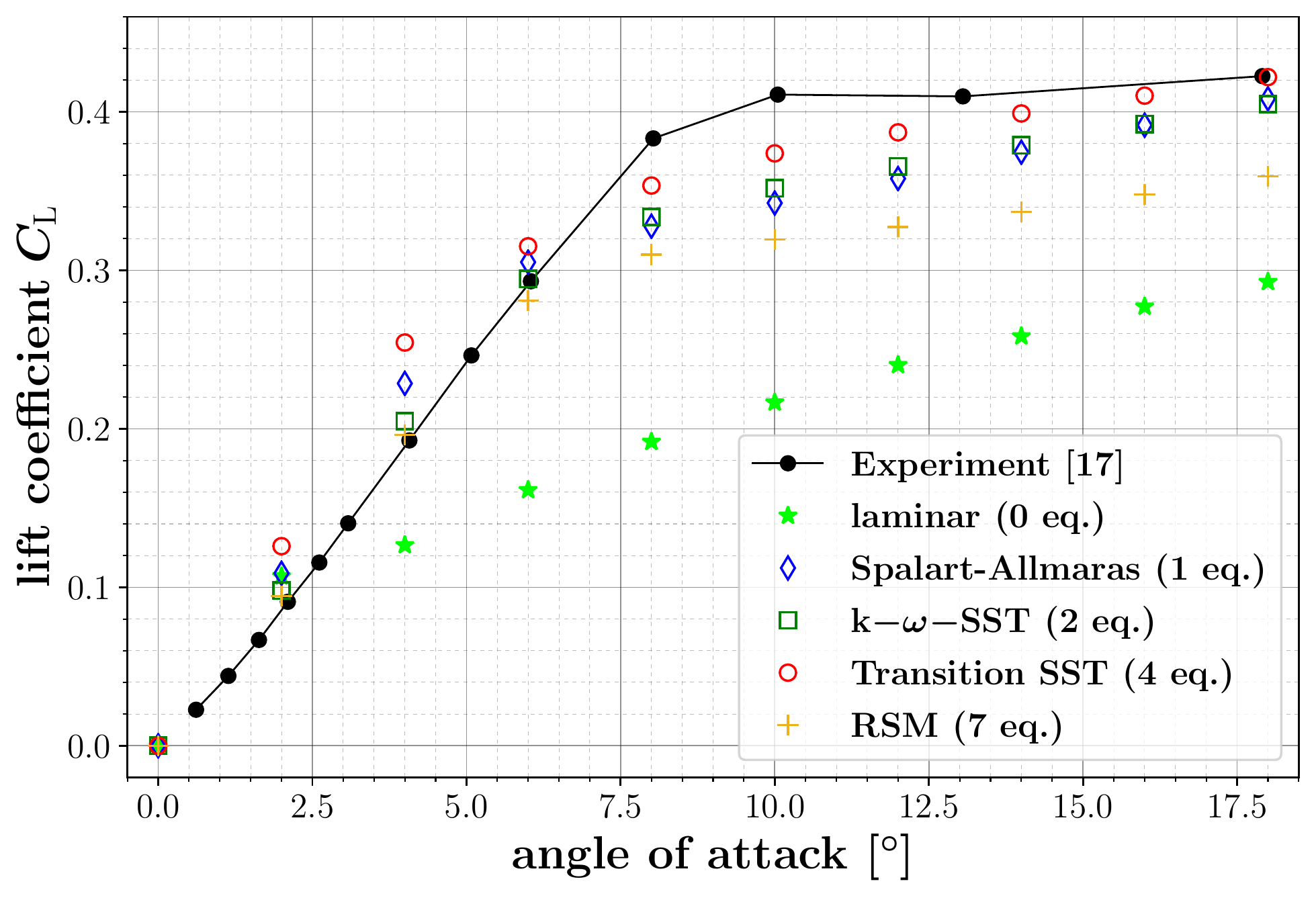}
				\caption{Comparison of the lift coefficient}
				\label{fig:Latoine_Lift}
			\end{subfigure}
			\hfill
			\begin{subfigure}{0.49\textwidth}
				\includegraphics[width=\textwidth]{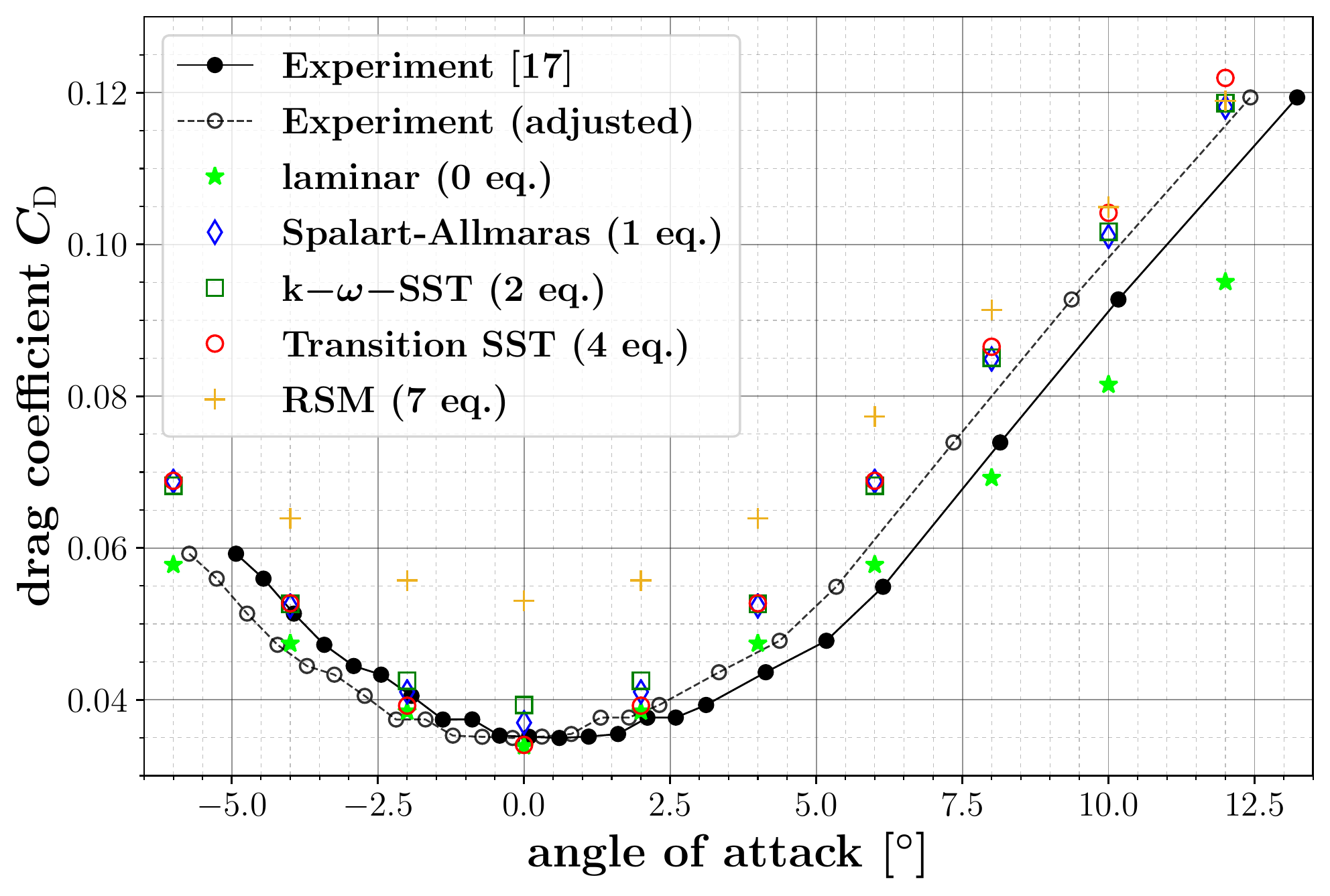}
				\caption{Comparison of the drag coefficient}
				\label{fig:Latoine_Drag}
			\end{subfigure}
			\caption{Comparison of the CFD results to the experimental results obtained by Latoine \parencite{LatoineRingWing}.} 
			\label{fig:Validation_Latoine}
		\end{figure}
	
		In figure \ref{fig:Validation_Latoine} the CFD results compared to the experimental data on the small hollow cylinder by Latoine \parencite{LatoineRingWing} are presented. The tested cylinder has a diameter of $\SI{0.152}{m}$ and a thickness of $\SI{0.51}{mm}$; the mesh consisted of 3.85 million cells. Wind tunnel experiments were performed with a wind speed of $\SI{12.2}{m/s}$ and a turbulence intensity of $0.02\%$; those parameters were also set in the simulation. From there 2000 time steps with a step size of $\SI{5e-4}{s}$ (CFL $\approx 10$) were calculated. Several turbulence models, which have great influence on the CFD solution, are compared. As the Reynolds number is very low, also a laminar solution is computed which fits the experimental data well for very small AoA up to \ang{2}. However, for larger AoA the solution diverges as small flow separations form, especially when comparing the lift coefficient. All simulations involving turbulence modeling can predict the large flow separation occurring at approximately \ang{8} onward, but overestimate the decrease in lift. The Transition SST (\textbf{S}hear \textbf{S}tress \textbf{T}ransport) model hereby is the best model while the \textbf{R}eynolds-\textbf{S}tress-\textbf{M}odel (RSM) greatly underestimates the lift force after separation. For smaller AoA this trend is reversed, RSM and k-$\omega$-SST can predict the lift slope accurately while Transition SST and \textbf{S}palart-\textbf{A}llmaras (SA) compute a lift slope far greater than captured in the experiment. In this low Reynolds regime the flow is not fully turbulent, which is an explanation for the deviation between the models. All models can predict the drag sufficiently well; except the RSM, which heavily over-predicts drag. The Transition SST model is especially good at capturing the drag at small AoA. It has to be noted that the drag should be fully axis-symmetric to an AoA of \ang{0} since the hollow cylinder investigated is symmetric, however the measured drag coefficient in the experiment showed a shift of approximately \ang{0.8} and is not symmetric. When adjusting this shift, the drag values fit the simulation data almost perfectly (see figure \ref{fig:Latoine_Drag}). From this validation case a laminar calculation as well as a simulation using the RSM turbulence model can be excluded. Therefore, the second validation case was only simulated using the remaining three turbulence models SA, k-$\omega$-SST and Transition SST.

\begin{figure}[h]
			\begin{subfigure}{0.49\textwidth}
				\includegraphics[width=\textwidth]{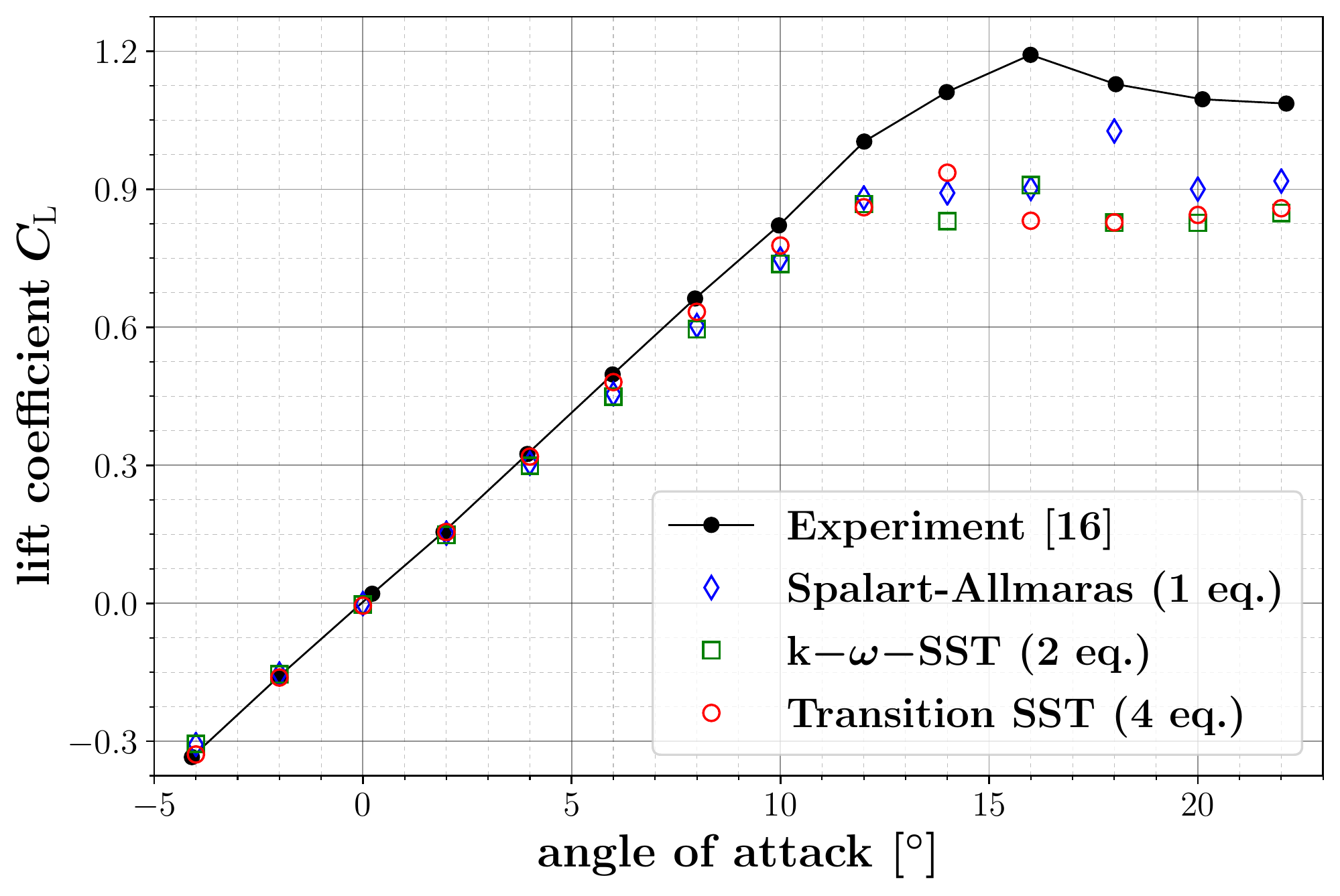}
				\caption{Comparison of the lift coefficient}
				\label{fig:Traub_Lift}
			\end{subfigure}
			\hfill
			\begin{subfigure}{0.49\textwidth}
				\includegraphics[width=\textwidth]{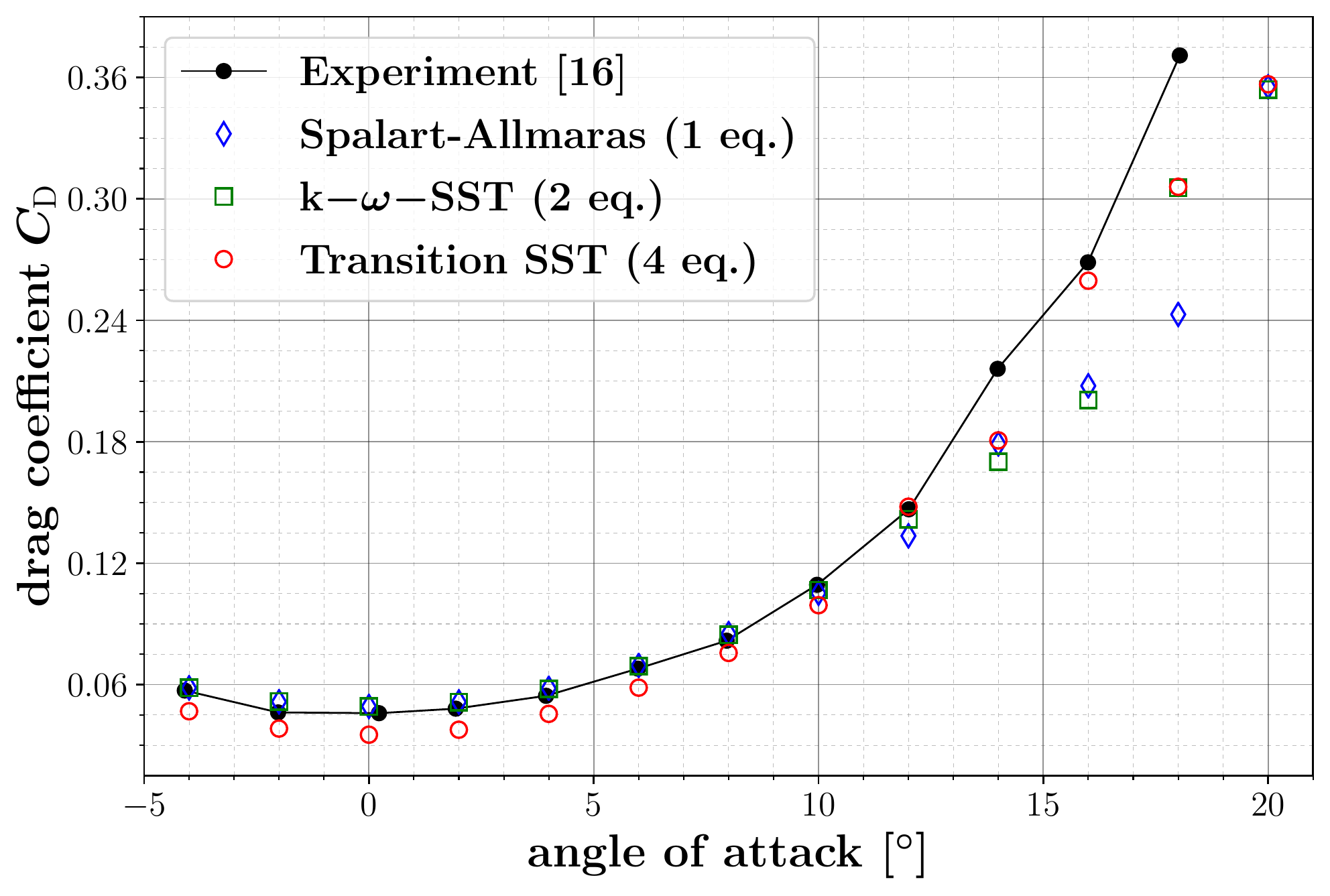}
				\caption{Comparison of the drag coefficient}
				\label{fig:Traub_Drag}
			\end{subfigure}
			\caption{Comparison of the CFD results to the experimental results obtained by Traub \parencite{TraubAnnularAirfoil2}.} 
			\label{fig:Validation_Traub}
		\end{figure}

		In the second validation case a closed NACA0012 revolution ($\lambda=1/2$, chord length $\SI{0.1}{m}$) including its mount was simulated at wind tunnel conditions of $\SI{40}{m/s}$ wind speed and a turbulence intensity of $0.24\%$ \parencite{TraubAnnularAirfoil2}. 3000 time steps with a step size of $\SI{1e-4}{s}$ (CFL $\approx 6$) were calculated. A finer mesh consisting of 8.21 million cells was used due to a more sophisticated geometry. The results are displayed in figure \ref{fig:Validation_Traub}. One can deduce that the setup works very well for low AoA where little flow separation is present. The lift slope deviates from the experimentally measured one by $8.9\,\%$ (SA), $10.2\,\%$ (k-$\omega$-SST) and $5.4\,\%$ (Transition SST). Calculating the average deviation of the drag coefficient up to an angle of \ang{10} one gets errors of $4.7\,\%$ (SA), $4.6\,\%$ (k-$\omega$-SST) and $15.9\,\%$ (Transition SST). For higher AoA all turbulence models struggle again due to large flow separations. 
			
		Both validations show, that the simulation can qualitatively predict the flow even with large flow separations occurring at higher AoA. From a quantitative standpoint the simulations are only valid for small AoA which is the expected behavior. Nonetheless, both SA and the $\text{k}-\omega-\text{SST}$ turbulence model produce good results in both validation cases. Transition SST showed good performance in specific situations (low AoA drag for low Reynolds number case), but was not as consistent. As the lift in the first validation case was best approximated using the k-$\omega$-SST model, it will be the choice for the simulations of the X-Zylo. It was later observed that for the case of the X-Zylo at small AoA, the Transition SST model predicts a drag coefficient 25\% smaller than that calculated with the k-$\omega$-SST model. As the drag coefficient in the small Reynolds number case was better approximated with Transition SST and the difference is significant, the calculations were repeated using the Transition SST model to compare both results with the found experimental behavior (see section \ref{sec:ComparisonResults}).
		
		As further validation a paper by Werle \parencite{WerleSummary} summarized most of the available experimental and CFD data to test the predictions by Weissinger \parencite{WeissingerRingWing}. It was shown that the lift slope was almost exactly predictable for all cases using Weissingers approximation \eqref{eq:liftWeissinger}. Therefore also this can be used as further validation for the obtained CFD results for the X-Zylo.

		\subsubsection{Mesh Independence Study}
		
		A small mesh independence study for the two SST turbulence models was done on four meshes with varying sizes, ranging from 2 to 17 million mesh cells, roughly doubling every step. The flow velocity in all further computations involving the X-Zylo was set to $\SI{17.35}{m/s}$ (Mach 0.05 at the pressure far-field) and a turbulence level of $1\%$ was used. Those values were estimated as the standard environment of an X-Zylo flying in a sports hall. For every mesh a CFL number of 1 was used in the calculation and \SI{0.5}{s} with an AoA of \ang{5} were simulated. The results obtained using the k-$\omega$-SST model are visible in table \ref{tab:MeshStudy}, the results for the Transition SST calculations are only summarized. It can be observed that especially drag and lift force can be well approximated using even the coarsest mesh. In contrast the error for the COP--- calculated from the finest mesh (\textbf{M4})---is still high using meshes \textbf{M1} or \textbf{M2}. Since the COP was seen less important than the aerodynamic forces, the coarse mesh (\textbf{M2}) was used in all further simulations as it was a good balance between simulation time and the mesh induced error. Later it was discovered that the COP is an extraordinary sensitive parameter and one should have opted for the fine mesh (\textbf{M3}), see section \ref{sec:SidewaysDrift}. The error for forces calculated using the medium mesh (\textbf{M2}) lie well below $5\,\%$ with both turbulence models ($0.3\,\%$ for k-$\omega$-SST, $1.4\,\%$ for Transition SST), therefore the influence of the turbulence model exaggerates the mesh induced error.
		\begin{table}[H]
			\centering
		\begin{tabular}{l||c|c|c|c|c}
			\thickhline
			\multirow{2}{*}{Mesh refinement} & \multirow{2}{*}{number of cells} & $C_\text{D}$ & $C_\text{L}$ & COP location & \multirow{2}{*}{\% error} \\ 
			&  & (\% error) & (\% error) & (\% error) & \\ \hline \hline
			\multirow{2}{*}{coarse (\textbf{M1})} & \multirow{2}{*}{2,268,018} & 0.08742 & 0.3714 & 22.99 \%& \multirow{2}{*}{3.69 \%} \\
			&  & (3.51 \%) & (0.57 \%) & (6.98 \%) &  \\ \hline
			\multirow{2}{*}{medium (\textbf{M2})} & \multirow{2}{*}{4,237,776} & 0.08474  & 0.3702  & 22.24 \% & \multirow{2}{*}{1.35 \%} \\
			&  & (0.33 \%) & (0.24 \%) & (3.49 \%)& \\ \hline
			\multirow{2}{*}{fine (\textbf{M3})} & \multirow{2}{*}{8,389,523} & 0.08404 & 0.3694 & 21.72 \%  &  \multirow{2}{*}{0.53 \%}\\
			&  & (0.50 \%) & (0.03 \%) & (1.07 \%) & \\ \hline
			very fine (\textbf{M4}) & 16,960,108  & 0.08446 & 0.3693 & 21.49 \% &  $-$ \\ \thickhline
		\end{tabular}
		\captionsetup{justification=justified}
		\caption{Results of the mesh independence study for the k-$\omega$-SST turbulence model. The error is calculated from the finest mesh (\textbf{M4}), the total error is the arithmetic average of the single deviations. The COP is given by the percentual location on chord. The same calculations were performed using the Transition SST Turbulence Model, obtaining slightly higher total error margins of $5.90\,\%$ (\textbf{M1}), $3.23\,\%$ (\textbf{M2}), and $1.16\,\%$ (\textbf{M3}). However, independent of the turbulence model, the largest error is always seen for the calculation of the COP.} 
		\label{tab:MeshStudy}
		\end{table}

		\subsubsection{Results}
		\label{sec:CFD_results}
		
		The CFD simulations for the X-Zylo were carried out using 5500 time steps (each \SI{4.4e-5}{s}, CFL 1.5) totaling a simulated flight time of about $\SI{0.24}{s}$. The mean over the last 500 time steps was calculated to average small numerical fluctuations as well as periodic effects. As noted before in section \ref{sec:CFD_validation}, both SST models show a significant discrepancy in their drag behavior for small AoA. Since neither solution can be disregarded using the validation cases, all computations were made for both turbulence models. While in both validation cases the drag estimates for small AoA using the Transition SST model were conceivably smaller than k-$\omega$-SST, in the small Reynolds number regime of the case by Latoine, this estimate fits the data better. As the Transition SST model is designed to operate well in the turbulence transition range where the X-Zylo mostly operates, it is not unlikely that it performs well in this case. The final conclusion can be drawn when the calculated trajectories using the CFD data are compared to the experimental results in section \ref{sec:ComparisonResults}.	
		\begin{figure}[H]
			\includegraphics[width=\textwidth]{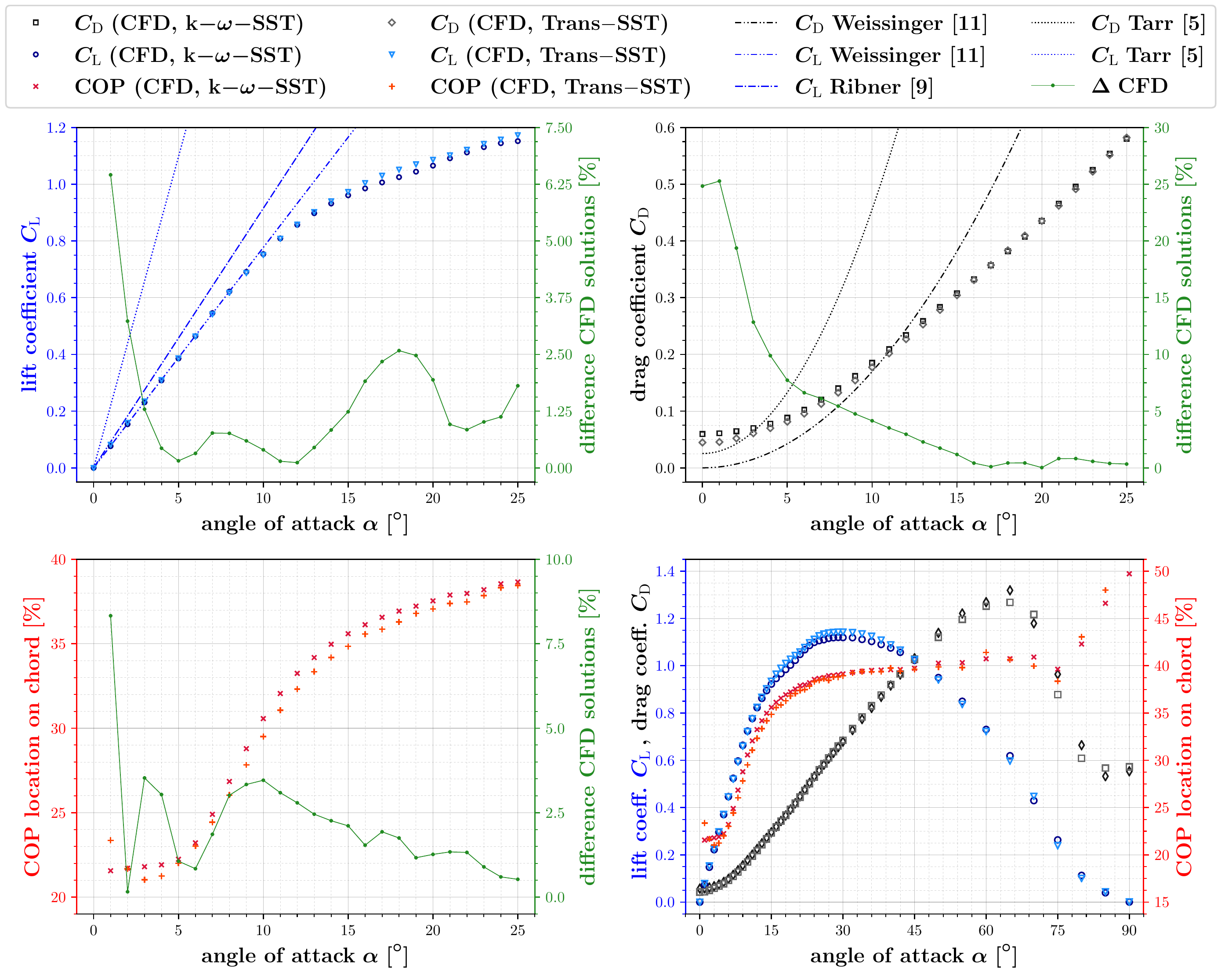}
			\captionsetup{justification=justified}
			\caption{Obtained CFD results using ANSYS Fluent \parencite{ANSYS_Fluent} in comparison to several analytic approximations. The dotted solid line graphs show the deviation of the Transition SST model to the k-$\omega$-SST model. All CFD data is interpolated using cubic splines to be used in the trajectory calculation in section \ref{sec:ComparisonResults}.}
			\label{fig:results}
		\end{figure}
		%\footnotetext[1]{Note that as Weissinger only calculates the induced drag of such a configuration the drag graph should start} 
		%\footnotetext{in the origin. For better comparison to the induced drag in the CFD simulations a viscous drag term equal to}
		%\footnotetext{that in the simulation is added.}
		The results of the simulations as well as the analytic approximations from several sources can be seen in 
		figure \ref{fig:results}. While the CFD results use a non-rotating X-Zylo with wavy trailing edge, the other sources are calculated for a thin ring wing with length $\SI{54.5}{mm}$ and diameter $\SI{97}{mm}$. As Hirata, et al. \parencite{HirataXZylo} calculated the drag and lift coefficient of a simplified model with a length of $\SI{60}{mm}$ and a diameter of $\SI{100}{mm}$ using motion analysis, the values are scaled accordingly to fit the real dimensions of an X-Zylo. It also has to be mentioned that source \parencite{HirataXZylo} only lists a single value for $C_\text{D}$ and $C_\text{L}$ at an AoA of \ang{2}. Therefore this data is not listed in figure \ref{fig:results}.
		
		At first, it can be seen that while both SST turbulence models have a different drag behavior at small AoA, they coincide well for higher AoA. The error for the lift force only exceeds 3\% for  AoA smaller than \ang{2}; the error for drag also plummets to under 10\% for AoA greater than \ang{4}. For the COP location also a small error of under 4\% for AoA greater than \ang{1} is seen. Altogether, the error gets smaller the higher the AoA gets with the exception being the lift coefficient. Here both turbulence models differ in the prediction of the flow separation, however not significantly. It can also be noted that the X-Zylo seems to behave like a traditional biplane wing with its lift curve showing two different slopes before and after large flow separation occurs.
		
		As can be seen the approach by Tarr \parencite{TarrXZylo}---equations \eqref{eq:liftTarr} and \eqref{eq:dragTarr}---greatly overestimates the lift and induced drag force generated by the X-Zylo, while underestimating viscous drag. For the lift slope the deviation to the CFD results is more than $180\%$, also the viscous drag is only half of what was computed (42\% for k-$\omega$-SST, 56\% for Transition SST). The approach by Hirata, et al. \parencite{HirataXZylo} also underestimates drag and lift forces. When comparing the sole data point given to the Fluent simulation a difference of $40\%$ for the lift force (both models) and $41\%$ (k-$\omega$-SST) or  $27\%$ (Transition SST) difference for the drag force is calculated. The analytical approximations of the lift force found by Ribner \parencite{RibnerRingAirfoil}, Pivko \parencite{PivkoRingfluegel} and especially Weissinger \parencite{WeissingerRingWing} coincide well with the simulation data obtained. The difference in lift slope of the refined formula \eqref{eq:liftWeissinger} by Weissinger and the CFD simulation is less than 1\% for both SST models which is remarkable. This also serves as additional validation since formula \eqref{eq:liftWeissinger} was found to be very precise before by Werle \parencite{WerleSummary}.
		Nonetheless, only Tarr approximates viscous drag of the X-Zylo, therefore the drag curve for the Weissinger approximation starts in the origin. The induced drag force calculated using formula \eqref{eq:dragPIVKO} with the correction term for $C_\text{L}$ by Weissinger also matches the induced drag calculated using CFD for an AoA smaller than \ang{5}, but diverges from the obtained simulation results for angles greater than \ang{5}.		
		
		\begin{figure}[H]
			\includegraphics[width=\textwidth]{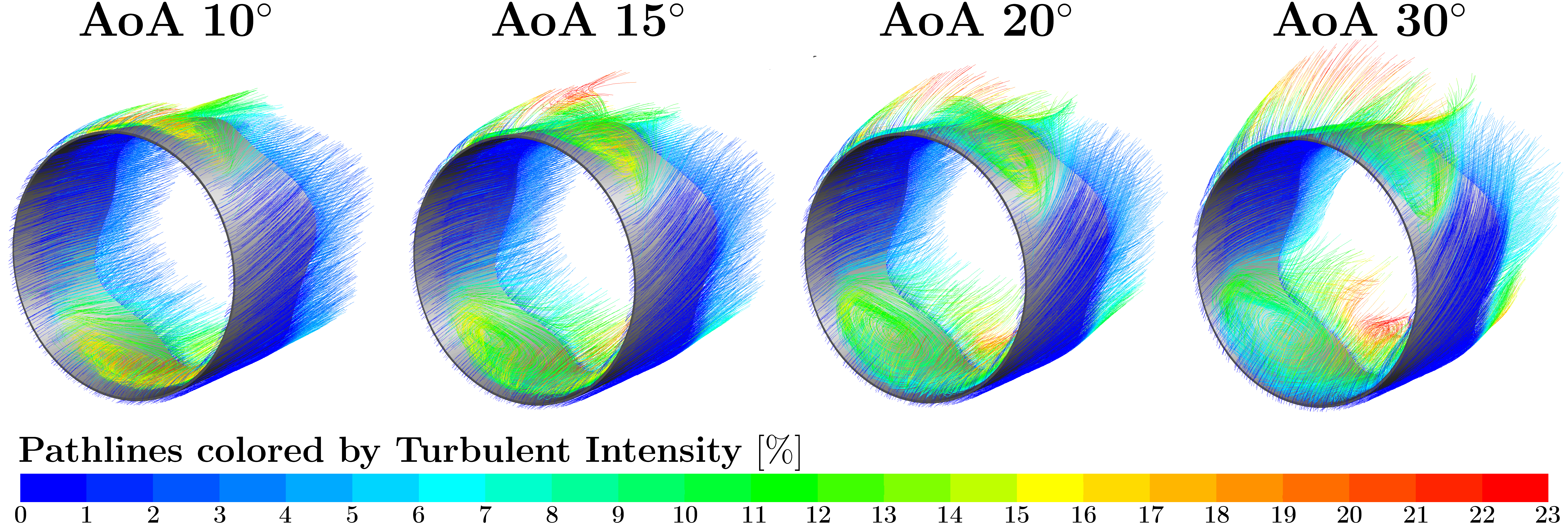}
			\captionsetup{justification=justified}
			\caption{Pathlines around a non-rotating X-Zylo showing the turbulent intensity along their path for different AoA. The re-circulation zones are clearly visible with their size increasing for rising AoA.}
			\label{fig:pathlines}
		\end{figure}
	
		Figure \ref{fig:pathlines} shows pathlines colored by turbulence intensity around the X-Zylo for high AoA to qualitatively capture the flow separation. It can be seen that as the flow slowly separates in the center of the upper and lower wing segment, the side flow swirls into this area due to the lower pressure region created. For AoA smaller than \ang{10} this effect is small and the separation bubble is only concentrated to the center portion of the wing while growing outwards for higher AoA. As the study of the flow separation is not a focus in this paper, this will not be investigated further.

		\subsubsection{Influence of the Model's Rotation}
		\label{sec:CFD_Rotation}
		
		%\noindent
		%\begin{minipage}[c]{0.5\textwidth}
		%	\raggedright
		%	To account for the rotation of the X-Zylo during flight a Sliding Mesh approach was taken to simulate the ring with different rotational frequencies. The simulation were conducted for an angle of attack of \ang{5} using the $\text{k}-\omega-\text{SST}$ turbulence model, the results can be seen in figure \ref{fig:Results_Rotation} . Especially the influence of the rotation towards the drag is not negligible for higher frequencies as the rotational frequency of an X-Zylo shot by the launch mechanism (section \ref{sec:????}) can exceed 40 revolutions per second. It is seen that the increase in drag stems from a longer distance traveled by the air over the chord as the air is deflected near the surface by the rotating ring. This also increases lift and shifts the center of pressure, however this influence is relatively small.\\
			
		%	\noindent
		%	The interpolation for $C_\text{D}$ and $C_\text{L}$ was done using a cubic polynomial, the COP fitting only uses a quadratic polynomial. This interpolation is then used in the program to simulate the trajectory.  
		%\end{minipage}
		%\hfill
		%\begin{minipage}[c]{0.45\textwidth}
		%	\begin{figure}[H]
		%		%\includegraphics[width=\textwidth]{Results_X_Zylo_Rotation}
		%			\caption{Results for a rotating X-Zylo using a Sliding Mesh approach}
		%		\label{fig:Results_Rotation}
		%	\end{figure}
		%\end{minipage}
		%\vspace{0.6cm}
		
\begin{wrapfigure}{r}{0.53\textwidth}
			\vspace{-10pt}
			\centering
			\includegraphics[width=0.5\textwidth]{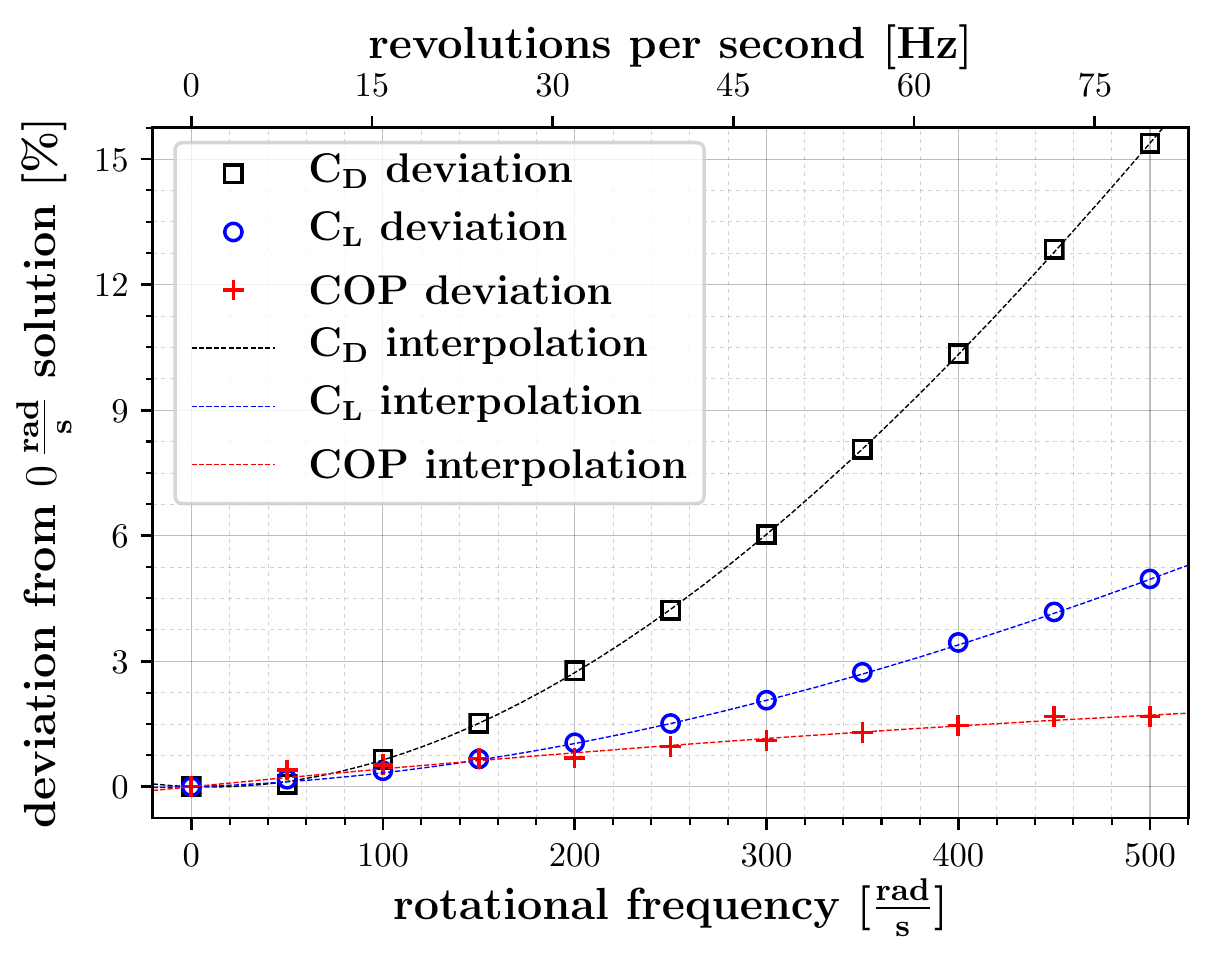}
			\vspace{-4pt}
			\centering
			\captionsetup{margin=0.3cm}
			\caption{Results for a rotating X-Zylo at an \\ AoA of \ang{5} using a Sliding Mesh approach.}
			%\vspace{-70pt}
			\label{fig:Results_Rotation}
		\end{wrapfigure}

		To account for the rotation of the X-Zylo during flight, a Sliding Mesh approach was used to simulate the ring with different rotational frequencies. The simulations were conducted for an AoA of \ang{5} using the k-$\omega$-SST model and a mesh containing 5.6 million cells, the results can be seen in figure \ref{fig:Results_Rotation}. The percentual difference between the non-rotating solution and the simulation with non-vanishing 
		rotational frequency is shown. Especially the influence of the rotation towards the drag is not negligible as the rotational frequency of an X-Zylo shot by the launch mechanism can exceed 50 revolutions per second (see table \ref{tab:LaunchComparisonStats}). It is seen that the increase in drag stems from a longer distance traveled by the air over the chord as the air is deflected near the surface by the rotating ring. This also increases lift and shifts the COP, however this influence is relatively small. The interpolation for $C_\text{D}$ and $C_\text{L}$ was done using a cubic polynomial, the COP fitting only uses a quadratic polynomial. Those interpolations are then used in the program to simulate the trajectory. 
		Moreover, the wall shear stress was captured for the rotating X-Zylo to calculate the decrease in angular velocity over time for different rotational frequencies (see equation \eqref{eq:RotDecreaseEquation}). \\

%------------------------------------------------------------------
\section{Launch Construction}
\label{sec:LaunchMechanism}
\vspace{-12pt}
\begin{figure}[h]
	\includegraphics[width=1\textwidth]{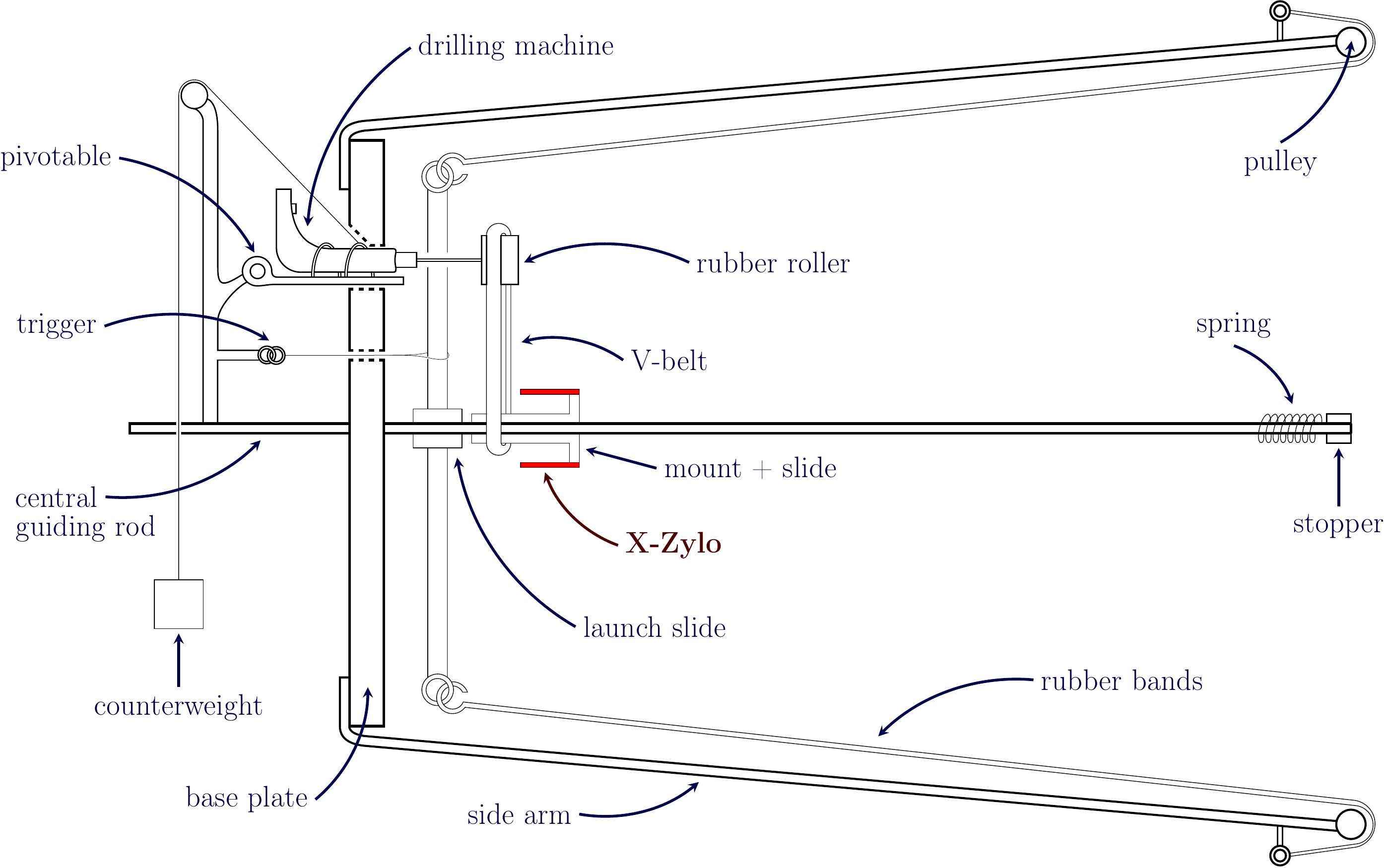}
	\caption{Schematic of the launch construction. The different parts are labeled as well as the actual object for launch marked in red. The functionality of the parts is covered in the text.}
	\label{fig:schematic}
\end{figure}
\begin{figure}[h]
	\includegraphics[width=\textwidth]{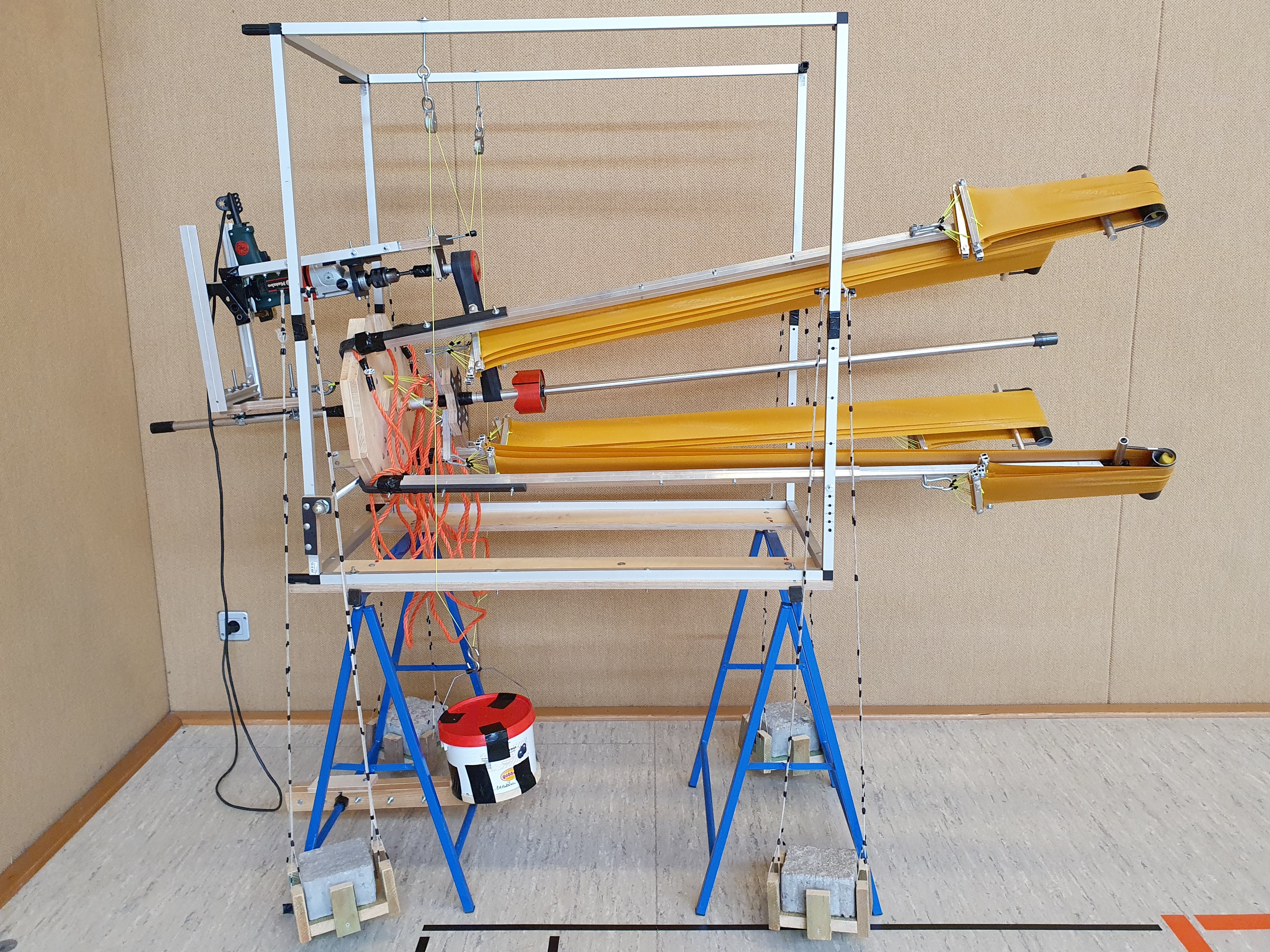}
	\caption{Picture of the cocked launch device which was developed and constructed for this project. The X-Zylo colored in red sits on a mount which is then accelerated forward by rubber bands. The rotation is achieved using a drill and a V-belt.}
	\label{fig:whole-Constr}
\end{figure}

\noindent
To be able to throw the ring in a reproducible and controlled fashion, a launch device was constructed, which can be seen in picture \ref{fig:whole-Constr}. Especially the swerving motion of the ring at launch when thrown by a human as well as a non-vanishing initial AoA (see section \ref{sec:HumanInducedLaunch}) complicate the early flight behavior.
\subsection{Explanation of the used Mechanism}
\vspace{-6pt}

\noindent
To explain the mechanism in a nutshell, schematic \ref{fig:schematic} shows the most important parts. The ring sits loosely on a mount which can be altered for different ring geometries. This mount can slide on a central guiding rod which is oriented in the direction the ring should be launched. The initial launch angle can be chosen between \ang{0} and \ang{25}. The whole mount will be spun by a V-belt connected to a drilling machine, a counterweight maintains tension in the belt. When the mount spins at the desired speed, the weight is lifted, releasing the tension in the belt. The trigger can be pulled to release the cocked launch slide, catapulting the launch slide as well as the mount forwards using rubber bands (Thera-Band Gold). A stopper at the end of the guiding rod stops the mount spontaneously and releases the X-Zylo; a spring dampens the hit on the stopper. To evaluate the velocity and angular frequency of the ring at launch, a slow motion video of the ring is captured (see section \ref{sec:ExperimentalCorrections}). A scale in the $xz$-plane is set up to calculate the initial velocity from the video footage. When fully cocked, the launch mechanism has a draw force of approximately $\SI{1100}{N}-\SI{1600}{N}$, depending on the number of rubber bands used on each side arm.

\subsection{Reproducibility of the Device}
\label{sec:Reproducibility}
	
\begin{figure}[H]
	\includegraphics[width=1\textwidth]{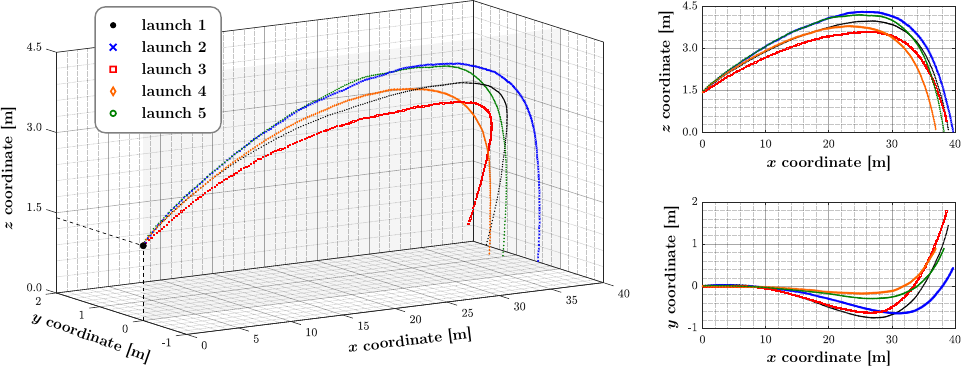}
	\caption{Comparison of the trajectory of five launches with the launch contraption. To test the reproducibility of the device all launches were done using the same settings. The positions shown are the COM locations.}
	\label{fig:LaunchComparison_tot}
\end{figure}

The reproducibility of the launch construction was tested by launching the X-Zylo five times using the same settings and comparing the initial launch values (table \ref{tab:LaunchComparisonStats}) as well as the observed trajectories (figure \ref{fig:LaunchComparison_tot}). Full reproducibility would be achieved if the standard deviation of the launch parameters would be smaller than the estimated uncertainty on each individual parameter
\noindent
evaluated using the camera setup (see section \ref{sec:ExperimentalCorrections}). This would then cut the uncertainty on initial parameters due to the reliability of the construction. 

From the trajectories it is visible that the goal of achieving reproducibility with the launch mechanism was not met. While the flight distance varies only slightly, especially the sideways drift is severely different for the launches. When comparing the data shown in table \ref{tab:LaunchComparisonStats}, it becomes obvious that the disagreement in sideways drift stems from the huge difference in the initial rotational frequency. This problem can be traced back to the belted motor drive mechanism. The counterweight has to be lifted manually in order to lower the drill, releasing the tension of the V-belt. In the same instance the trigger has to be pulled. However, this is hardly possible as the trigger mechanism is poorly operable. Therefore, the tension in the V-belt is often released before the trigger can be pulled, decreasing the rotation of the mount rapidly. Table \ref{tab:LaunchComparisonStats} shows that good reproducibility is almost achieved for the launch velocity, the launch angle still shows a high standard deviation due to the outlier launch 3.
\begin{table}[h]
	\centering
	\begin{tabular}{l||c|c|c|c|c}
		\thickhline
		& initial velocity& launch angle & init. rot. frequency& launch& time \\ 
		& $v_\text{launch}$ [m/s]& $\alpha_\text{launch}$ [$^\circ$]&  $\omega(t_0)/(2\pi)$ [Hz] & distance [m] & aloft [s] \\  \hline \hline
		launch 1 & $15.9 \pm 0.2$ & $11.8\pm 0.3$ & $18.5\pm0.4$  & $38.9\pm 0.2$ & $4.33\pm 0.07$ \\  \hline 
		launch 2 & $16.6\pm 0.4 $ & $12.3\pm 0.5$ & $38\pm2$ & $39.7\pm 0.3$ & $4.47\pm 0.07$\\  \hline 
		launch 3 & $15.8\pm 0.3 $ & $10.2\pm 0.4$ & $16\pm1$ & $38.6\pm 0.3$ & $4.27\pm 0.07$\\  \hline 
		launch 4 & $15.4\pm 0.4 $ & $12.6\pm 0.5$ & $63\pm3$ & $37.0\pm 0.3$ & $4.43\pm 0.07$\\  \hline 
		launch 5 & $16.1\pm 0.2 $ & $13.0\pm 0.4$ & $60\pm3$ & $38.3\pm 0.3$ & $4.47\pm 0.07$\\  \hline \hline 
		median & $16.0\pm 0.4 $ & $12.0\pm 1.1$ & $39\pm22$ & $38.5\pm 1.0$ & $4.39\pm 0.09$\\  \hline 
		\thickhline
	\end{tabular}
	\captionsetup{justification=justified}
	\caption{Comparison of five launches using the launch contraption with the same settings. The initial values for the velocity, the launch angle and the rotational frequency were determined with uncertainty. The median of all five launches as well as the standard deviation is given for comparison with the uncertainty estimated for each observation. Note that as launch 1 was studied in detail (see section \ref{sec:DetailedResultsShot3}), the initial parameters were determined with greater precision.} 
	\label{tab:LaunchComparisonStats}
\end{table}
\noindent
However, the difference in the launch angle for the other four launches is small enough to be satisfactory. All in all the reproducibility fails due to the rotational velocity (standard deviation more than ten times the individual uncertainty), where the V-belt mechanism has to be improved. Nevertheless, the contraption is still more reliable compared to a human induced launch and moreover satisfies the constraint of an initial AoA close to \ang{0}. This will prove necessary later, see section \ref{sec:HumanInducedLaunch}.

%------------------------------------------------------------------

\section{Experimental Procedure and Trajectory Evaluation}
\label{sec:ExperimentalProcedure}

\subsection{Preparatory Measurements}
\label{sec:PrepMeasurements}
Before conducting the experiments, the mass of the X-Zylo was measured using a high-resolution balance. Measuring two traditional X-Zylo's, an average mass of $\SI{22.73\pm0.16}{g}$ was found. Another X-Zylo was carefully disassembled and the plastic and metal part measured independently. With the mass of the metal ring being $\SI{16.42\pm0.01}{g}$ and the plastic weighing $\SI{6.27\pm0.01}{g}$  one can deduce the COM being $\SI{12.2\pm 0.1}{mm}$ behind the leading edge. The glue mass between the metal ring and the plastic hull was negligible. Comparing the results to the measurements done by Tarr \parencite{TarrXZylo} shows good agreement in the X-Zylo's mass. A larger discrepancy for the COM location is seen, which was found to be ``at an axial distance of $\SI{1.3}{cm}$ from the leading edge'' \parencite{TarrXZylo}.

For the experiments the X-Zylo was colored red for better visibility and a black line was added to calculate the rotational frequency from the slow motion footage. It was however found in earlier tests that the additional mass from the red paint was not negligible and shifted the COM. By weighing the painted X-Zylo it was found that the mass increased by \SI{1.24}{g} ($10.5\%$) and the COM shifted from \SI{12.2}{mm} to \SI{13.0}{mm} behind the leading edge which significantly changed the observed drifting behavior (see section \ref{sec:SidewaysDrift}).

\subsection{Camera Setup and Corrections}
\label{sec:ExperimentalCorrections}

\begin{figure}[h]
	\includegraphics[width=\textwidth]{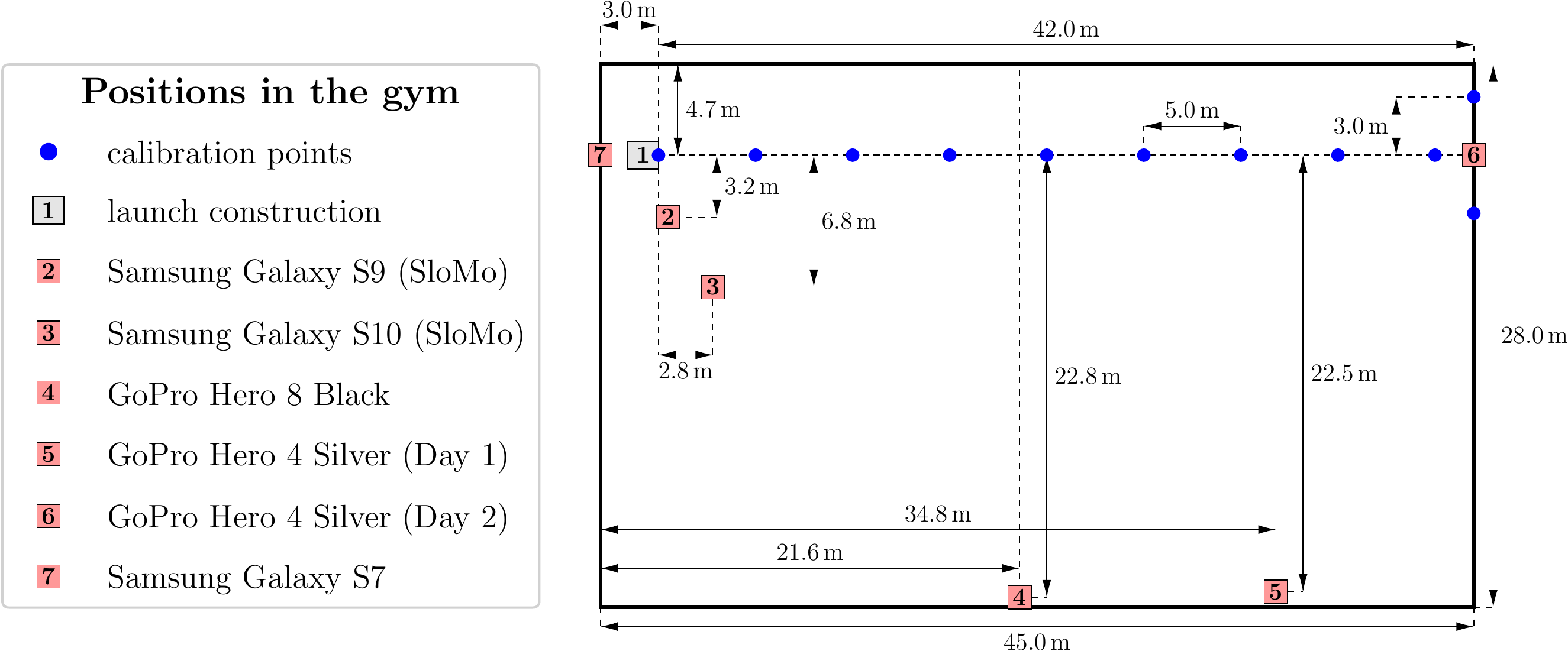}
	\caption{Sketch (to scale) of the camera and calibration setup in the school gym for the final experiments. All camera positions shown were fixed, only the Samsung Galaxy S9 was moved throughout the experiments.}
	\label{fig:camera_setup_gym}
\end{figure}

The trajectory of the ring in the $xz$-plane was captured using a GoPro HERO 8 Black in linear mode (4k, 60fps). Linear mode uses the dedicated GoPro-intern software to eliminate the barrel distortion typically encountered with their cameras. Therefore only insignificant warp effects are still visible which interfere with a qualitative analysis of the flight. Additionally, a Samsung Galaxy S10 and S9 are both used as close-up high-speed cameras capturing 960fps at a resolution of 720p. Another Samsung Galaxy S7 (4k, 30fps) is used to capture the sideways drift of the X-Zylo during flight in the $yz$-plane. Finally, a GoPro HERO 4 Silver (linear mode, 1080p, 30fps) was used as a backup camera to probe different locations during flight. The camera setup with their positions can be found in figure \ref{fig:camera_setup_gym}. From now on almost exclusively the abbreviations S10, S9, S7, GoPro8 and GoPro4 are used instead of the full camera names mentioned above.

Using the open-source software Tracker \parencite{Tracker}, the trajectories could be manually evaluated frame by frame. The experiments were conducted in a school gym to reduce external factors, former outside tests revealed bad flight behavior due to even slightly windy conditions. To quantitatively capture the trajectory, both the $xz$- and $yz$-plane were equipped with calibration points marking different distances. Additionally, a two-meter-long colored calibration bar was used for the slow motion footage to accurately compute the launch velocity (see figure \ref{fig:Shot_TimeSeries}). 

However, just observing the trajectory is not enough. Several effects have to be accounted for to correct different imaging errors. As camera distortions are hard to deal with and are thought to have negligible impact since they are already internally corrected for the GoPro footage, they are ignored. Only purely geometrical corrections independent of the camera are discussed in the following.

\subsubsection*{Ring outside the Plane of Measurement}

\begin{figure}[h]
	\centering
	\includegraphics[width=\textwidth]{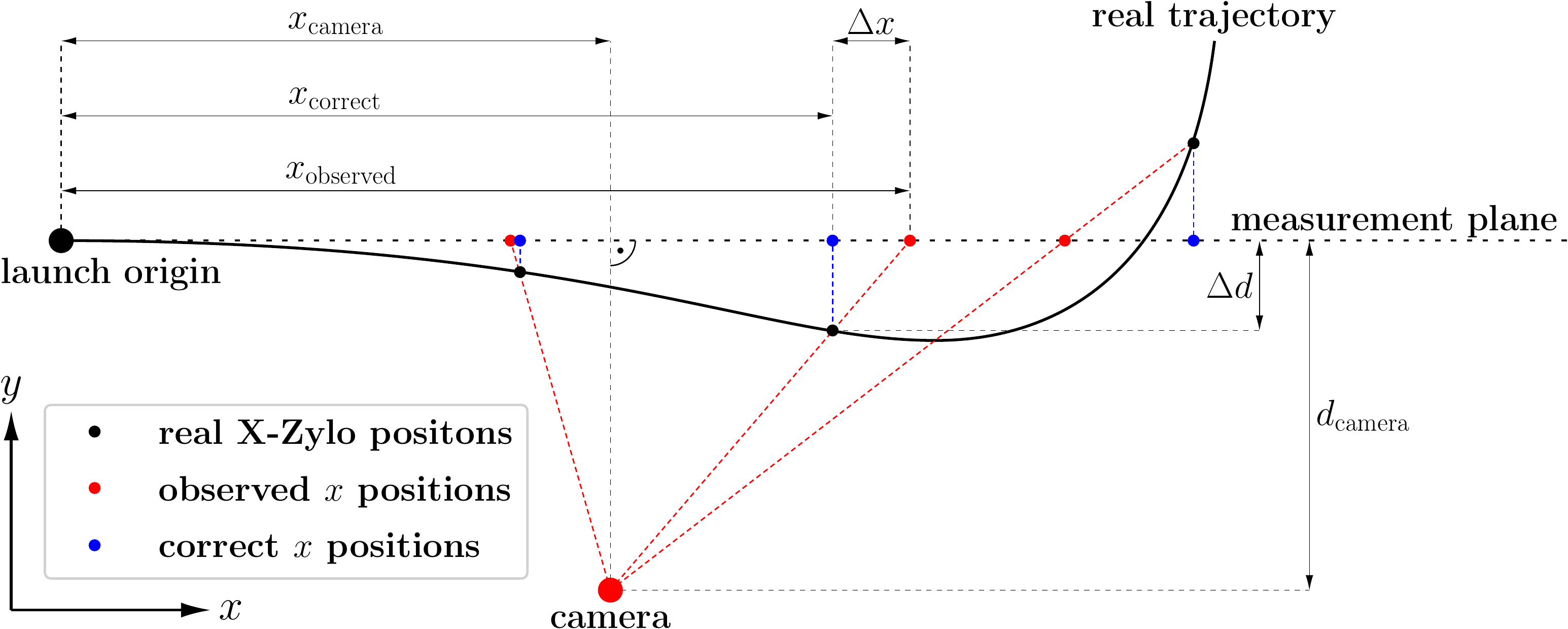}
	\captionsetup{justification=justified}
	\caption{Using the intercept theorem and equation \eqref{eq:x_correction} one can correct the observed X-Zylo positions. This is necessary as the X-Zylo leaves the measurement plane and therefore only a projection is measured, which is illustrated in this figure.}
	\label{fig:intercept_theorem_far}
\end{figure}

\noindent
The plane of measurement is here defined to be the plane perpendicular to the camera's view, in which the calibration points are located. Therefore the camera only captures the projection of the ring onto this plane of measurement. 
If the ring is not located in this plane, one has to account for this via the intercept theorem, which is demonstrated in figure \ref{fig:intercept_theorem_far}.

%\sidecaptionvpos{figure}{c}
%\begin{SCfigure}
%	\centering
%	\includegraphics[width=0.7\textwidth]{intersect_theorem_far.pdf}
%	\caption{Using the intercept theorem and equation \eqref{eq:x_correction} one can correct the observed X-Zylo positions. This is necessary as the X-Zylo leaves the measurement plane and therefore only a projection is measured, which is illustrated in this figure.}
%	\label{fig:intercept_theorem_far}
%\end{SCfigure}

As the ring swerves during flight, it leaves the $xz$-plane captured by the GoPro8. This results in a correction term (see equation \eqref{eq:x_correction}) which has to be applied to the data. Mind that both $\Delta x$ and $\Delta d$ are signed quantities. In figure \ref{fig:intercept_theorem_far} $\Delta d$ is negative and $\Delta x$ positive.
\begin{equation}
x_\text{correct}=x_\text{observed}- \Delta x=x_\text{observed}+ (x_\text{observed}-x_\text{camera})\cdot\frac{\Delta d}{d_\text{camera}}\;.
\label{eq:x_correction}
\end{equation}
As the S7 films the sideways drift while the GoPro8 is capturing the $xz$-projection, both cameras together can recreate the trajectory corrected for the swerving motion. The S7 however is also subject to the intercept theorem since the calibration points are set up at the opposite end of the gym. Therefore the image corrections of both cameras are coupled. The rectifications are applied sequentially; first the S7 footage is adjusted with the $x$-coordinate of the GoPro8 video, then the GoPro's footage is corrected with the swerving motion of the S7. One could apply the corrections again, however those higher order terms are of negligible magnitude.\\

\subsubsection*{Ring in close Proximity to the Camera}

\begin{figure}[h]
	\centering
	\includegraphics[width=\textwidth]{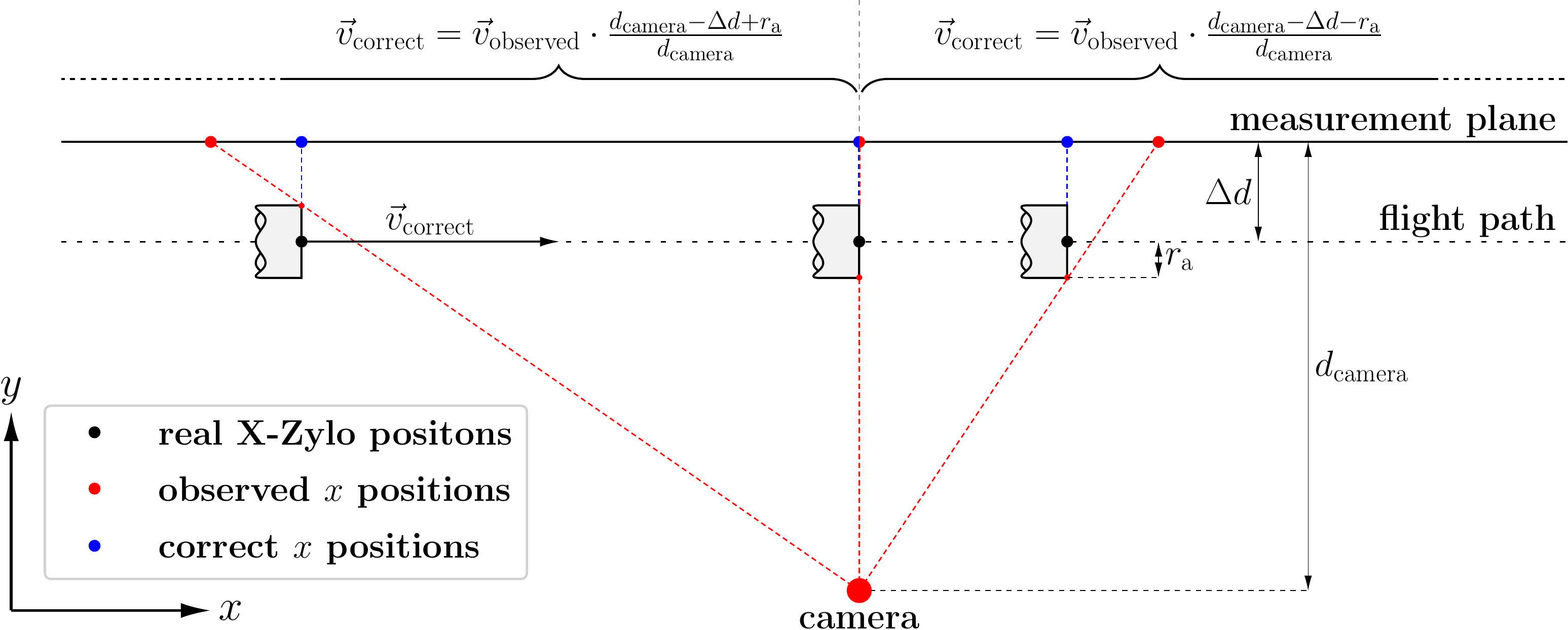}
	\captionsetup{justification=justified}
	\caption{The intercept theorem also applies if the plane of measurement is not exactly located in the flight path of the X-Zylo. If the camera is very close to the flight path, one has an additional effect of observing different parts of the ring during flight.}
	\label{fig:intercept_theorem_near}
\end{figure}

%\sidecaptionvpos{figure}{c}
%\begin{SCfigure}
%	\centering
%	\includegraphics[width=0.7\textwidth]{intersect_theorem_near.pdf}
%	\caption{The intercept theorem also applies if the plane of measurement is not exactly located in the flight path of the X-Zylo. If the camera is very close to the flight path, one has an additional effect of observing different parts of the ring during flight.}
%	\label{fig:intercept_theorem_near}
%\end{SCfigure}

\begin{figure}[h]
	\includegraphics[width=1\textwidth]{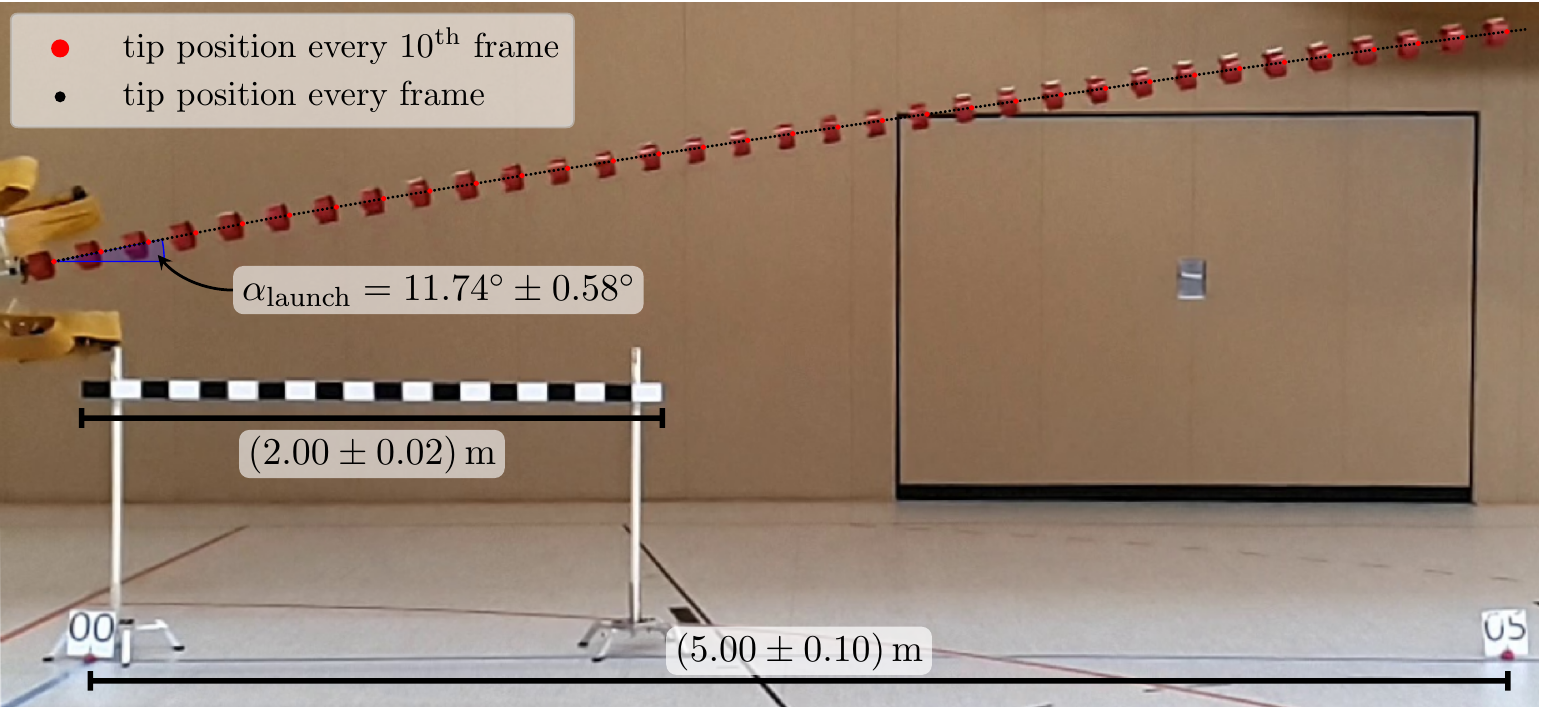}
	\captionsetup{justification=justified}
	\caption{Time Series of an X-Zylo immediately after launch. Every frame the position is manually evaluated using Tracker \parencite{Tracker}, which are the small black points seen in the figure. The larger red points show the tracked X-Zylo position every $10^\text{th}$ frame. The uncertainty on the calibration scales stems from being in slightly different measurement planes (see figure \ref{fig:Shot_yz_plane}) as well as an estimated scaling error due to camera distortions (see section \ref{sec:error_trajectory}).}
	\label{fig:Shot_TimeSeries}
\end{figure}

From the S7 footage it was seen that the X-Zylo trajectory was offset from the measurement stick set up (see figure \ref{fig:Shot_yz_plane}). This results again in an application of the intercept theorem for the close-up cameras, illustrated in figure \ref{fig:intercept_theorem_near}. However, this was not the only effect observed. When looking closely at picture \ref{fig:Shot_TimeSeries}, which shows a time series of the tracked X-Zylo seen from the S10's slow motion footage, one spots that the X-Zylo is tracked at different locations throughout the flight. As the object is probed almost precisely at its foremost point every frame, the spot tracked with this method changes. This is also illustrated in figure \ref{fig:intercept_theorem_near}. Before passing the camera, the side facing the measurement plane is tracked while after that the opposite side facing the camera is tracked. As only the velocities in the slow motion footage were important, a simple application of the intercept theorem results in the corrected velocities. This correction is important even if the difference between the tracked points is only the diameter of the ring, therefore circa \SI{5}{cm}. This comes due to the close proximity of the camera to the flight path. As for example the S9 is located only about \SI{3.3}{m} from the measurement plane, one gets an offset of approximately $1.5\%$ which---for standard launch conditions with a velocity magnitude of \SI{16}{m/s}---results in an error of \SI{0.24}{m/s}, that can not be neglected.\\

%------------------------------------------------------------------

\section{Comparison between Theory and Experiment}
\label{sec:ComparisonResults}

To compare the experimental results to the predictions made by the theoretical model, one launch is studied in great detail. Furthermore, the influence of the launch angle is investigated and compared to the model. At last, the sideways drifting behavior as well as the impact a human induced launch causes are studied. 

	\subsection{Detailed Results for a single Launch}
	\label{sec:DetailedResultsShot3}
	
\begin{wrapfigure}{l}{0.4\textwidth}
		\vspace{-5pt}
		\centering
		\includegraphics[width=0.38\textwidth]{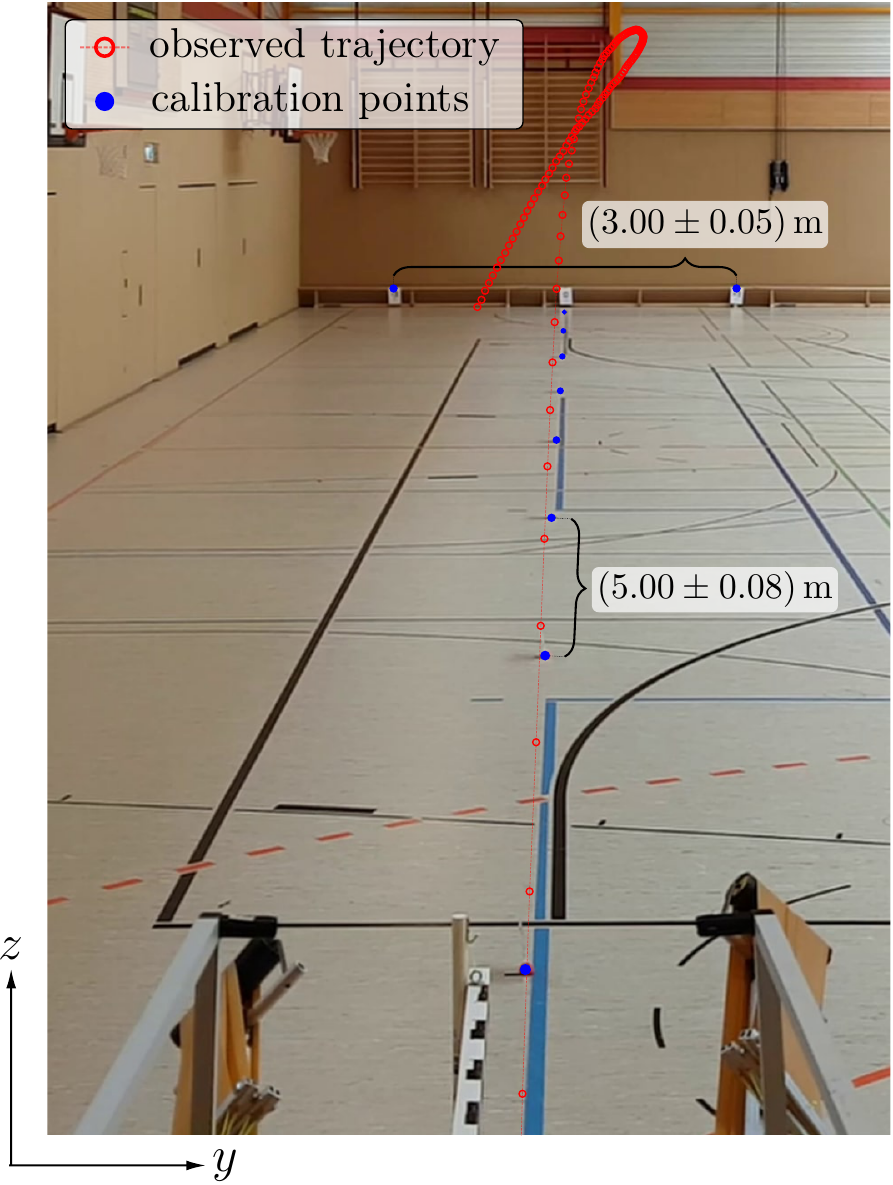}
		\vspace{-4pt}
		\centering
		\captionsetup{margin=0.3cm}
		\caption{Examined launch in the $xy$-plane (raw footage of S7).}
		\label{fig:Shot_yz_plane}
		\vspace{-70pt}
	\end{wrapfigure}

\begin{figure}[H]
		\includegraphics[width=1\textwidth]{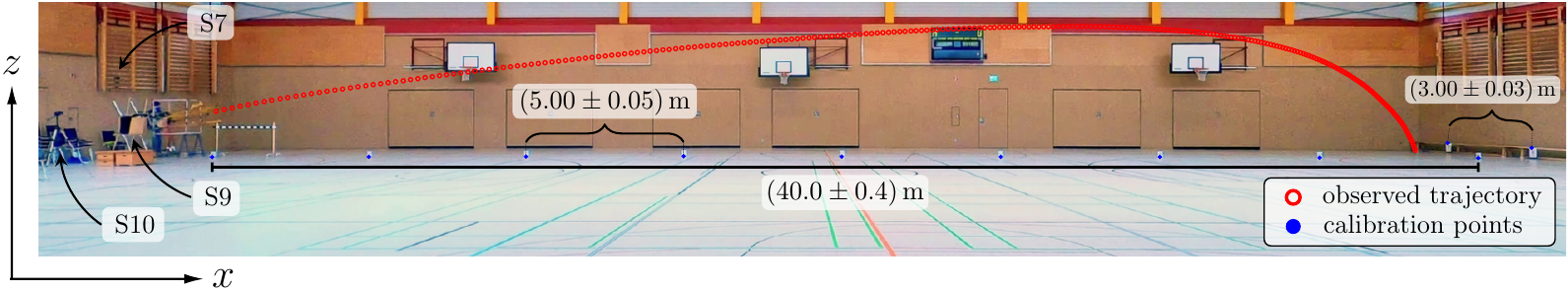}
		\captionsetup{justification=justified}
		\caption{Examined launch in the $xz$-plane (raw footage of GoPro8). The close up cameras are marked as well as the calibration distances shown with uncertainty.}
		\label{fig:Shot_xz_plane}
	\end{figure}

	Launch 1 seen in section \ref{sec:Reproducibility} is used for the detailed study of a single launch. This is useful as the initial rotational frequency is sufficiently low to observe the drifting behavior in detail. In addition, the launch angle is optimal to achieve a long distance shot while also gaining enough height for greater insight into the vertical components. Figure \ref{fig:Shot_TimeSeries} shows a time series of this launch observed by the S10 camera, while figures \ref{fig:Shot_yz_plane} and \ref{fig:Shot_xz_plane} present the trajectory of the X-Zylo. The calibration points---as already seen in figure \ref{fig:camera_setup_gym}---are also marked in the pictures. Using the scale uncertainty due to camera distortions, the lengths are afflicted with an error using the values from table \ref{tab:ErrorRatesR}. Mind that the trajectories seen show only the raw footage and are therefore not adjusted to the errors mentioned in section \ref{sec:ExperimentalCorrections}.
	
	To get the initial launch parameters one can view the footage of the different cameras. As the parameters change rapidly after launch, only the first five frames of each video are used. From those frames the median as well as the standard deviation for the different values are computed using the 
	Student's t-distribution assuming statistical scattering of the values. Additionally, a systematic uncertainty is estimated for each camera. This yields 
	\begin{align*}
	v_\text{launch} &=(16.15\pm0.03\pm0.30)\;\text{m/s}\qquad \text{(S9)}\; ,\\ &=(15.84\pm0.15\pm0.15)\;\text{m/s}\qquad \text{(S10)}\; ,\\ &=(15.5\;\,\:\!\:\!\underbrace{\pm\:0.2\;\,\:\!}_{\text{stat}}\:\!\underbrace{\pm\:0.5\;\,\:\!}_{\text{sys}})\;\text{m/s}\qquad \text{(GoPro8)}\; .
	\end{align*}
	Using those values the weighted mean can be calculated, resulting in a launch velocity of $\SI{15.94\pm0.21}{m/s}$. No difference between systematic and statistical error is made after applying the weighted mean. Repeating the procedure for the launch angle one gets
	\begin{align*}
	\alpha_\text{launch}&=(11.86\pm0.29\pm0.20)\,\degree\qquad \text{(S9)}\; ,\\ &=(11.74\pm0.43\pm0.15)\,\degree\qquad \text{(S10)}\; ,\\ &=(11.6\;\,\:\!\pm0.3\;\,\:\!\pm0.5\;\,\:\!)\,\degree\qquad \text{(GoPro8)}\; ,
	\end{align*}
	which then results in $\alpha_\text{launch}=(11.77\pm0.34)\,\degree$. The initial rotational frequency after applying the weighted average is $\omega(t_0)/(2\pi)=\SI{18.5\pm0.4}{Hz}$. Additionally, more insignificant uncertainties are set 
	
	\noindent
	for the ring mass, the location of the COM, the launch height, the gravitational constant and the ring thickness. Using those uncertainties one can calculate the theoretical trajectory with uncertainty as described in section \ref{sec:error_calculation}. However, the error margins for the CFD simulations also have to be specified, e.g. using the mesh study (see table \ref{tab:MeshStudy}). The uncertainties for the Transition SST model were set to be $\pm2\%$ for $C_\text{D}$ and $C_\text{L}$, and $\pm8\%$ for the location of the COP; the values for k-$\omega$-SST are set to $\pm1\%$ and $\pm5\%$ respectively. For both turbulence models the uncertainty for the mean wall shear stress $\tau_\text{w}$ was set to $\pm15\%$ as equation \eqref{eq:RotDecreaseEquation} is only approximately correct. 
	\begin{figure}[H]
		\includegraphics[width=1\textwidth]{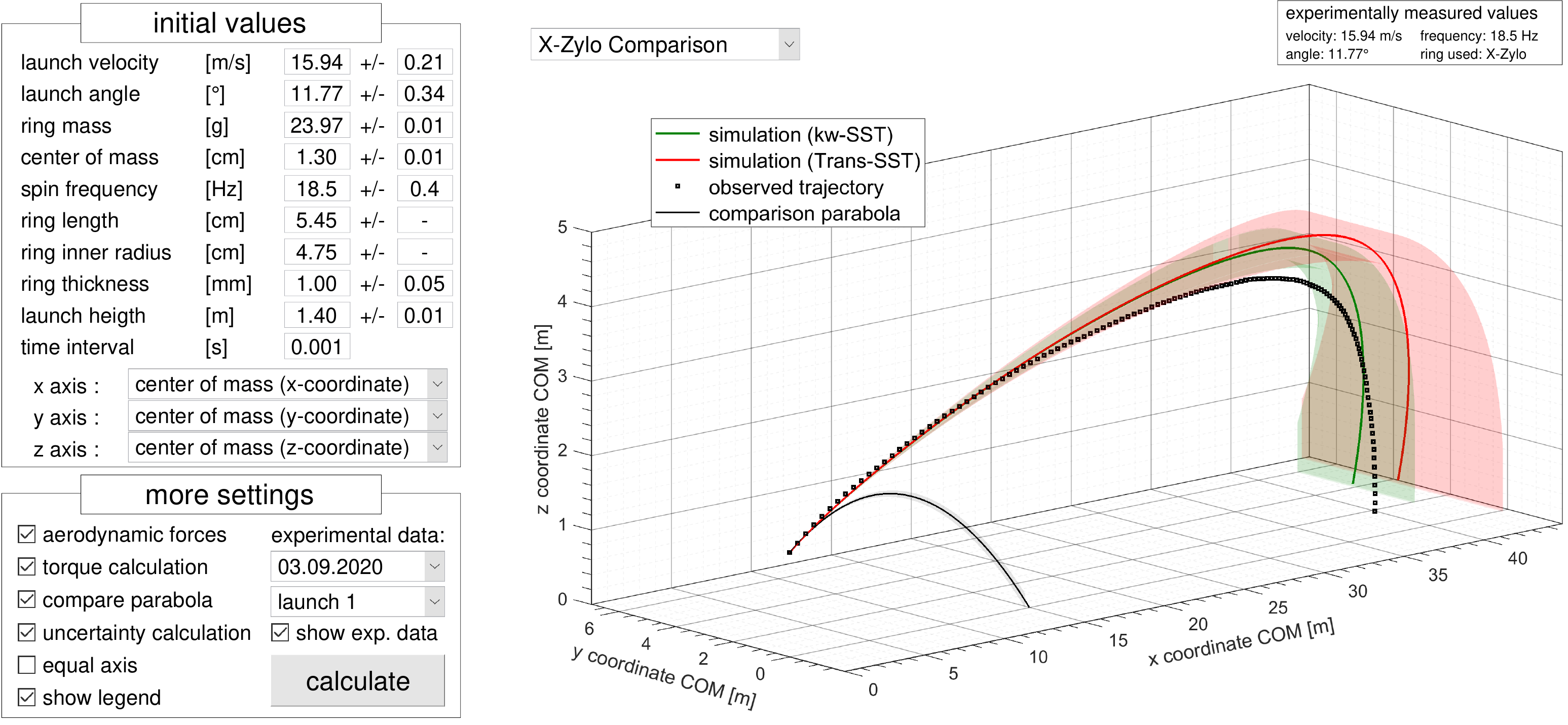}
		\captionsetup{justification=justified}
		\caption{3D Trajectory of the X-Zylo in comparison with the theoretical predictions. The interface seen here is an interactive MATLAB GUI written to easily probe the rings flight in detail. Using the drop-down menus enables the user to plot almost all combinations of variables calculated. For reasons of space from now on only the graph---then plotted in Python---is shown without the GUI.}
		\label{fig:TrajGUI}
	\end{figure}

\begin{figure}[H]
		\includegraphics[width=1\textwidth]{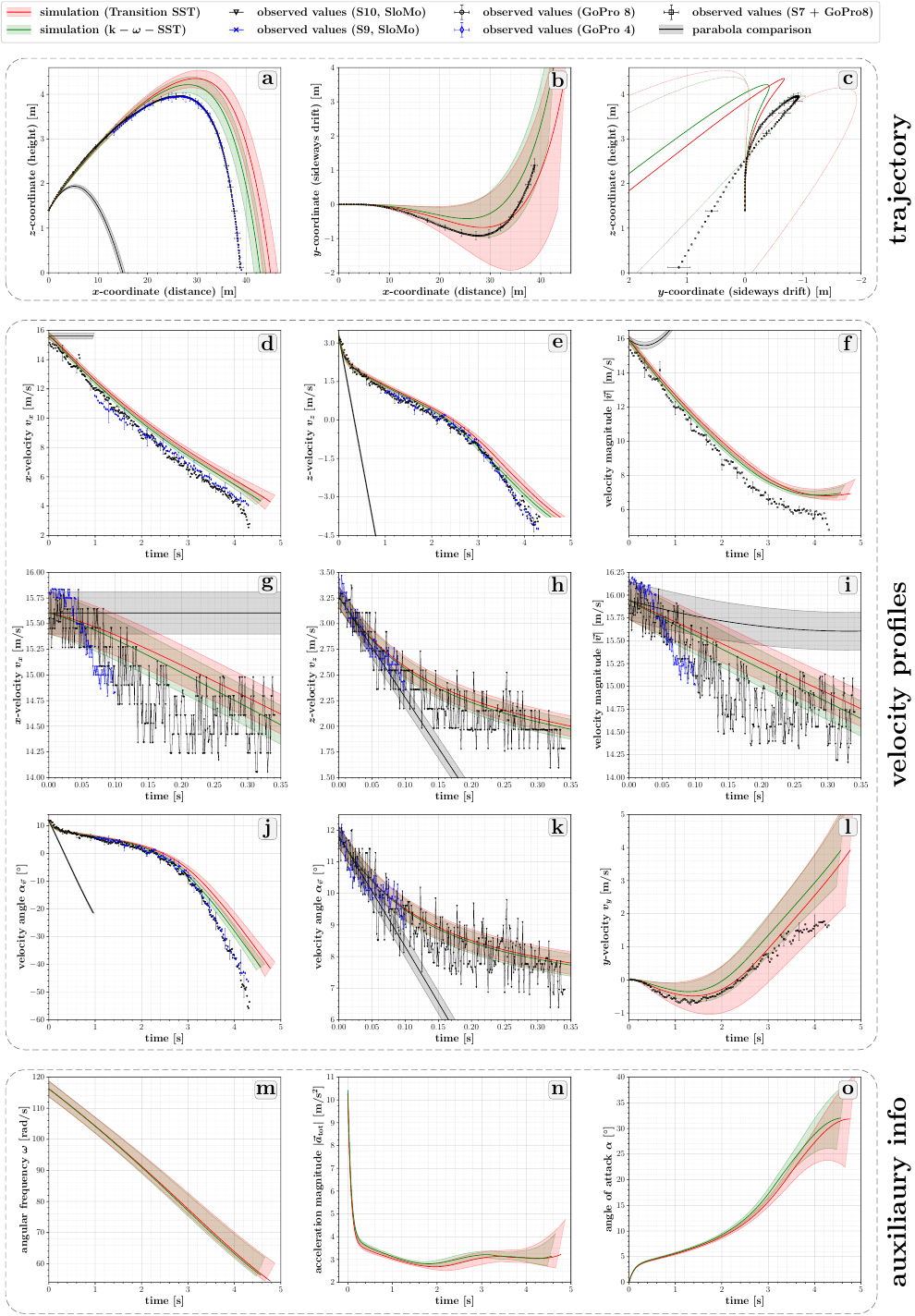}
		\captionsetup{justification=justified}
		\caption{Comparison of the theoretical quantities with the observed trajectory data for a single launch. Plots a-c show the trajectory in the $xz$-, $xy$- and $yz$-plane while plots d-l show different velocity profiles. In plots m-o additional information not probed for the real trajectory is visible.}
		\label{fig:allResultPlots}
	\end{figure}

	The theoretical trajectories for both turbulence models with their confidence interval are seen in figure \ref{fig:TrajGUI}. The observed trajectory is plotted without uncertainty for better visibility. The interactive GUI developed for the program shows the different initial values with uncertainty as well as the different calculation settings. As this 3D view of the trajectory is only lightly helpful, figure \ref{fig:allResultPlots} shows the different quantities as 2D plots for better comparison. 
	
	Figure \ref{fig:allResultPlots} presents the observed trajectory with derived quantities as well as the predictions by both turbulence models. Additionally, for comparison a conventional flight parabola without fluid forces is shown in several subplots. The uncertainty on the simulated quantities is calculated using the uncertainty on the initial launch values and the procedure described in section \ref{sec:error_calculation}. The error margins of the observed values are calculated using the procedure in section \ref{sec:error_trajectory}, in which the camera induced error is estimated. In the following every subplot of figure \ref{fig:allResultPlots} is individually evaluated. 
	
	\begin{enumerate}[listparindent=\parindent]
		\item[{\hyperref[fig:allResultPlots]{\ref*{fig:allResultPlots}a})}] The projection of the observed trajectory into the $xz$-plane is shown. As can be seen, after all corrections are applied to the raw footage as described in section \ref{sec:ExperimentalCorrections}, both camera perspectives (GoPro8 and GoPro4) coincide really well in the given confidence interval.\\
		\indent
		When comparing the experimental data with the theoretical predictions, it can be seen that during early flight, a good match could be achieved. However, later during flight the paths diverge outside the joint error margin. The decrease in height (second drop) is also more rapid in the observation compared to both predictions. This is only lightly visible with the given aspect ratio of the plot, when having equal axis this becomes more obvious. Nonetheless, the overall shape of the trajectory is in good agreement with the theory. Mind that the $z$-axis is heavily scaled, therefore the trajectory seems not as flat as it actually is.\\
		\indent
		Lastly when comparing both turbulence models, it is seen that k-$\omega$-SST gives better predictions since it predicts a higher drag value for low AoA. Therefore the distance covered by the ring is smaller which better resembles reality. Also it is seen that even though both models differ quite significantly for low AoA drag prediction, the influence is rather slim. This comes due to the first drop, as small AoA are only present for a short amount of time, see figure \hyperref[fig:allResultPlots]{\ref*{fig:allResultPlots}o}.
		
		\item[{\hyperref[fig:allResultPlots]{\ref*{fig:allResultPlots}b})}] When looking at the projection of the trajectory in the $xy$-plane, it is visible that the qualitative behavior of the X-Zylo can be modeled well. The X-Zylo first swerves to the right, then turns mid-flight and rapidly swerves left at the end of flight, as seen in figure \ref{fig:Shot_yz_plane}. This can be explained using the model described in section \ref{sec:TheoreticalModel}. As the COP moves upstream, it passes the COM of the ring at an AoA of approximately \ang{6.5}. Consequently, the torque acting on the ring changes direction in mid-flight when the COM is passed by the COP which results in a direction change of the swerving motion. This torque sign change is further denoted as \emph{torque flip}. It can be seen that the sideways drift is stronger than predicted, however the ring hits the ground earlier in the experiment. Therefore, the X-Zylo is not able to drift as far at the end of flight as in the simulation. The sideways drift also influences the observation in the $xz$-plane, as all spatial directions are coupled. An increase in $y$-velocity reduces the other velocity directions accordingly and is therefore visible in the $xz$-projection.\\
		\indent
		The uncertainty on the theoretical prediction is huge in this case due to the high error margin imposed on the COP location ($\pm5\%$ for k-$\omega$-SST, $\pm8\%$ for Transition SST). In addition to the uncertainty for the mean wall shear stress---which describes the decrease in rotational velocity---one gets this large uncertainty. It is also visible that the additional $3\%$ error on the Transition SST's COP data make a big difference as the uncertainty almost doubles due to nonlinear behavior. It is later seen that the location of COP and COM are very sensitive flight parameters, see section \ref{sec:SidewaysDrift}. 
		
		\item[{\hyperref[fig:allResultPlots]{\ref*{fig:allResultPlots}c})}] The $yz$-projection is less interesting on its own, but it completes the whole picture. For better visibility only the uncertainty bounds are shown. One can see the behavior also observed in the real world, seen in figure \ref{fig:Shot_yz_plane}. Mind that the values shown in the plot are corrected for perspective, therefore both figures do not look perfectly alike. The qualitative behavior is predicted correctly, however another time it is seen that the swerving to the right is more pronounced than expected. Using this representation it is also better visible that the ring looses height faster compared to the simulation at the end of flight. Even though the observed sideways drift is more rapid at the end of flight, the slope late flight shows a steeper descend.
		 
		\item[{\hyperref[fig:allResultPlots]{\ref*{fig:allResultPlots}d})}] As can be expected the $x$-velocity decreases over time due to drag acting on the ring. Again the behavior seen in the observed trajectory is also seen in the theoretical prediction. The disparity seen seems to stem from a difference in the launch velocity. The theoretical prediction shows an almost constant offset from the observed velocity, indicating that the used launch velocity was overestimated using the slow motion footage. Even though the smaller launch velocity captured by the GoPro8 was accounted for in the calculation of the initial parameters, due to a really large uncertainty the influence of that measurement made a negligible influence in the weighted average. This shows that the uncertainty of the launch velocity for the other cameras was likely underestimated.\\
		\indent
		Another detail that can be observed in the plot is that the $x$-velocity decreases more rapidly shortly before hitting the ground as the sideways drift of the X-Zylo accelerates. This can also be seen in the theoretical prediction where this behavior is starting to form just as the ring hits the ground, therefore a bit later as in the observed trajectory. The parabolic trajectory shows a constant $x$-velocity as no drag forces are applied.
		
		\item[{\hyperref[fig:allResultPlots]{\ref*{fig:allResultPlots}e})}] The $z$-velocity shows an interesting behavior. At launch the vertical velocity is quite large, but as expected by the theoretic prediction in section \ref{sec:TheoreticalModel}, the ring at first loses height rather quickly (first drop). The asymptotic behavior for $t\rightarrow 0$ is seen to fit the parabolic trajectory, which was predicted. Only after about \SI{0.3}{s} this initial drop is absorbed by the increasing lift force due to the rising AoA. The second drop is visible later during flight starting from about \SI{2.8}{s} onward. There the ring begins to lose height faster than seen before, which can be explained by the flow separation forming at higher AoA. Later in plot \hyperref[fig:allResultPlots]{\ref*{fig:allResultPlots}o} this becomes more obvious when looking at the AoA. The mentioned equilibrium phase of the flight in between both drop phases can be seen quite well using this representation.\\
		\indent
		Comparing theory and experiment, both curves match remarkably well until the second drop phase sets in. During the second drop the decrease in $z$-velocity is underestimated by the simulation, showing that the impact of the flow separation occurring was underrated.
		
		\item[{\hyperref[fig:allResultPlots]{\ref*{fig:allResultPlots}f})}] The velocity magnitude follows a similar pattern as the $x$-velocity. However, due to the increasing negative $z$-velocity at the end of flight the velocity magnitude plateaus with a tendency to rise again. A similar behavior is seen in the observed trajectory. Due to additional effects stemming from the $x$- and $y$-velocity, the observed velocity magnitude drops again shortly before hitting the ground. This effect is not visible in the theoretical predictions and is not fully understood.
		
		\item[{\hyperref[fig:allResultPlots]{\ref*{fig:allResultPlots}g})}] This plot shows the observed $x$-velocity obtained by the slow motion footage (see figure \ref{fig:Shot_TimeSeries}). As the velocity range is small and the  uncertainties are fairly big, no error-bars were put on the experimental data. When looking at the S10's data, the theoretical behavior was approximately met, however a dip is seen when the X-Zylo is in the center of the camera's field of view. This effect most probably occurs as some additional camera effects are not corrected or are underestimated. As most probably the velocities observed in the center of the field of view are more accurate, this would also backup the claim that the launch velocity could be overestimated in the slow motion footage.\\
		\indent
		The data obtained by the S9 is even more questionable as a far faster drop in velocity than expected is observed. Here the error stems from the position of the camera, as it was seen in the footage that the camera was slightly tilted sideways, therefore producing a skewed image. As both cameras are very close to the trajectory, the camera distortion produces a relatively big error.\\
		\indent
		Another problem can be seen when looking at the plot itself. The experimental data shows a quantized manner as only discrete velocities are seen, which is non-physical. The resolution of the camera produces this pattern, as a pixelated image can only entail discrete length differences. Therefore, one has to track the X-Zylo almost pixel-perfect, which induces even more error.\\
		\indent
		It is quite obvious that the used method---capturing the launch of the X-Zylo with cameras---to calculate the initial values accurately is not sufficient. Even though correction terms are applied, the camera induced errors still dominate the obtained results. The closer the camera is to the trajectory, the stronger those effects become.
		
		\item[{\hyperref[fig:allResultPlots]{\ref*{fig:allResultPlots}h})}] The $z$-velocity matches the theory better as camera errors like barrel distortion and the sideways tilt of the S9 mostly influence the $x$-coordinate. While the X-Zylo flight spans almost the whole field of view horizontally, it only covers a small distance vertically. The vertical components are only inflicted with serious errors for high launch angles.\\
		\indent
		The mentioned asymptotic behavior for $t\rightarrow 0$ is best visible using this representation. While the parabolic trajectory shows a linear decrease in $z$-velocity, the X-Zylo's increasing AoA dampens this decrease until a force balance is formed. This is the transition between the first drop into the equilibrium phase.  
		
		\item[{\hyperref[fig:allResultPlots]{\ref*{fig:allResultPlots}i})}]
		The dip in $x$-velocity observed by the S10 as well as the fast drop for the S9 are still visible in the velocity magnitude plot. As the $x$-component of the velocity dominates during early flight, the plot does not significantly alter from the observed early-time behavior of the $x$-velocity itself seen in figure {\hyperref[fig:allResultPlots]{\ref*{fig:allResultPlots}g}}.
		
		\item[{\hyperref[fig:allResultPlots]{\ref*{fig:allResultPlots}j})}] One can also look at the $xz$-projection of the velocity vector. The angle between the $x$-axis and this projection is defined by $\alpha_{\vec{v}}=\arctan(v_z/v_x)$ and will be denoted as \emph{velocity angle}. Using this quantity, the different flight phases are nicely visible, especially the transition from the equilibrium phase towards the second drop. The first drop is less pronounced as in plot {\hyperref[fig:allResultPlots]{\ref*{fig:allResultPlots}e}} showing the $z$-velocity, however still visible. In this representation it can also be spotted that the observed data only really diverges after flow separation occurs and the second drop is initiated. The difference between theory and experiment in the plots {\hyperref[fig:allResultPlots]{\ref*{fig:allResultPlots}d+f}} across the whole flight duration can be traced back to the velocity magnitude, which was likely overestimated from the slow motion footage. Using this representation the velocity magnitude has no influence, only the ratio of $x$- to $z$-velocity is important. Therefore this representation is most likely more accurate.
		
		\item[{\hyperref[fig:allResultPlots]{\ref*{fig:allResultPlots}k})}]
		When looking at the velocity angle in the slow motion footage, the first drop is observed really well with both cameras. The theory and the experiment coincide almost perfectly and again the asymptotic behavior for $t\rightarrow 0$ matches free fall.
		
		\item[{\hyperref[fig:allResultPlots]{\ref*{fig:allResultPlots}l})}] The $y$-velocity also follows the expected contour, however there the first qualitative difference between theory and experiment is seen. Instead of the $y$-velocity increasing linearly during the second drop phase, the acceleration decreases and tends to zero, so that the velocity plateaus. It is unknown if the velocity remains stable or reverses a second time, which would then need a new explanation. With the given data no conclusion can be found for this issue as especially the sideways drift is hard to model. The Transition-SST model is seen to fit better to the real drift behavior than the k-$\omega$-SST turbulence model.
		
		\item[{\hyperref[fig:allResultPlots]{\ref*{fig:allResultPlots}m})}] The angular frequency was only measured at launch and not during the rest of the flight. Therefore, it is unknown whether the simplified model yields accurate results. Mind that even though the behavior seems linear, this is not the case, especially for a longer flight duration.
		
		\item[{\hyperref[fig:allResultPlots]{\ref*{fig:allResultPlots}n})}] The experimental values for the acceleration magnitude are not shown since they scatter significantly as even small imperfections in the tracking amplify when deriving the quantities. In the theoretical prediction the first drop is perfectly seen. As the X-Zylo does not generate lift at start, the acceleration magnitude is $\SI{9.81}{m/s}$ (plus additional drag force). In the first $\SI{0.2}{s}$ this magnitude drops rapidly as the AoA of the ring increases. The behavior late in flight is more complex and the different components of the acceleration would need to be reviewed independently for a deeper analysis. Those are however not included here. 
		
		\item[{\hyperref[fig:allResultPlots]{\ref*{fig:allResultPlots}o})}] At last the theoretical AoA of the ring is shown. Here also the different phases are illustrated quite well. The ring starts off with an AoA of \ang{0}; this condition was preimposed and an assumption made for the launch with the launch device. In section \ref{sec:HumanInducedLaunch} this assumption is discussed further. It is seen that as expected the AoA increases rapidly shortly after launch, then increases only slightly during the equilibrium phase. After that the second drop sets in, apparent from the rapid rise in AoA. This rise sets in after an AoA of approximately \ang{10} is exceeded due to the flow separation forming, which was seen in figure \ref{fig:allResultPlots}. It is perfectly visible that the effects seen in the CFD data necessarily also show in the simulated trajectory.  
	\end{enumerate}

All in all the behavior of the X-Zylo was qualitatively predicted very well. The different phases mentioned in section \ref{sec:TheoreticalModel} are clearly distinguishable in both the theory and the experiment. Moreover, the duration of the different phases could also be predicted well. This in turn shows that the CFD simulation was able to predict the flow separation qualitatively as already seen in the validation cases.  However, it is seen that even though the qualitative flight was predicted correctly, especially the velocity of the ring at launch seems to be overestimated from the slow motion footage. This can be seen when looking at the trajectory in the $xz$-plane ({\hyperref[fig:allResultPlots]{\ref*{fig:allResultPlots}a}}) and the velocity magnitude plot for the whole trajectory ({\hyperref[fig:allResultPlots]{\ref*{fig:allResultPlots}f}}), which shows an offset for the whole flight duration. It is theorized that amplified camera errors due to the proximity of the slow motion cameras to the trajectory interfere with an accurate measurement of the initial parameters. Hints for such behavior are seen in figure {\hyperref[fig:allResultPlots]{\ref*{fig:allResultPlots}g}} with the observed dip in the center of the field of view. 

An important detail is the asymptotic behavior seen for $t\rightarrow 0$. Both the simulation and the observed data asymptotically match the free fall simulation at launch, confirming a first drop. Not only does this confirm the qualitative theory (see section \ref{sec:EarlyFlight}), but also becomes very important when studying a human induced launch. Resolving the additional effects one has to account for in a human induced launch (see section \ref{sec:HumanInducedLaunch}), it will become clear that without the launch contraption, a useful analysis of the flight would likely be impossible as exactly this asymptotic behavior is violated.   

	\subsection{Influence of the Launch Angle}
	\label{sec:LaunchAngleInfluence} 
	
\begin{figure}[H]
		\includegraphics[width=\textwidth]{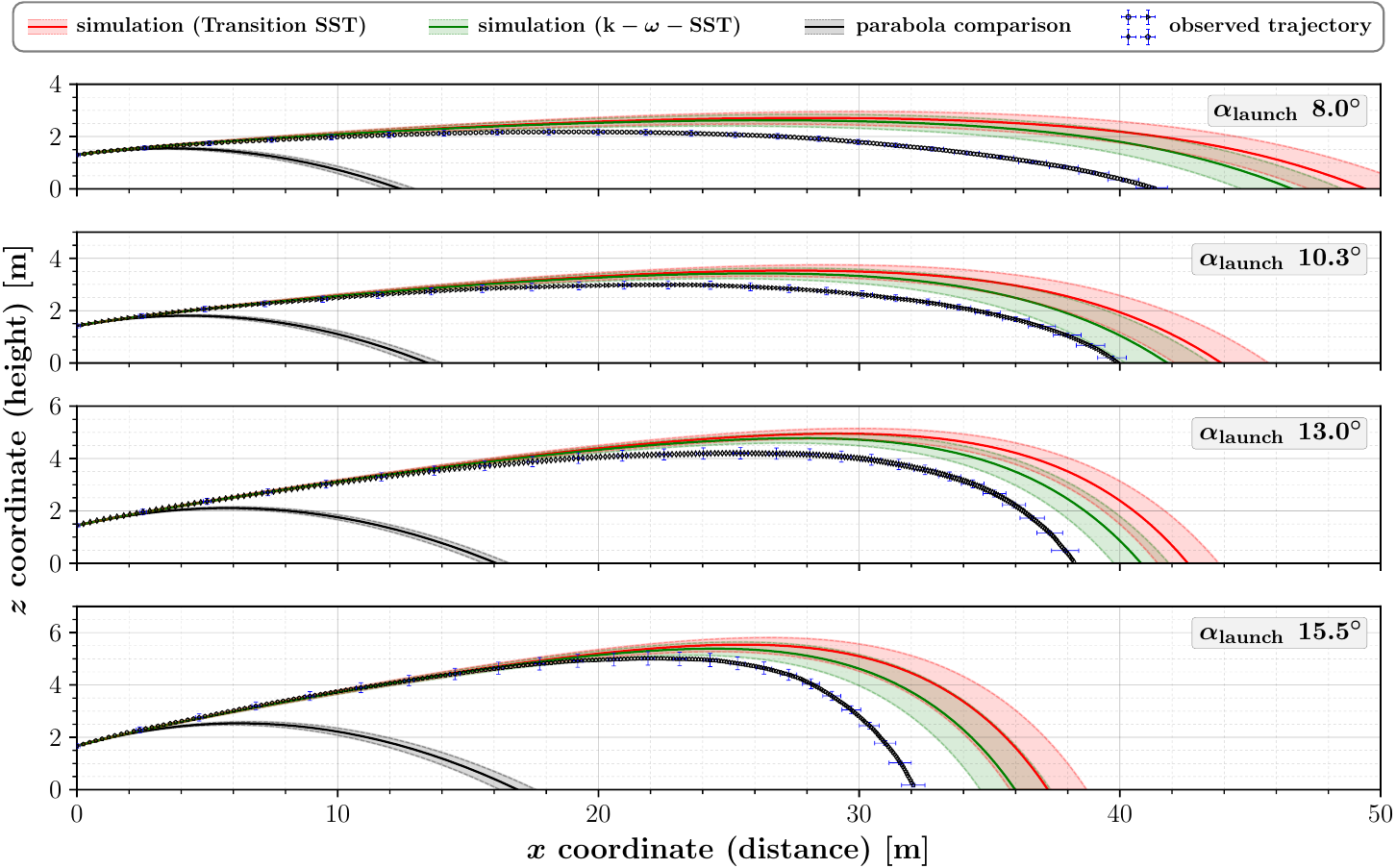}
		\captionsetup{justification=justified}
		\caption{Comparison of the trajectories for different launch angles using equal axis for a better view on the flight. Due to this scale the error-bars in $z$ direction are scaled by a factor of 3 for better visibility.}
		\label{fig:AoAComparison_xz}
	\end{figure}

\begin{figure}[h]
		\includegraphics[width=\textwidth]{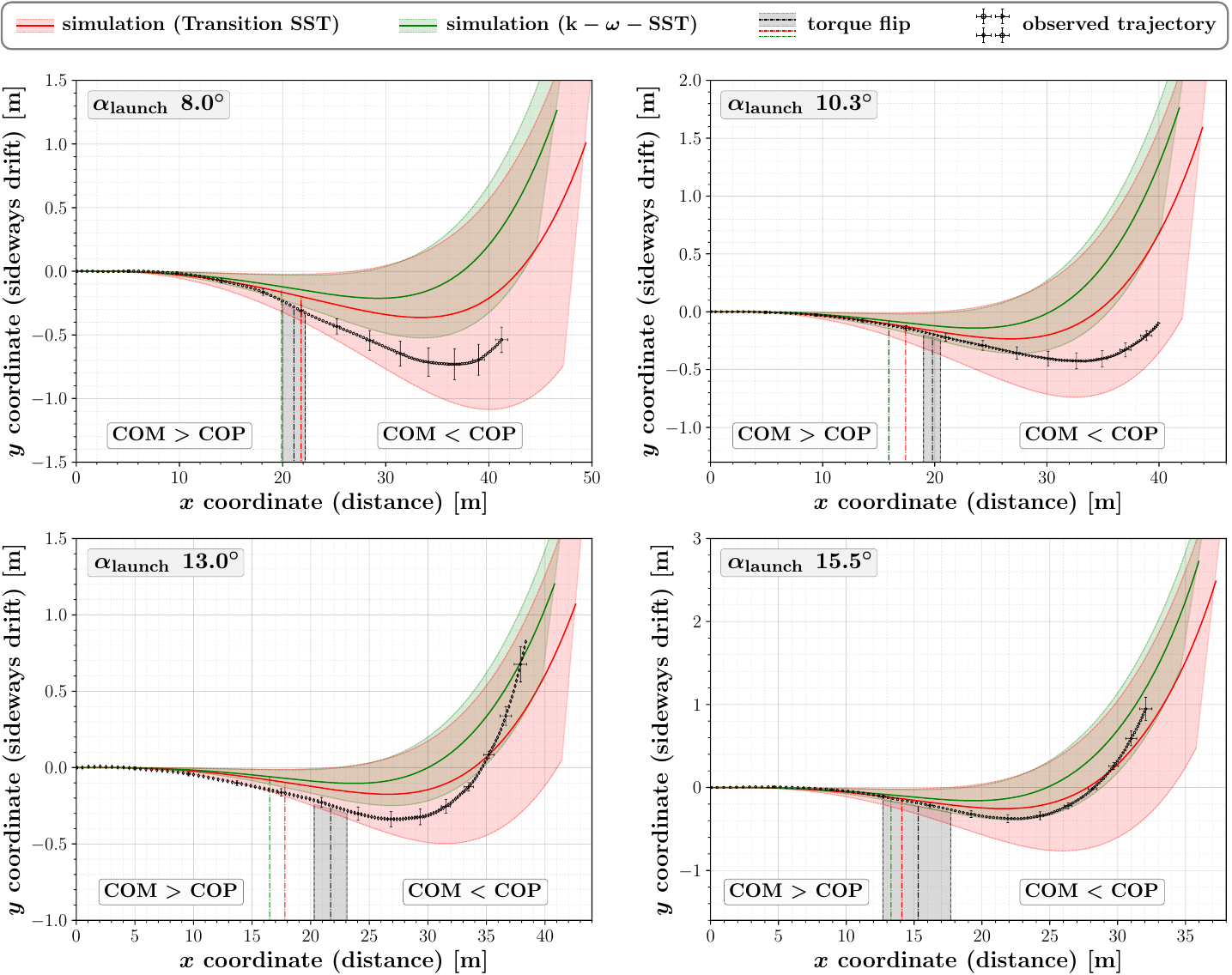}
		\captionsetup{justification=justified}
		\caption{Sideways drift behavior of the X-Zylo in flight for different launch angles. It is marked where the COP moves past the COM on the chord and therefore the drift behavior changes. It is shortly noted with $\text{COM} > \text{COP}$ when the COM is in front of the COP, and vice versa. Note that it is implicitly assumed that the COP always lies on the symmetry axis of the X-Zylo. This assumption was used for all computations and is thought to sufficiently depict the real situation.}
		\label{fig:AoAComparison_xy}
	\end{figure}

	As the launch angle proves to be an essential parameter in the flight of an X-Zylo and the launch mechanism is build especially to control that factor, several launches with different initial launch angles are compared. The initial launch values taken from the slow motion footage can be read from table \ref{tab:AoAComparisonStats}.
	\begin{table}[h]
		\centering
		\begin{tabular}{l||c|c|c|c|c}
			\thickhline
			& launch angle & initial velocity & init. rot. freq.& launch& time \\ 
			& $\alpha_\text{launch}$ [$^\circ$] & $v_\text{launch}$ [m/s]& $\omega(t_0)/(2\pi)$ [Hz] & distance [m] & aloft [s] \\  \hline \hline
			flat (\textbf{L1})& $\;8.0\pm 0.5 $ & $15.9\pm 0.3$ & $47\pm3$ & $41.3\pm 0.3$ & $4.02\pm 0.07$\\  \hline 
			medium (\textbf{L2}) & $10.3\pm 0.4 $ & $15.4\pm 0.3$ & $42\pm3$ & $40.0\pm 0.3$ & $4.23\pm 0.07$\\  \hline
			steep (\textbf{L3})& $13.0 \pm 0.4$ & $16.1\pm 0.2$ & $60\pm3$  & $38.3\pm 0.2$ & $4.47\pm 0.07$ \\  \hline 
			steepest (\textbf{L4}) & $15.5\pm 0.5 $ & $15.4\pm 0.3$ & $25\pm2$ & $32.1\pm 0.3$ & $4.00\pm 0.07$\\  \hline 
			\thickhline
		\end{tabular}
		\captionsetup{justification=justified}
		\caption{Initial launch values with uncertainty for the compared launches taken from the slow motion footage. The launch distances and the flight time are included for comparison.} 
		\label{tab:AoAComparisonStats}
	\end{table}
	As seen the launch angles varied from \ang{8} to \ang{15.5} with almost equal step size. The initial velocity however shows a variance depending on the launch. Also the rotational frequency is very different for the launches which was already seen in section \ref{sec:Reproducibility}. The predicted trajectories as well as the experimental data can be seen in figure \ref{fig:AoAComparison_xz} and \ref{fig:AoAComparison_xy}. At first, it is seen that the predicted trajectory shows a farther flight distance than observed in all cases. As seen in section \ref{sec:DetailedResultsShot3}, several effects play a major role for the underestimated flight distance. In particularly an underrated flow separation for high AoA, a miscalculation of the launch velocity and a difference in drift behavior are the most prone errors. The steepest launch \textbf{L4} represents the underrated flow separation really well, as the trajectory only deviates from the theory shortly before the second drop sets in. The sideways drift behavior observed in the medium flat launch \textbf{L2} seen in figure \ref{fig:AoAComparison_xy} shows, that the drift has non-negligible influence on the drop. As the ring drifts less significantly into the positive $y$-direction compared to the simulated trajectory, the observed drop is less rapid than simulated. Modeling the sideways drift more accurately could in fact also improve the overall performance of the prediction as all flight directions are strongly coupled. 
	
	%This effect would however always lead to a sharper drop at the end of flight. In contradiction to that it can be seen in the medium flat launch (\textbf{L2}) that here the observed drop is less rapid than theorized. One has to consult the sideways drift shown in figure \ref{fig:AoAComparison_xy} to gain more insight. As the sideways drift is seen to be stretched longer, the $y$-velocity of the ring at the end of flight is considerably smaller than expected, hence the $x$-velocity is greater. It is therefore seen that the observed drop speed is also dependent on the sideways drift behavior. This could also explain the deviation seen early in flight for the flat launch (\textbf{L1}), as here the drift behavior differs from the theoretical values a lot early on. 

	For the sideways drift of the trajectory (see figure \ref{fig:AoAComparison_xy}), the observed values are still in the confidence interval of the prediction made by the Transition-SST turbulence model, however only due to the huge uncertainties given by the model. As additional information it is shown where the $y$-acceleration vanishes and the torque on the ring flips (vertical lines). This happens when the COP passes through the COM of the ring. For the observed trajectory an additional confidence interval is shown as the torque flip is not sharply visible in the data. One can deduce that the torque flip is mostly theorized too early during flight. The ring generally drifts stronger into the negative $y$-direction before turning, showing that either the COM or COP data is erroneous. Even minimal shifts in either parameter results in visible differences of the sideways drift (see section \ref{sec:HumanInducedLaunch}). 

\begin{figure}[H]
		\includegraphics[width=\textwidth]{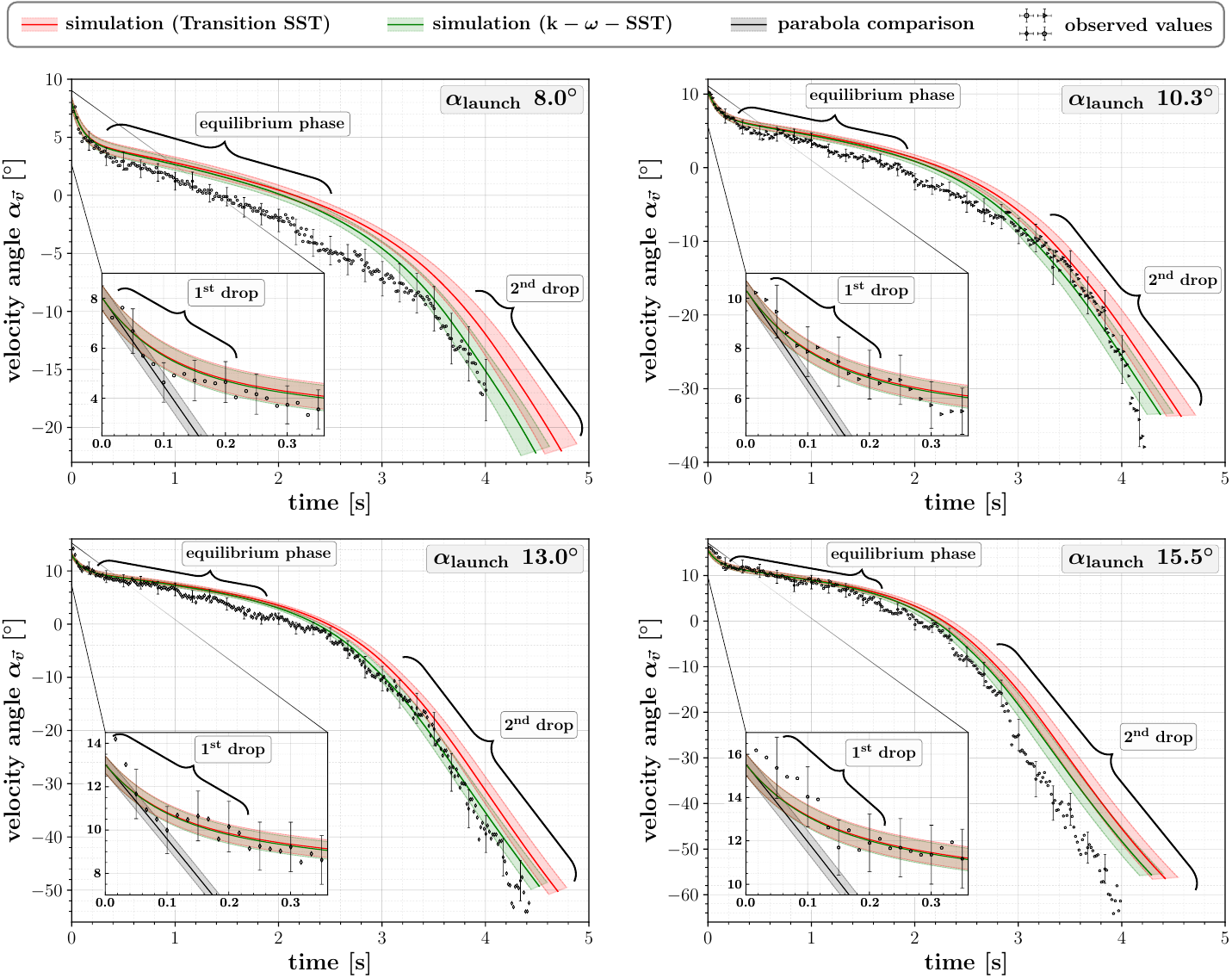}
		\captionsetup{justification=justified}
		\caption{Comparison of the velocity angle of the ring during flight for different launch angles. The first \SI{0.35}{s} after launch are magnified and the different flight phases marked.}
		\label{fig:AoAComparison_vz}
	\end{figure}

	Figure \ref{fig:AoAComparison_vz} represents the experimental observation of the velocity angle for all four launches. The theoretical prediction again shows good agreement, however the deviation seen is larger than formerly observed in figure {\hyperref[fig:allResultPlots]{\ref*{fig:allResultPlots}e}}. This is especially the case for the equilibrium phase of the low launch angle cases (\textbf{L1}, \textbf{L2}). Here the first drop lasts slightly longer than expected, therefore creating a faster decline in the velocity angle than predicted. A good explanation was not found as it would be expected that for lower AoA the prediction would be better as the CFD simulations should yield more accurate results. In the case \textbf{L4} the faster drop during late flight can be seen again in the steeper slope of the velocity angle during the second drop phase.
	
	Concerning the time duration of the different phases one clearly observes that as expected the equilibrium phase gets shorter for higher launch angles. Here it comes into play that the X-Zylo generates less lift in the $z$-direction for a steeper climb, therefore the AoA to support the weight has to be higher. This increases the equilibrium AoA and therefore the drag on the ring during the whole equilibrium phase. This can be seen as the velocity angle in the equilibrium phase is higher for higher launch angles, which directly translates to the AoA as the ring axis vector is virtually constant during the first two flight phases and only changes in the third phase. The transition phase leading to the second drop is seen to be longer for small launch angles, with the second drop phase being longer for higher launch angles as the ring gained more height during flight. The equilibrium phase independently of the launch angle shows an almost uniform duration. A notable detail in figure \ref{fig:AoAComparison_vz} is that in all launches, the asymptotic behavior of the velocity angle for $t\rightarrow 0$ again matches that of a parabolic trajectory. The deviation seen for the first frames of the high launch angle cases is also small enough to still be in the given uncertainty bounds. Therefore, the preimposed condition of an \ang{0} AoA is approximately met.
	
\begin{wrapfigure}{l}{0.55\textwidth}
		\vspace{-5pt}
		%\centering
		\includegraphics[width=0.48\textwidth]{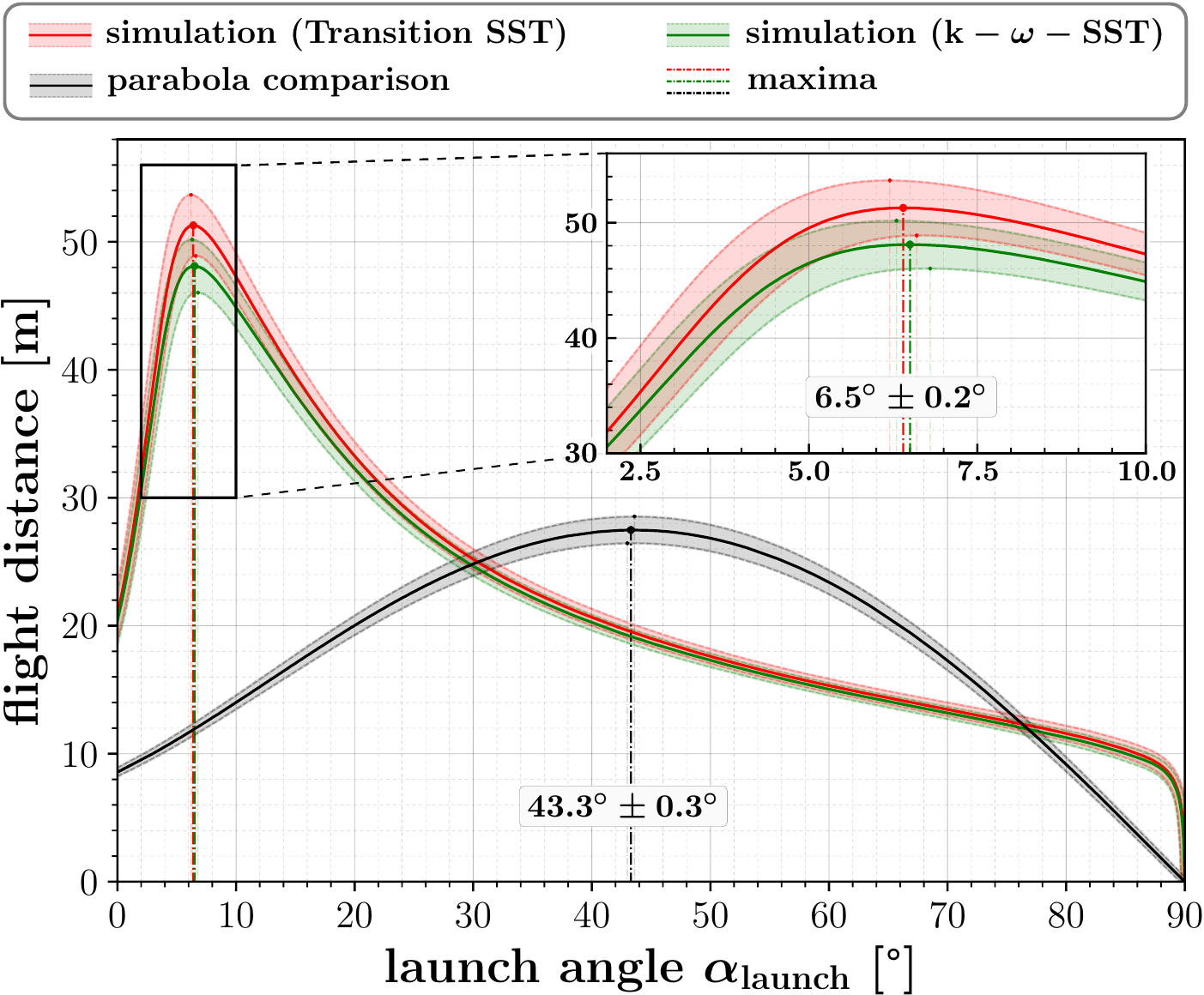}
		\vspace{-4pt}
		\centering
		\captionsetup{margin=0.3cm}
		\captionsetup{justification=justified}
		\caption{Parameter analysis of the flight distance in respect to the launch angle for both turbulence models and a traditional parabolic trajectory.}
		\label{fig:LaunchAngle_ParameterAnalysis}
		\vspace{-20pt}
	\end{wrapfigure}

	Figure \ref{fig:LaunchAngle_ParameterAnalysis} shows a parameter analysis of the flight distance of an X-Zylo with set initial parameters and variable launch angle. The flight distance with respect to the launch angle is plotted for launch angles in the range [\ang{0}, \ang{90}]. For comparison the curve for a parabolic trajectory is shown, as well as the error margins for the flight distances. It is visible that instead of the flight distance rising with higher launch angle, a small launch angle is favorable. 
	
	\vspace{-10pt}

	\noindent
	The initial values used for the parameter analysis were set to be
	\begin{align*}
		v_\text{launch}&=\SI{16\pm0.3}{m/s}\;,\\
		\omega(t_0)/(2\pi)&=\SI{50\pm3}{rad/s}\;,\\
		h(t_0)&=\SI{1.40\pm0.01}{m}\;.
	\end{align*} 
	The uncertainty for the launch angle was set to $\pm\:$\ang{0.4}. All other parameters were set as if the colored X-Zylo was used. As seen in figure \ref{fig:AoAComparison_xz}, the flight distance decreases for larger launch angles, however the maximum is predicted to be at a still smaller launch angle of \ang{6.5}$\:\pm\:$\ang{0.2}. Unfortunately no launch angles smaller than \ang{8} were experimentally recorded during the final test run, therefore no comparison can be made for those cases. The parabolic trajectory shows a deviation from the expected optimal launch angle of \ang{45} due to the height offset at launch. It can be seen that the X-Zylo operates well only in a small launch angle interval of about [\ang{3},\ang{15}]. It is also seen that the flight distance is much more sensitive to launch angles lower than the optimal value than to higher angles. This is due to two effects. On one hand a higher launch angle yields a smaller equilibrium AoA, decreasing drag over an elongated flight period. However, for small launch angles the ring cannot cover a long flight distance as it looses height during the first drop. This leads to a trajectory that curves to the ground early in flight and therefore cannot fully utilize the equilibrium phase. The optimal trade-off strongly varies depending on the ring mass, the launch height and the initial velocity magnitude. As an in depth parameter analysis would go beyond the scope of this paper, this is not covered in more detail.   
	
	\subsection{Sideways Drift}
	\label{sec:SidewaysDrift}
	
\begin{wrapfigure}{l}{0.6\textwidth}
		\vspace{-5pt}
		\centering
		\includegraphics[width=0.55\textwidth]{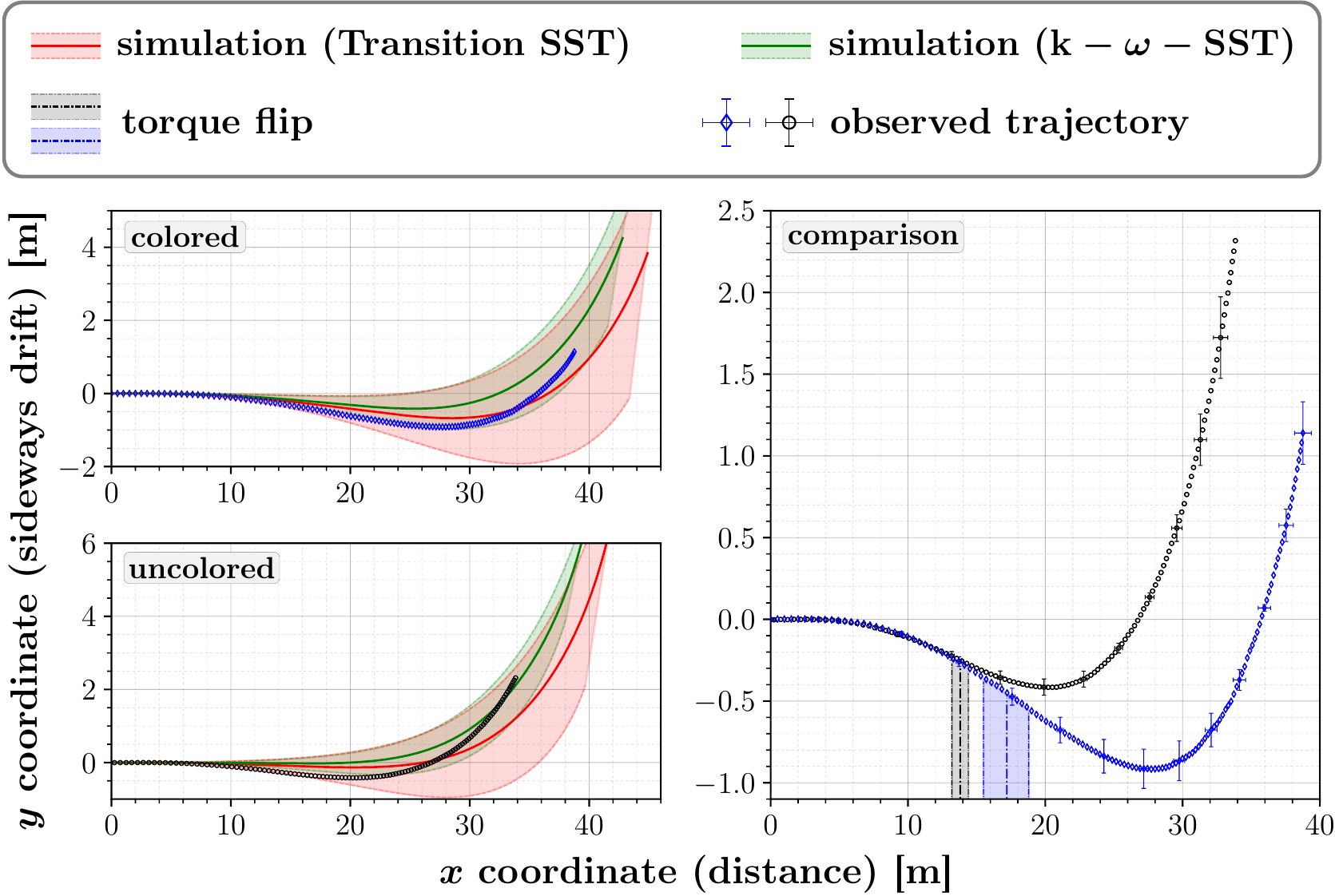}
		\vspace{-4pt}
		\centering
		\captionsetup{justification=justified}
		\captionsetup{margin=0.3cm}
		\caption{Sideways drift behavior for the flight of a colored X-Zylo compared with a launch of an unaltered X-Zylo.}
		\label{fig:ComparisonColorDrift}
		\vspace{-10pt}
	\end{wrapfigure}

	The sideways drift is heavily dependent on the location of the COM. This is shown in figure \ref{fig:ComparisonColorDrift}, where two different X-Zylo launches are shown. The initial conditions for both launches only differ slightly ($\Delta v_\text{launch}=2.8\%$, $\Delta \alpha_\text{launch}=1.4\%$, $\Delta \omega(t_0)=13.5\%$), however one X-Zylo is colored uniformly while the second one is unaltered. As stated in section \ref{sec:PrepMeasurements}, coloring the X-Zylo shifted the  
	\noindent
	COM by $6\%$ downstream, which has a big  influence on the observed flight pattern. The unaltered ring turns significantly earlier, which is an effect of the COM being further upstream. As predicted by the CFD data shown in figure \ref{fig:results}, the COP moves upstream with increasing AoA. The positions in flight where a change in $y$-acceleration is seen are marked with the respective uncertainty. These are the positions where the COP moves past the COM and the torque on the ring flips. An earlier torque flip for the uncolored X-Zylo is seen as would be expected. 
	\noindent
	As can be seen in figure \ref{fig:LaunchComparison_tot}, the flight is not very sensitive for the rotational velocity of the ring, only the drift amplitude shows an almost linear dependence on the initial rotational frequency. It is seen that the location of the COM is a much more sensitive parameter.

	\subsection{Human Induced Launch}
	\label{sec:HumanInducedLaunch}
	
\begin{figure}[H]
		\includegraphics[width=\textwidth]{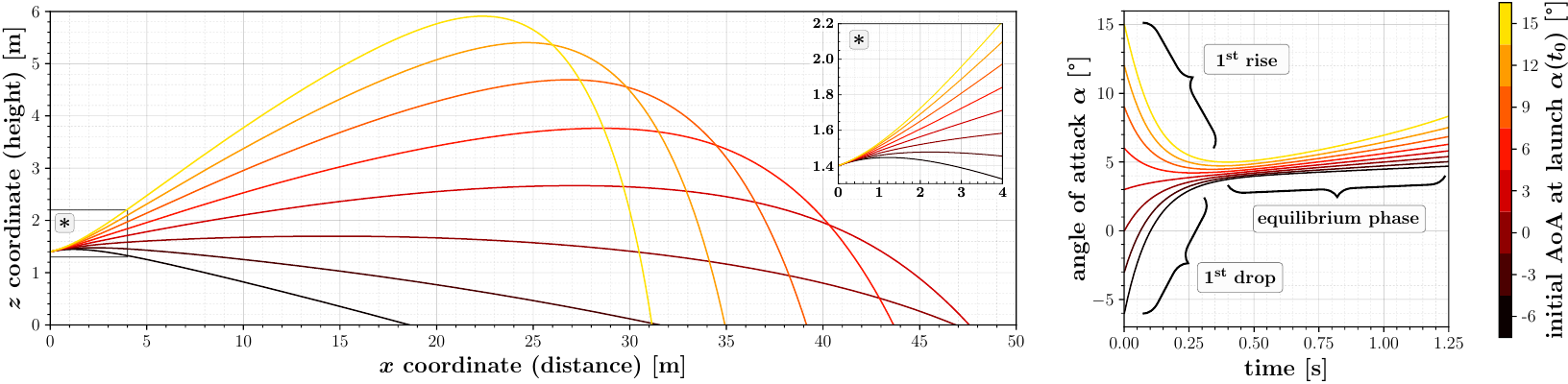}
		\captionsetup{justification=justified}
		\caption{Comparison of trajectories with different initial AoA. A \ang{0} initial AoA is the core assumption of all former calculations. However, when the ring is launched by a human, this assumption must not hold, causing different phenomena to occur.}
		\label{fig:HumanInducedLaunch}
	\end{figure}

	A core assumption made in all theoretical calculations is that the initial AoA of the ring is \ang{0}. However, it is seen in the work by Hirata, et. al. \parencite{HirataXZylo} that this assumption does not hold for a launch by a human. To evaluate how the trajectory and the AoA behave for different initial parameters, figure \ref{fig:HumanInducedLaunch} shows trajectories for an initial launch angle of \ang{5} and an initial velocity magnitude of \SI{16}{m/s}. The initial AoA is varied from \ang{-6} to \ang{15} in discrete steps and the resulting $xz$-projection of the theoretical trajectory shown. Additionally, the AoA is shown as it covers the early flight in more detail. 
	
	It is seen that the trajectory of the ring bends upwards early in flight when the initial AoA is very high. This can be explained using the model developed in section \ref{sec:EarlyFlight}. As the AoA at launch is higher than the equilibrium AoA, this produces a larger lift force than the opposite gravitational pull and therefore the ring is accelerated upward. This then results in a smaller AoA and returns the ring to the equilibrium AoA over time. The first flight phase is therefore seen to be a first rise of the ring instead of a drop. It is visible that the ring rapidly moves towards an equilibrium AoA no matter the initial AoA. From there on the flight is normal again, only the first flight phase is altered. However, it can be seen that even a change in initial AoA of \ang{3} gives tremendously different results in the trajectory. Moreover, one can observe that the equilibrium AoA is different for the different initial AoA values. This effect is due to the different velocity magnitudes the ring has when encountering the equilibrium. As the drag is approximately rising quadratically for higher AoA, the trajectory with an initial AoA of \ang{15} shows a higher equilibrium AoA as the ring lost more speed early on in the first rise. Additionally, with the trajectory bending upward, again the lift does contribute less towards maintaining a straight flight as the $z$-component is diminished. As already stated in section \ref{sec:LaunchAngleInfluence}, it is again seen that the first flight phase before an equilibrium is achieved lasts approximately $\SI{0.25}{s}$ no matter the initial AoA or the launch angle. Similar to a pendulum a stronger deviation from the equilibrium position results in a greater restoring force. 
	
\begin{wrapfigure}{l}{0.6\textwidth}
		\vspace{-5pt}
		\centering
		\includegraphics[width=0.55\textwidth]{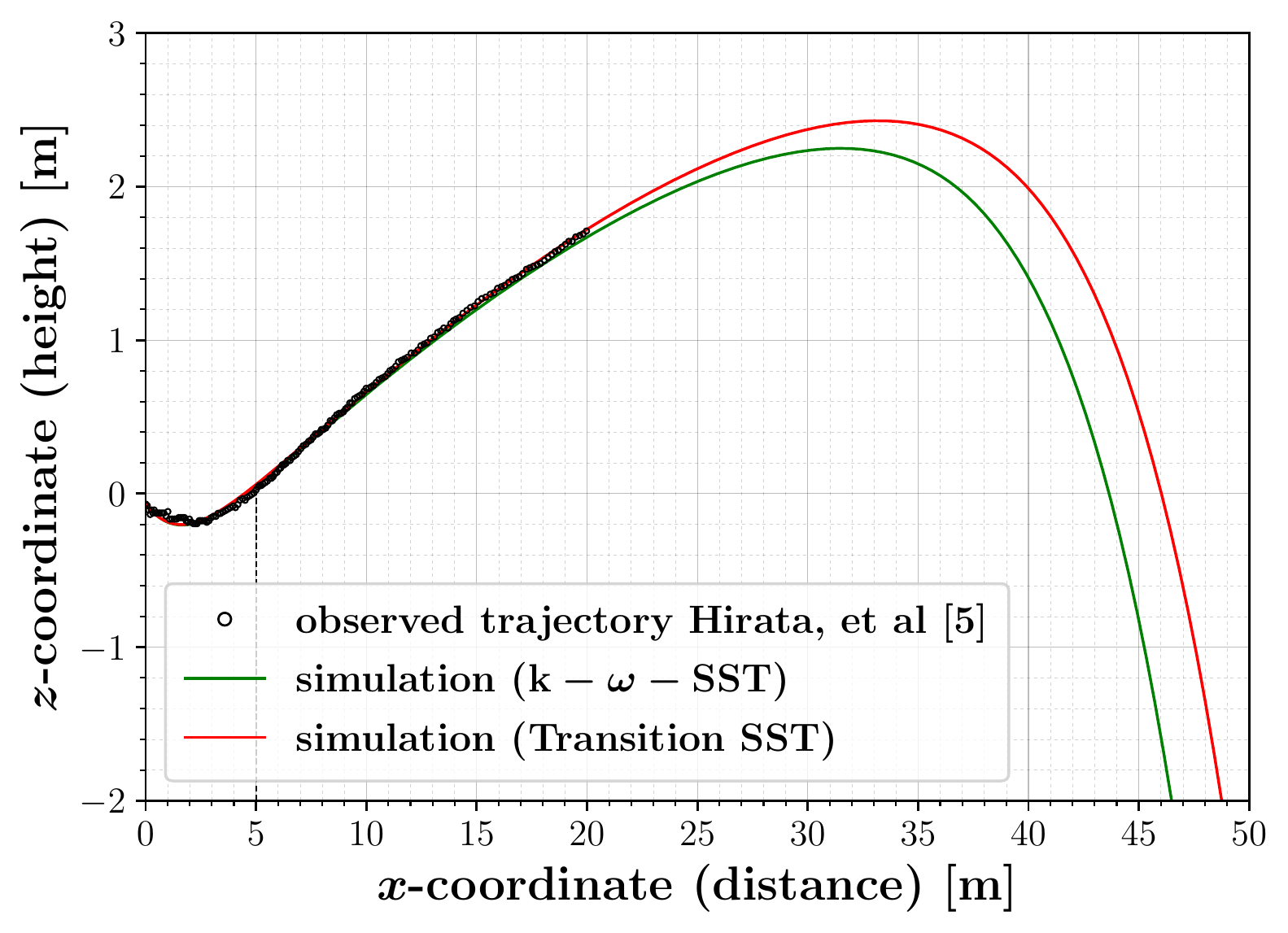}
		\vspace{-4pt}
		\centering
		\captionsetup{margin=0.3cm}
		\captionsetup{justification=justified}
		\caption{Launch by a human recorded by Hirata, et. al. \parencite{HirataXZylo}. The launch can be modeled with a high initial AoA of approximately \ang{21} and a negative launch angle of about $-\ang{10}$.}
		\label{fig:HirataLaunch}
		\vspace{-12pt}
	\end{wrapfigure}

	A real human-induced launch is visible in figure \ref{fig:HirataLaunch}, carried out by Hirata, et. al. \parencite{HirataXZylo}. It is visible that the flight behavior is remarkably different from the observed behavior with the launch
	\noindent
	\noindent
	device. The X-Zylo rises after launch before encountering the equilibrium phase, just as seen in figure \ref{fig:HumanInducedLaunch}. When trying to reconstruct the found trajectory one finds good agreement for a negative launch angle of $-10^\circ$ as well as for a high initial AoA of about $21^\circ$. Those values---even if only approximate---show the significance of a launch device, with which the initial AoA can be controlled. As seen in figure \ref{fig:HumanInducedLaunch}, an initial AoA of more than \ang{10} would lead to an extremely fast rise of the X-Zylo. Because of that, experimentally evaluating the initial parameters would become impossible, as the velocity vector used to find the launch angle changes drastically in the first $\SI{0.1}{s}$ of flight. Additionally, the initial AoA of the ring itself is hard to determine. As significant initial parameters cannot be determined accurately, a quantitative comparison between theory and experiment is unlikely to be of good quality. One could only use a maximum likelihood analysis to determine the most suitable initial parameters, however without knowing the experimental parameters, a comparison is still not possible.
	
	It has to be emphasized that as in all launches conducted using the dedicated launch device, the asymptotic behavior for $t\rightarrow 0$ was approximately that of a parabolic trajectory. However, the implications of a non-zero initial AoA as seen in figure \ref{fig:HumanInducedLaunch} would be strongly visible. This shows that the condition of an initial \ang{0} AoA is suitably met with the launch apparatus, as even small deviations would result in a different asymptotic behavior. 

	\subsection{Open Issues}
	\label{sec:OpenIssues}
	Several problems still prevail, which have to be considered for further analysis of the flight. A non-comprehensive list with some open issues is described in the following:
	
	\begin{itemize}
		\item[i)] \textbf{CFD Data:} The CFD Data is seen to be the weak point of the theoretical model. Even though the model was validated and even rotational effects were taken into consideration, still a deviation can be seen especially for the high AoA flight. Here more sophisticated methods as for example LES simulations would be needed for better coverage of the effects. Also in contrary to what was thought before, the COP is a very sensible parameter and has to be known with high precision. Therefore, the medium \textbf{M2} mesh with its $3.5\%$ (k-$\omega$-SST) or $6.9\%$ (Transition SST) error from the finest mesh is not accurate enough. A better resolution here would be needed for this parameter as well as an additional validation with experimental COP data. Moreover, the assumption that $C_\text{D}$ and $C_\text{L}$ do not depend on the flow velocity has to be investigated. As was seen in figure \hyperref[fig:allResultPlots]{\ref*{fig:allResultPlots}f}, the velocity magnitude during flight ranges from \SI{6}{m/s} to \SI{16}{m/s}. In contrast, the CFD analysis was performed for a flow velocity of \SI{17.15}{m/s}, which could yield more slight deviations that add up over time. Also, the influence of the model's rotation was only calculated for an AoA of \ang{5}, a different behavior could be present for other AoA.
		
		\item[ii)] \textbf{Launch Device:} To alter the launch device for higher reproducibility would be far fetched, only the belted motor drive mechanism could be improved easily to retain a more stable rotational frequency. However, as the initial values are calculated for every launch individually, reproducibility is not seen as a key factor needed for better experiments. It is more important to satisfy the initial \ang{0} AoA condition needed to make good predictions. In former iterations of the launch mechanism, this condition was not met and strong effects as seen in figure \ref{fig:HumanInducedLaunch} were visible. Those issues were resolved, however it is not known what the deviation from the condition at this final point really is. As it is not feasible to observe an initial AoA of less than \ang{1}, the only possibility would be to further improve the device to ensure the condition is met accordingly all the time. 
		
		\item[iii)] \textbf{Tracking the Trajectory:} As five different cameras were used with non-identical lenses and intern software it is impractical to correct the trajectory reliable. The used corrections as stated in \ref{sec:ExperimentalCorrections} only concern geometrical corrections, camera induced errors were neglected. At best one would use two synchronized high resolution cameras that observe the trajectory. With additional markers on the X-Zylo one would then be able to use more advanced tracking software with computer vision algorithms to probe the flight accurately. This would then also ensure reproducibility as a set algorithm performs the tracking and correction procedure rather than a human tracking frame by frame manually. However, this would need a dedicated camera setup currently not available to the author. Future improvements in LiDAR (\textbf{Li}ght \textbf{D}etection \textbf{A}nd \textbf{R}anging) technology could also be a suitable tracking alternative, bypassing some imaging errors one encounters with cameras.
		
		\item[iv)] \textbf{Initial flight parameters:} The initial flight parameters as well as their uncertainty are key factors for the theoretical prediction of the flight behavior and are seen as the biggest error source involved. Using close-up cameras, only the rotational frequency is reliably accessible; the initial launch velocity and the launch angle can only be estimated roughly. Several factors as for example the close proximity of the cameras to the trajectory and tracking the object manually hinder this process. If one tracks the X-Zylo frame by frame in the slow motion footage, the time resolution is very high. However, as the distance traveled is only some pixels wide, the velocity deviates much and shows a discrete spectrum. If however only every tenth frame is probed, the time resolution is to small to observe the first drop in its full extent since the initial quantities change rapidly after launch. For example in the experiments the initial launch angle changed about \ang{1.5} in the first $\SI{0.1}{s}$ aloft (see figure \ref{fig:AoAComparison_vz}). Additionally, camera distortions and other geometric considerations mentioned in section \ref{sec:ExperimentalCorrections} due to the proximity of the camera further increase the uncertainty on the gained values. Therefore, it is necessary to either use a dedicated high-speed camera setup for the launch or to use alternate means of probing the initial values. One could think of several light barriers in close distance to the launch origin to probe the velocity reliably. The launch angle could theoretically be measured using the central guiding rod of the launch mechanism. For now this was not done as the stability of the device does not allow for accurate measurements using the apparatus itself.
 		
	 	\item[v)] \textbf{Incomplete Theoretical Model:} The theoretical model is known to be incomplete as minor effects were not accounted for, for example a small Magnus force when drifting sideways, therefore having a small velocity component perpendicular to the ring's axis. It was also preimposed that the COP is always located on the ring's symmetry axis, which does not have to be true. As those additional effects should have insignificant influence on the flight, they were neglected. However, it is not certain if for different flight scenarios those effects become important, e.g. for crosswinds outside. There could also be additional forces and torques at play that were not considered.  
	\end{itemize}

	%The used comparison between experiment and theory is straightforward with the initial parameters being inserted into the theoretical model. A second possibility however would be to use a maximum likelihood analysis to fit the model into the observed trajectory. This yields the most likely initial parameters predicted by the theoretical model, which could then be compared to the observed parameters. One could then gain insight on the error of the observation. However, it is unknown whether this technique could yield good results as not only the observation of the initial parameters but also CFD data is still inflicted with greater error than expected.
%------------------------------------------------------------------

\section{Conclusion}
It was found that the conceptual explanation of the flight as well as the theoretical simulation can predict and explain the observed flight behavior in great detail. Only minor details like the plateauing $y$-velocity during late flight are not yet fully understood. However, it is not unlikely that those remaining anomalies are an effect of the imperfect camera setup used. The flight of an X-Zylo was seen to be very sensitive to several initial parameters, especially the initial launch angle, the initial AoA, and the location of the COM. Particularly the initial AoA, which can be kept close to \ang{0} with a launch device, makes a quantitative analysis of the whole flight almost impossible when launching the X-Zylo by hand. Therefore, even without good reproducibility, the launch mechanism still is a key part of the experiment as it allows to neglect the impact of a critical parameter. 

From a quantitative point of view the flight was found to be modeled sufficiently precise, however several areas have to be improved upon. At first the second drop cannot be captured in its full extent as the flow separation strongly impacts the second half of the flight. It is seen that the CFD data can predict the moment of separation correctly, however the impact of the separation is underestimated. Therefore, most simulations show a longer flight distance and a less significant drop at the end of flight compared to the observed data. Another observation is that the sideways drift is stronger than predicted, therefore the COM and COP locations have to be calculated more accurately. Another weak point is the used camera setup. A more dedicated setup with additional calibration points and an automated tracking procedure would be needed to advance the experimental measurements. Especially during early flight it is hard to capture the velocity and the initial launch angle with great detail, as both values change rapidly in the first $\SI{0.2}{s}$ of flight.

All in all it is expected that advancements in the CFD data and the camera setup would entail a remarkable accordance between theory and experiment. While a qualitative parameter analysis can indeed be made with the given setup (see figure \ref{fig:LaunchAngle_ParameterAnalysis}), a detailed quantitative analysis can not yet be achieved. Therefore, the goal of this work was only partially achieved with further work being necessary.
\subsection*{Acknowledgments}
I want to thank the whole SciComp research group at the TU Kaiserslautern, especially Ole Burghardt and Tim Albring, for their correspondence and 
indispensable help to start this project. Additionally, I would like to thank CADFEM as well as Dr. Thomas Frank and the LRZ for providing the license for the ANSYS software and access to High Performance Computing resources. Furthermore, advice given by Prof. Dr. Nicolas Gauger and Prof. Dr. Christian Breitsamter regarding the publication of this work was greatly appreciated. I wish to extend my special thanks to Hermann Steffen and Christopher Reinbold for their insightful comments on earlier versions of this manuscript.
\vspace{0.4cm}

%------------------------------------------------------------------

\section{References}
\printbibliography[heading=none]
\vspace{0.4cm}

\section{Appendix}

\subsection{Error Approximation for the Simulation}
\label{sec:error_calculation}

An important step in the trajectory simulation is calculating an uncertainty corridor in which the X-Zylo is predicted, using the given uncertainties measured in the experiment. Therefore, a rough error approximation is necessary, as all the initial parameters for the launched X-Zylo as well as the drag and lift data are afflicted with error.

The trajectory is calculated using a function $f$, which inputs are all the initial parameters $\{a, b, c, ...\}$ and which output is the trajectory (and auxiliary information):
\begin{equation*}
\vec{P}_\text{com}(x(t), y(t), z(t))=f(a, b, c, ...)\; .
\end{equation*}
\noindent
Let us consider an initial uncertainty $\Delta a$ for the parameter $a$. The used approach is to call $f$ three times, once with the parameter $a$ and then twice with parameters $a\pm\Delta a$. This yields three different trajectories:
\begin{align*}
\Big[\vec{P}_\text{com}\Big]_0&=f(a, b, c, ...)\; ,\\
\Big[\vec{P}_\text{com}\Big]_{\pm\Delta a}&=f(a\pm\Delta a, b, c, ...)\; .
\end{align*}

This is done for every parameter with an inscribed uncertainty while the time step $\Delta t$ is kept constant so that one ends up with many trajectories $\big[\vec{P}_\text{com}\big]_0, \big[\vec{P}_\text{com}\big]_{\pm\Delta a}, \big[\vec{P}_\text{com}\big]_{\pm\Delta b},\ldots$ . 
Therefore, it is now possible to calculate a vectorial difference between the baseline trajectory $\big[\vec{P}_\text{com}\big]_0$ and the ones induced with error for each time step $t_\text{n}$
\begin{equation*}
\Big[\Delta \vec{P}_\text{com}(t_n)\Big]_{\pm\Delta a} =\Big[\vec{P}_\text{com}(t_n)\Big]_{\pm\Delta a}-\Big[\vec{P}_\text{com}(t_n)\Big]_0=\begin{pmatrix} \Delta x(t_n)\\\Delta y(t_n)\\ \Delta z(t_n) \end{pmatrix}_{\pm\Delta a}\; ,
\end{equation*}
which is also calculated for every other trajectory, labeled with corresponding subscripts. Now the approximate maximum uncertainty at time $t_\text{n}$---assuming independent uncertainties for the initial parameters---for e.g. the $x$-coordinate is calculated to be: 
\begin{equation*}
%\Big[\Delta x(t_n)\Big]_{\pm\text{tot}} = \sqrt{\Big(\big[\Delta x(t_n)\big]_{\pm\Delta a}\Big)^2+\Big(\big[\Delta x(t_n)\big]_{\pm\Delta b}\Big)^2 +\Big(\big[\Delta x(t_n)\big]_{\pm\Delta c}\Big)^2+ ... }=\sqrt{\sum_{i\in \{a,b,c, ...\}}{\Big(\big[\Delta x(t_n)\big]_{\pm\Delta i}\Big)^2} }>0\;.
\Big[\Delta x(t_n)\Big]_{\pm\text{tot}} = \sqrt{\Big(\big[\Delta x(t_n)\big]_{\pm\Delta a}\Big)^2+\Big(\big[\Delta x(t_n)\big]_{\pm\Delta b}\Big)^2 + ... }=\sqrt{\mathlarger{\mathlarger{\sum}}_{i\in \{a,b,c, ...\}}{\Big(\big[\Delta x(t_n)\big]_{\pm\Delta i}\Big)^2} }>0\;.
\end{equation*}

To not mix between trajectories over- or undershooting the baseline trajectory $\big[\vec{P}_\text{com}\big]_0$, it is checked that all differences contributing to the errors $\Delta_{\pm \text{tot}}$ are of the same sign. Consequently
\begin{wrapfigure}[16]{r}{0.5\textwidth}
	\vspace{-5pt}
	\centering
	\includegraphics[width=0.47\textwidth]{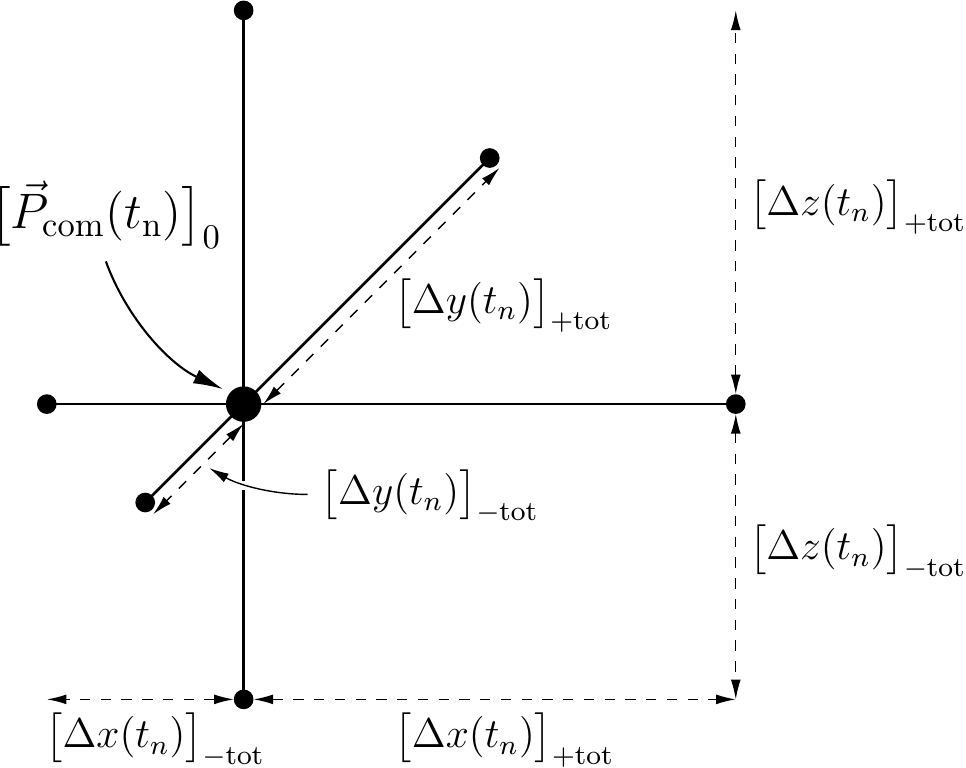}
	\vspace{-5pt}
	\centering
	\captionsetup{margin=0.3cm}
	\caption{Visualization of the uncertainty approximation at time $t_\text{n}$.}
	\label{fig:error_approx}
	\vspace{3pt}
\end{wrapfigure} 
$x(t_n) \pm\big[\Delta x(t_n)\big]_{\pm\text{tot}}$ are upper and lower bounds for the $x$-coordinate of the X-Zylo at time $t_\text{n}$. Figure \ref{fig:error_approx} shows all six limiting points that get calculated and in which bounds the X-Zylo is most likely to be found during flight. One could now lay a generalized ellipsoid through those points, however for the sake of simplicity the bounding box around those points is used as the approximation. In the 2D case of looking at the trajectory in the $xz$-plane, the upper and lower trajectory limits will be the points 
\small
\begin{align*}
\Big[\vec{P}_\text{com}(t_n)\Big]_\text{upper}&=\Big[\vec{P}_\text{com}(t_n)\Big]_{0} + \Big[\Delta \vec{P}_\text{com}(t_n)\Big]_{+\text{tot}}\;,\\
\Big[\vec{P}_\text{com}(t_n)\Big]_\text{lower}&=\Big[\vec{P}_\text{com}(t_n)\Big]_{0} - \Big[\Delta \vec{P}_\text{com}(t_n)\Big]_{-\text{tot}}\; 
\end{align*}
\normalsize
as they are the corners of the bounding box. 

This approach can be generalized beyond the trajectory to other parameters that are outputted by the simulation, however those cases are easier due to them being plotted against time rather then other parameters also inflicted with error.

%$$\bigg(\big[\Delta x(t_n)\big]_{+\text{tot}}-\big[\Delta x(t_n)\big]_{-\text{tot}}\bigg) \times \bigg(\big[\Delta y(t_n)\big]_{+\text{tot}}-\big[\Delta y(t_n)\big]_{-\text{tot}}\bigg) \times \bigg(\big[\Delta z(t_n)\big]_{+\text{tot}}-\big[\Delta z(t_n)\big]_{-\text{tot}}\bigg)\;.$$

\subsection{Error Approximation for the experimental Trajectory Observation}
\label{sec:error_trajectory}

Not only the simulation is inflicted with error, also the usage of cameras to observe the flight's trajectory gives erroneous information. In section \ref{sec:ExperimentalCorrections}, it is touched upon how the observed trajectory is corrected and which imaging errors are considered. However, as some camera induced errors were left out of the calculation and the applied corrections still contain small uncertainties, the influence of those effects has to be approximated. The calculation is only shown for the X-Zylo's position measurements, the velocity errors were calculated similarly. Several error sources for the location of the X-Zylo are seen as independent, each afflicted with a certain error:

\begin{itemize}
	\item[i)] \textbf{Uncertainty while tracking:} This error is purely human-induced. As the X-Zylo is only seen as a small black blob in the tracking software, it is hard to accurately track the ring. Especially when the background changes color---e.g. when the ring flies in front of a basketball net--- the position is only vaguely visible. In addition, manual tracking is not perfect and not the same spot is tracked every frame, which further increases the uncertainty. Still this error is relatively small compared to other uncertainties. All in all a \SI{1}{cm} uncertainty is added due to this tracking procedure for all coordinates.
	
	\item[ii)] \textbf{Barrel distortion:} Even though both GoPro's shoot in linear mode, a very small barrel distortion is still visible which affects the calibration scale as well as the observed position of the X-Zylo. The error on the scale is discussed further in point iii), only the error on the tracked position is approximated here. 
	
	As the barrel distortion effects increase towards the edges of the camera's field of view, the added uncertainty depends on the difference between the position of the camera $x_\text{camera}$ and the tracked object $x_\text{correct}$. A linear error term is then added 
	\begin{equation}
	\delta x_\text{distortion}=\left\lvert x_\text{correct}-x_\text{camera}\right\rvert \cdot R_\text{distortion}
	\end{equation}
	with the error rate $R_\text{distortion}$. This rate has to be approximated for every camera used as well as every coordinate due to the barrel distortion acting differently for different coordinates. The approximated rates can be read from table \ref{tab:ErrorRatesR}.
	
	\item[iii)] \textbf{Scale error:} As already mentioned in point ii), the small barrel distortion adds an uncertainty to the calibration scale used. From measuring each calibration segment it was found that the uncertainty is under $1\%$ for the GoPro8 and about $4\%$ for the GoPro4. However, it was also seen that the larger error of the GoPro4 does not stem only from the scale error. Only about $1\%$ again is from the scale error, the rest is due to error v). Using this scale uncertainty an additional linear error term
	\begin{equation}
	\delta x_\text{scale}= x_\text{correct}\cdot R_\text{scale}
	\end{equation}
	is added to the whole uncertainty. The scale error $R_\text{scale}$ is dependent on the camera itself and on the rotation of the camera.
	
	\item[iv)] \textbf{Error due to X-Zylo's drift:} As can be seen in equation \eqref{eq:x_correction} the $x$-correction involves---additionally to the observed X-Zylo position---the different camera locations seen in figure \ref{fig:camera_setup_gym}. As those were measured beforehand, even a large measurement error of about \SI{0.1}{m} would add a negligible error, however the value $\Delta d$ is inflicted with greater error as it has to be taken from the measurement of another camera (S7). As no dedicated equipment was used and both cameras were not synchronized, an error of $10\%$ for $\Delta d$ is used.
	
	\item[v)] \textbf{Camera rotation:} It has a big influence if the camera is rotated slightly and is not set perfectly perpendicular to the flight path. This also influences the scale, again due to the intercept theorem. Therefore, if it is visible in the footage that the camera is turned slightly, the scale error factor $R_\text{rotation}$ has to be added to the value $R_\text{scale}$ as seen in iii). 
	
\end{itemize}

\noindent
%All error rates have to be approximated for both GoPros as well as the S7 as those all capture the trajectory itself. The values used for the $3^\text{rd}$ launch, which is studied in detail in section \ref{sec:DetailedResultsShot3}, are listed in table \ref{tab:ErrorRatesR}. 
Summarized we get a total uncertainty on the position of the X-Zylo, e.g. the $x$-coordinate is afflicted with an approximate error of
\small
\begin{align}
	\delta x_\text{total}&=\delta x_\text{tracking}+\delta x_\text{distortion}+\delta x_\text{drift}+\delta x_\text{scale}+\delta x_\text{rotation} \\ 
	&\;\resizebox{.92\hsize}{!}{$=\SI{1}{cm}+\left\lvert x_\text{correct}-x_\text{camera}\right\rvert \cdot R_\text{distortion}+\left\lvert x_\text{correct}-x_\text{camera}\right\rvert\cdot\frac{\lvert\Delta d\rvert}{d_\text{camera}}\cdot (0.1+R_\text{rotation})+x_\text{correct}\cdot R_\text{scale}$}\, . \nonumber 
	\label{eq:ErrorPosition}
\end{align}
\normalsize

Note that all errors and free parameters are only estimated based on the footage to make an educated approximation of the real uncertainty. A better consideration of the uncertainties is needed. 
\begin{table}[h]
	\centering
	\begin{tabular}{l||c|c|c||c|c|c||c|c|c}
		\thickhline
		 & $R_\text{dist,x}$ & $R_\text{dist,y}$& $R_\text{dist,z}$& $R_\text{scale,x}$ & $R_\text{scale,y}$& $R_\text{scale,z}$& $R_\text{rot,x}$&$R_\text{rot,y}$&$R_\text{rot,z}$ \\ \hline \hline
		GoPro8 &0.005&0.005&$-$&0.010&0.010&$-$&0.05&0.01&$-$\\ \hline
		GoPro4 &0.008&0.008&$-$&0.015&0.015&$-$&0.50&0.10&$-$\\ \hline
		S7 &$-$&$-$&0.005&$-$&$-$&0.015&$-$&$-$& 0.05\\ \thickhline
	\end{tabular}
	\caption{Estimated values for all error rates. ($R_\text{dist}=R_\text{distortion}$ and $R_\text{rot}=R_\text{rotation}$)} 
	\label{tab:ErrorRatesR}
\end{table}

\end{document}